\documentclass[twocolumn]{aastex631}

\usepackage{bmpsize}
\usepackage{units}
\usepackage{comment}
\usepackage{amsmath}
\usepackage{afterpage}
\usepackage{macro_id}

\newcommand{\ISEF}{ISEF International Fellowship}
\newcommand{\NEF}{NASA Einstein Fellow}

\begin{document}

\title{The JADES Transient Survey: Discovery and Classification of Supernovae in the JADES Deep Field}

\author[0000-0002-4781-9078]{Christa DeCoursey}
\affiliation{Steward Observatory, University of Arizona, 933 N. Cherry Ave, Tucson, AZ 85721, USA}

\author[0000-0003-1344-9475]{Eiichi Egami}
\affiliation{Steward Observatory, University of Arizona, 933 N. Cherry Ave, Tucson, AZ 85721, USA}

\author[0000-0002-2361-7201]{Justin D. R. Pierel} 
\altaffiliation{\NEF}
\affiliation{Space Telescope Science Institute, 3700 San Martin Drive, Baltimore, MD 21218, USA}

\author[0000-0002-4622-6617]{Fengwu Sun}
\affiliation{Center for Astrophysics $|$ Harvard \& Smithsonian, 60 Garden St., Cambridge MA 02138 USA}

\author[0000-0002-4410-5387]{Armin Rest}
\affiliation{Space Telescope Science Institute, 3700 San Martin Drive, Baltimore, MD 21218, USA}
\affiliation{Physics and Astronomy Department, Johns Hopkins University, Baltimore, MD 21218, USA}

\author[0000-0003-4263-2228]{David A. Coulter}
\affiliation{Space Telescope Science Institute, 3700 San Martin Drive, Baltimore, MD 21218, USA}

\author[0000-0003-0209-674X]{Michael Engesser}
\affiliation{Space Telescope Science Institute, 3700 San Martin Drive, Baltimore, MD 21218, USA}

\author[0000-0003-2445-3891]{Matthew R. Siebert}
\affiliation{Space Telescope Science Institute, 3700 San Martin Drive, Baltimore, MD 21218, USA}

\author[0000-0003-4565-8239] {Kevin N.\ Hainline}
\affiliation{Steward Observatory, University of Arizona, 933 N. Cherry Ave, Tucson, AZ 85721, USA}

\author[0000-0002-9280-7594] {Benjamin D.\ Johnson}
\affiliation{Center for Astrophysics $|$ Harvard \& Smithsonian, 60 Garden St., Cambridge MA 02138 USA}

\author[0000-0002-8651-9879] {Andrew J.\ Bunker}
\affiliation{Department of Physics, University of Oxford, Denys Wilkinson Building, Keble Road, Oxford OX1 3RH, UK}

\author[0000-0002-1617-8917] {Phillip A. Cargile}
\affiliation{Center for Astrophysics $|$ Harvard \& Smithsonian, 60 Garden St., Cambridge MA 02138 USA}

\author[0000-0003-3458-2275] {Stephane Charlot}
\affiliation{Sorbonne Universit\'e, CNRS, UMR 7095, Institut d'Astrophysique de Paris, 98 bis bd Arago, 75014 Paris, France}

\author[0000-0003-1060-0723]{Wenlei~Chen} 
\affiliation{Department of Physics, Oklahoma State University, 145 Physical Sciences Bldg, Stillwater, OK 74078, USA}

\author[0000-0002-2678-2560] {Mirko Curti}
\affiliation{European Southern Observatory, Karl-Schwarzschild-Strasse 2, 85748 Garching, Germany}

\author[0009-0003-7427-9614] {Shea DeFour-Remy}
\affiliation{Steward Observatory, University of Arizona,
933 N. Cherry Ave, Tucson, AZ 85721, USA}

\author[0000-0002-2929-3121] {Daniel J.\ Eisenstein}
\affiliation{Center for Astrophysics $|$ Harvard \& Smithsonian, 60 Garden St., Cambridge MA 02138 USA}

\author[0000-0003-2238-1572]{Ori D. Fox}
\affiliation{Space Telescope Science Institute, 3700 San Martin Drive, Baltimore, MD 21218, USA}

\author[0000-0003-3703-5154]{Suvi Gezari}
\affiliation{Space Telescope Science Institute, 3700 San Martin Drive, Baltimore, MD 21218, USA}

\author[0000-0001-6395-6702]{Sebastian Gomez}
\affiliation{Space Telescope Science Institute, 3700 San Martin Drive, Baltimore, MD 21218, USA}

\author[0000-0001-5754-4007]{Jacob Jencson} 
\affiliation{Physics and Astronomy Department, Johns Hopkins University, Baltimore, MD 21218, USA}

\author[0000-0002-7593-8584]{Bhavin A.~Joshi} 
\affiliation{Physics and Astronomy Department, Johns Hopkins University, Baltimore, MD 21218, USA}

\author[0009-0005-5102-9668]{Sanvi Khairnar}
\affiliation{Steward Observatory, University of Arizona,
933 N. Cherry Ave, Tucson, AZ 85721, USA}

\author[0000-0002-6221-1829]{Jianwei Lyu}
\affiliation{Steward Observatory, University of Arizona,
933 N. Cherry Ave, Tucson, AZ 85721, USA}

\author[0000-0002-4985-3819] {Roberto Maiolino}
\affiliation{Kavli Institute for Cosmology, University of Cambridge, Madingley Road, Cambridge CB3 0HA, UK}
\affiliation{Cavendish Laboratory, University of Cambridge, 19 JJ Thomson Avenue, Cambridge CB3 0HE, UK}
\affiliation{Department of Physics and Astronomy, University College London, Gower Street, London WC1E 6BT, UK}

\author[0000-0003-1169-1954]{Takashi J. Moriya}
\affiliation{National Astronomical Observatory of Japan, National Institutes of Natural Sciences, 2-21-1 Osawa, Mitaka, Tokyo 181-8588, Japan}
\affiliation{Graduate Institute for Advanced Studies, SOKENDAI, 2-21-1 Osawa, Mitaka, Tokyo 181-8588, Japan}
\affiliation{School of Physics and Astronomy, Monash University, Clayton, Victoria 3800, Australia}

\author[0000-0001-9171-5236]{Robert~M.~Quimby}
\affiliation{Department of Astronomy/Mount Laguna Observatory, San Diego State University, 5500 Campanile Drive, San Diego, CA 92812-1221, USA}
\affiliation{Kavli Institute for the Physics and Mathematics of the Universe (WPI), The University of Tokyo Institutes for Advanced Study, The University of Tokyo, Kashiwa, Chiba 277-8583, Japan}

\author[0000-0003-2303-6519]{George H. Rieke}
\affiliation{Steward Observatory, University of Arizona,
933 N. Cherry Ave, Tucson, AZ 85721, USA}

\author[0000-0002-7893-6170] {Marcia J. Rieke}
\affiliation{Steward Observatory, University of Arizona, 933 N. Cherry Ave, Tucson, AZ 85721, USA}

\author[0000-0002-4271-0364] {Brant Robertson}
\affiliation{Department of Astronomy and Astrophysics, University of California, Santa Cruz, 1156 High Street, Santa Cruz CA 96054, USA}

\author[0000-0002-9301-5302]{Melissa Shahbandeh} 
\affiliation{Space Telescope Science Institute, 3700 San Martin Drive, Baltimore, MD 21218, USA}

\author[0000-0003-2238-1572]{Louis-Gregory Strolger} 
\affiliation{Space Telescope Science Institute, 3700 San Martin Drive, Baltimore, MD 21218, USA}

\author[0000-0002-8224-4505] {Sandro Tacchella}
\affiliation{Kavli Institute for Cosmology, University of Cambridge, Madingley Road, Cambridge CB3 0HA, UK}
\affiliation{Cavendish Laboratory, University of Cambridge, 19 JJ Thomson Avenue, Cambridge CB3 0HE, UK}

\author[0000-0001-5233-6989]{Qinan Wang} 
\affiliation{Physics and Astronomy Department, Johns Hopkins University, Baltimore, MD 21218, USA}

\author[0000-0003-2919-7495] {Christina C. Williams}
\affiliation{NSF's National Optical-Infrared Astronomy Research Laboratory, 950 North Cherry Ave, Tucson, AZ 85719, USA}

\author[0000-0001-9262-9997] {Christopher N.\ A.\ Willmer}
\affiliation{Steward Observatory, University of Arizona, 933 N. Cherry Ave, Tucson, AZ 85721, USA}

\author[0000-0002-4201-7367]{Chris Willott}
\affiliation{NRC Herzberg, 5071 West Saanich Rd, Victoria, BC V9E 2E7, Canada}

\author[0000-0002-0632-8897]{Yossef Zenati}
\altaffiliation{\ISEF}
\affiliation{Physics and Astronomy Department, Johns Hopkins University, Baltimore, MD 21218, USA}
\affiliation{Space Telescope Science Institute, 3700 San Martin Drive, Baltimore, MD 21218, USA}

\correspondingauthor{Christa DeCoursey}
\email{cndecoursey@arizona.edu}

\begin{abstract}
The JWST Advanced Deep Extragalactic Survey (JADES) is a multi-cycle JWST program that has taken among the deepest near-/mid-infrared images to date (down to $\sim$\,30 ABmag) over $\sim$\,25 arcmin$^2$ in the GOODS-S field in two sets of observations with one year of separation. This presented the first opportunity to systematically search for transients, mostly supernovae (SNe), out to $z$\,$>$\,2. We found 79 SNe: 38 at $z$\,$<$\,2, 23 at 2\,$<$\,$z$\,$<$\,3, 8 at 3\,$<$\,$z$\,$<$\,4, 7 at 4\,$<$\,$z$\,$<$\,5, and 3 with undetermined redshifts, where the redshifts are predominantly based on spectroscopic or highly reliable JADES photometric redshifts of the host galaxies. At this depth, the detection rate is $\sim$\,1--2 per arcmin$^2$ per year, demonstrating the power of JWST as a supernova discovery machine. We also conducted multi-band follow-up NIRCam observations of a subset of the SNe to better constrain their light curves and classify their types. Here, we present the survey, sample, search parameters, spectral energy distributions (SEDs), light curves, and classifications.  Even at $z$\,$\geq$\,2, the NIRCam data quality is high enough to allow SN classification via multi-epoch light-curve fitting with confidence.  The multi-epoch SN sample includes a Type\,Ia SN at $z_{\mathrm{spec}}$\,$=$\,2.90, Type\,IIP SN at $z_{\mathrm{spec}}$\,$=$\,3.61, and a Type Ic-BL SN at $z_{\mathrm{spec}}$\,$=$\,2.83. We also found that two $z$\,$\sim$\,16 galaxy candidates from the first imaging epoch were actually transients that faded in the second epoch, illustrating the possibility that moderate/high-redshift SNe could mimic high-redshift dropout galaxies.
\end{abstract}


\section{Introduction} \label{sec:intro}

The high-redshift (1.5\,$\lesssim$\,$z$\,$\lesssim$\,5) transient universe is still a relatively unexplored field of astrophysics due to the vast amount of resources required to discover supernovae (SNe) at $z\gtrsim1$. The Cosmic Assembly Near-Infrared Deep Extragalactic Legacy Survey (CANDELS) was a multi-cycle Treasury Hubble Space Telescope (HST) program consisting of $\sim$\,900 orbits executed over three years \citep{Grogin2011CANDELS:Survey}, and yet only $4$ SNe were discovered at $z$\,$>$\,2, with two spectroscopic and two photometric host redshifts \citep{Rodney2014TypeUniverse}. This $z$\,$>$\,2 sample was classified entirely photometrically, one as Type\,Ia (SN\,Ia; the thermonuclear explosion of a white dwarf) and three as core-collapse Type\,II (CC\,SNe; explosions of dying massive stars of $\gtrsim8 M_\odot$). Additionally, the Cluster Lensing and Supernova Survey with Hubble (CLASH) program performed 524 orbits over 25 cluster fields and parallel fields to search for SNe \citep{Postman2012}. They discovered 27 SNe in the non-lensed parallel fields, including 13 Type\,Ia SNe \citep{Graur2014}. Four of these Type\,Ia SNe were at $z$\,$>$\,1.2, but two of the redshifts were photometric host galaxy redshifts. Their highest redshift SN was $z_{\mathrm{phot}}$\,$=$\,1.68\,$\pm$\,0.15, and their sample was classified fully photometrically. Although this was a significant step forward, it illustrates the challenges of not only finding high-redshift SNe but also classifying them. As a result, there still exist substantial uncertainties associated with using SNe~Ia as high-redshift probes for dark energy \citep{Brout2022TheConstraints} and with using CC\,SN rates at $z$\,$>$\,1.5 to constrain the initial mass function (IMF) and cosmic star formation rate density \citep{Strolger2015TheSurveys}.

While there are many studies of high-redshift transients using ground-based telescopes \citep[]{Cooke2012Superluminous3.90, Moriya2019FirstProperties, Curtin2019FirstProperties, Smith2018StudyingTwo, Pan2017DES15E2mlf:Bang} and using HST \citep[]{Grogin2011CANDELS:Survey, Rodney2014TypeUniverse, Postman2012, Graur2014, OBrien2024}, the landscape of the high-redshift transient universe is changing dramatically because of James Webb Space Telescope (JWST) \citep[e.g.,][]{Frye2023TheG165.7+67.0, Yan2023JWSTsField, Pierel2024LensedGalaxy}. The unprecedented depth and spatial resolution of JWST Near-Infrared Camera (NIRCam) images have increased our efficiency for finding faint and distant transient sources considerably, which allows us to conduct a much more complete census of transients at high redshift. 

The Great Observatories Origins Deep Survey (GOODS; e.g., \citealt{Giavalisco2004TheImaging}) includes the most well-studied blank fields for many telescopes (Hubble, Spitzer, Herschel, etc.), motivating follow-up observations with JWST. During Sep--Oct of 2022 and 2023, the JWST Advanced Deep Extragalactic Survey (JADES; \citealt{Eisenstein2023OverviewJADES}) obtained $\sim$\,230 hours of deep NIRCam images over part of the CANDELS GOODS-South (GOODS-S) field containing the Hubble Ultra Deep Field (HUDF).  This area, called the JADES-Deep field, covers an area of $\sim$\,25 arcmin$^2$, and has been imaged with seven wide-band and two medium-band filters. We used this extensive NIRCam imaging dataset to search for transients by differencing the 2022 and 2023 data, thereby creating the JADES Transient Survey.


Even with the 2022 data alone, it was possible to identify a significant number of transient sources by comparing with the deep HST images from the Hubble Legacy Fields\footnote{https://archive.stsci.edu/prepds/hlf} (HLF). For example, 8 transients were discovered in the GOODS-S field, likely SNe at $z$\,$=$\,0.665--1.764 \citep{DeCoursey2023a}, and 17 were discovered in the GOODS-N field, likely SNe at $z$\,$=$\,0.520--2.325 \citep{DeCoursey2023b}.  With the 2023 data becoming available, the number of transient detections increased to $\sim$\,40 per epoch,  with $z_{\mathrm{median}}$ $\sim$\,2 and containing multiple $z$\,$>$\,3 candidates.  We therefore submitted a Director's Discretionary Time (DDT) proposal (PID 6541; PI Egami; 18 hours) to follow up a subset of these SN candidates with two additional NIRCam imaging epochs and Near-Infrared Spectrograph/Micro-Shutter Assembly (NIRSpec/MSA) spectroscopy.  This paper includes the analysis of these follow-up DDT data, presenting multi-epoch light curves and attempting to classify the SN types via light curve fitting. The spectroscopy results will be presented in E. Egami et al.\ (in preparation).

While the JADES Transient Survey is the deepest systematic transient survey of its kind and represents significant progress in transient science, the overarching motivator for the JADES program is discovering the earliest galaxies. The large number of filters available with the JADES NIRCam data allows excellent sampling of galactic spectral energy distributions (SEDs), producing highly reliable photometric redshifts \citep{Hainline2023TheGOODS-N}. Transient SEDs, however, can mimic high-redshift galaxy SEDs when they have only long-wavelength emission. Therefore, we also describe the importance of multiple imaging epochs in high-redshift galaxy surveys to reduce transient contamination through two examples in our sample.

This paper is organized as follows. In Section \ref{sec:obs}, we describe the multi-epoch NIRCam observations and detail the image processing procedure. Section \ref{sec:detection} outlines the transient/SN selection criteria, provides an analysis of the discovery epoch detection limit, describes PSF photometry measurement methods, and details host-galaxy subtraction. The results, which include an overview of the sample and its various properties, are presented in Section \ref{sec:results}. We also highlight the $z$\,$\geq$\,4 SN candidates as well as two $z$\,$\sim$\,16 galaxy candidates from the first imaging epoch that were actually fading transients. Section \ref{sec:discussion} details our attempt to classify the SNe as Type\,Ia, II, or Ibc via light curve fitting, and briefly lists out the non-SN transients that we discovered. It also analyzes possible implications of an overestimation of high-redshift galaxy abundance due to transient contamination in single-epoch high-redshift galaxy surveys. Lastly, in Section \ref{sec:conclusion}, we summarize the paper and highlight the main takeaways. 

Throughout this paper, we express magnitudes using the AB system \citep{Oke1983SecondarySpectrophotometry.} and adopt a flat $\Lambda$CDM cosmology with the following parameters: H$_0$\,$=$\,70 km s$^{-1}$ Mpc$^{-1}$, $\Omega _{tot}$\,$=$\,1.0, $\Omega_\Lambda$\,$=$\,0.7, and $\Omega_m$\,$=$\,0.3.


\section{Observations and Data Processing} \label{sec:obs}

\subsection{The JADES Deep Survey}


The JADES NIRCam Deep Prime field located in GOODS-S (hereafter JADES Deep) was observed through the JADES program (PID: 1180; PI: Eisenstein). The JADES observing strategy is described fully in \citet{Eisenstein2023OverviewJADES}. We conducted two epochs of 9-band JWST/NIRCam imaging in the JADES Deep Field, separated by one year, with similar observing configurations.
Each epoch included 4 wide-band short-wavelength (SW) photometric filters (F090W, F115W, F150W, F200W), 3 wide-band long-wavelength (LW) photometric filters (F277W, F356W, F444W), and 2 medium-band LW photometric filters (F335M and F410M). 

The first epoch (hereafter Epoch1) was taken UT 2022 September 29 - October 5, and the second epoch (hereafter Epoch2) was taken UT 2023 September 28 - October 3. The Epoch1 and Epoch2 program information is summarized in Table \ref{tab_obslog}. The Epoch1 data have been publicly released and are described in detail by \citet{Rieke2023JADESImaging}.  It covers an area of $\sim$\,25 arcmin$^2$ with an exposure time of at least 3.4 hours/pixel, achieving a 5-$\sigma$ depth of $\sim$\,29.5-30 mag.  The coverage is significantly deeper for areas where the dither/mosaic patterns overlap ($>$\,6.5 hours over at least 17 arcmin$^2$, $>$\,9.7 hours over at least 6.3 arcmin$^2$).  The deepest coverage achieves an exposure time of $\sim$\,17 hours, with the F115W filter over at least 6.7 arcmin$^2$.  With the Epoch2 data, this coverage has doubled. We list the Epoch1 and Epoch2 filters, readout pattern, groups, dithers, and exposure times in Table \ref{tab_obslog2}.

There was an additional single-pointing 9-band JWST/NIRCam observation (Obs 219) taken on UT 2023 November 15 that covered the southern portion of the JADES Deep Field (hereafter referred to as Epoch3). This was the re-execution of the failed Observation 19 of PID 1180, which fortuitously provided an extra epoch for some of the transient sources.  The same 9 NIRCam bands were used as in Epoch1 and Epoch2, but since Obs 219 is part of the NIRCam Medium Prime survey, the depth is shallower. Also, some observing parameters were slightly changed from those of Obs 19.  The Epoch3 program information is listed in Table \ref{tab_obslog} and the Epoch3 observing parameters are listed in Table \ref{tab_obslog2}. 

Similarly, three more single-pointing observations were carried out on 2024 January 1 as the re-execution of the failed Observations 20 and 23, now named Observations 220/222 and 223. Obs220/222, which we refer to as Epoch5.1, uses only a single SW and LW filter -- F200W and F277W. Obs223, which we refer to as Epoch5.3, uses the same 9 NIRCam filters as Epoch1 and Epoch2 but with shallower depths and slightly different observing parameters. The Epoch5.1 and Epoch5.3 program information and observing parameters are listed in Tables \ref{tab_obslog} and \ref{tab_obslog2}, respectively. 

The JADES Deep footprint (Epoch1 and Epoch2) as well as those of Obs219 (Epoch3), Obs220/222 (Epoch5.1), and Obs223 (Epoch5.3) are shown in Figure~\ref{footprint}.

\begin{deluxetable*}{cccccc}[t]
\tablecaption{Log of NIRCam Observations \label{tab_obslog}}   
\tablewidth{0pt}
\tablehead{
\colhead{Epoch} & \colhead{Program} & \colhead{PID} & \colhead{Observation} & \colhead{Date} & \colhead{Ref.}
}
\startdata
1 & JADES Deep   & 1180 & 7, 10, 11, 15, 17, 18 & 2022-09-29 -- 2022-10-05 & 1, 2 \\
2 & JADES Deep   & 1180 & 8, 9, 12, 13, 14, 16  & 2023-09-28 -- 2023-10-03 & 1 \\
3 & JADES Medium & 1180 & 219                   & 2023-11-15 & 1 \\
4 & DDT Follow-up & 6541 & 2, 3                  & 2023-11-28 & This work \\
5.1 & JADES Medium & 1180 & 220, 222         & 2024-01-01 & 1 \\
5.2 & DDT Follow-up & 6541 & 4, 5            & 2024-01-01 & This work \\
5.3 & JADES Medium & 1180 & 223              & 2024-01-01 & 1 \\
\enddata
\tablerefs{(1) \citet{Eisenstein2023OverviewJADES}; (2) \citet{Rieke2023JADESImaging}}
\end{deluxetable*}
\begin{deluxetable*}{cllccc}[t]
\tablecaption{NIRCam Observing Parameters \label{tab_obslog2}}   
\tablewidth{0pt}
\tablehead{
\colhead{Epoch} & \colhead{Filters} & \colhead{Readout Pattern} & \colhead{Groups} & \colhead{Dithers} & \colhead{Exp Time} \\
\colhead{} & \colhead{} & \colhead{} & \colhead{} & \colhead{} & \colhead{(hrs/filter)}
}

\startdata
1 & F090W F115W F150W F200W F277W F335M F356W F410M F444W & DEEP8    & 7 & 9 & 3.4  \\
\hline
2 & F090W F115W F150W F200W F277W F335M F356W F410M F444W & DEEP8    & 7 & 9 & 3.4  \\
\hline
3 & F090W F115W(x2) F150W F335M F356W F410M F444W         & DEEP8    & 5 & 6 & 1.6  \\
3 & F200W F277W                                           & MEDIUM8  & 8 & 6 & 1.4  \\
\hline
4 & F115W F150W F200W F277W F356W F444W                   & SHALLOW4 & 8 & 3 & 0.35 \\
\hline
5.1 & F200W F277W                                         & MEDIUM8  & 8 & 6 & 1.4  \\
\hline
5.2 & F150W F200W(x2) F277W F356W F444W                   & SHALLOW4 & 8 & 3 & 0.35 \\
\hline
5.3 & F090W F115W(x2) F150W F335M F356W F410M F444W       & DEEP8    & 5 & 6 & 1.6  \\
5.3 & F200W F277W                                         & MEDIUM8  & 8 & 6 & 1.4  \\
\enddata
\tablecomments{Filters with (x2) next to them have double the exposure time listed.}
\end{deluxetable*}

\subsection{DDT Follow-Up Observations} \label{subsec:follow_up}

We conducted additional follow-up JWST/NIRCam observations of the JADES Deep Field through DDT Program ID 6541 (PI: Egami). 
The V3 position angle (PA) was selected according to the available scheduling windows.  For each epoch, two pointings, connected with an interruptible sequence, were carefully designed to cover as many high-priority transient candidates as possible (e.g., those at high redshift or exhibiting significant brightening). 

Program 6541 epoch 1, taken on UT 2023 November 28 (hereafter referred to as Epoch4; V3PA\,$=$\,14 deg), used the NIRCam filters F115W, F150W, F200W, F277W, F356W, and F444W with the SHALLOW4 readout pattern with 8 groups and 3 standard subpixel dithers. The exposure time was 21~minutes/filter, achieving a 5-$\sigma$ depth of $\sim$\,27.9--28.6 mag.  To maximize the area of uniform depth, no primary dither was used, but this also leaves gaps between the SW detectors. Therefore, the NIRCam pointings were carefully adjusted to ensure that no high-priority transient targets would fall in these gaps. 

Program 6541 epoch 2, taken on UT 2024 January 1 (hereafter referred to as Epoch5.2; V3PA\,$=$\,45.425 deg), used NIRCam filters F150W, F200W, F277W, F356W, and F444W.  The same exposure parameters were used except for F200W.  Since many of the transients became undetectable in the F115W images of the Epoch4 data, the F115W observation was changed to that of F200W, doubling the exposure time with this filter (42~minutes).  The V3PA was set to that of the accompanying NIRSpec/MSA observation.  Epoch4 and 5.2 are separated by 34 days, which corresponds to $\sim$\,11 days in the rest-frame at $z$\,$=$\,2 and $\sim$\,7 days at $z$\,$=$\,4, allowing a good sampling of the light curves of high-redshift transients. 

The Epoch4 and Epoch5.2 program information and observing parameters are summarized in Tables \ref{tab_obslog} and \ref{tab_obslog2}, respectively, and the footprints of these DDT NIRCam imaging observations are shown in Figure~\ref{footprint}.  As part of DDT program 6541, we also obtained NIRSpec spectra of a subsample of the transients and their hosts. E. Egami et al. (in preparation) will describe these spectra.



\subsection{Data Processing} \label{subsec:processing}

The step-by-step NIRCam data processing procedure for the 2022 JADES Deep data were presented by \citet{Rieke2023JADESImaging}.  The 2023 JADES Deep data, as well as those obtained by the DDT follow-up program, were processed in a similar way.  A detailed description of the NIRCam imaging data reduction process will be presented in S. Tacchella et al. (in preparation).  We note that the final mosaic images were produced with a pixel scale of 0\farcs03/pixel for both SW and LW images, meaning that the pixel scale of the LW images is approximately half of the native value, which is 0\farcs06/pixel. These images are used for transient detections as described in Section \ref{sec:detection}.
However, because LW images are oversampled by a factor of two with respect to the native pixel scale, photometry in LW bands would be heavily affected by correlated noise of pixels. Therefore, we also produced LW mosaic images with a pixel scale of 0\farcs06/pixel for PSF photometry (Section \ref{subsec:psf_photometry}).

We generated difference images by subtracting the Epoch1 images from the Epochs 2-5.3 images in each respective overlapping filter. For the Epoch2-Epoch1 subtraction, the observing setup and observatory PA were very similar (i.e., the PSFs in the mosaics were effectively the same). The Epoch1 and Epoch2 images were projected onto the same pixel grid, so we were simply able to subtract the Epoch1 images from the Epoch2 images to get PSF matched, pixel registered difference images. For the follow-up images (Epochs 3-5.3), the PA was different from the Epoch1 images. So, while the mosaics were still projected onto the same pixel grid, the PSF was rotated in those images with respect to the Epoch1 data. This has deteriorated host-galaxy subtraction, but note that these difference images were used only for photometry and not for transient detection, which utilized the Epoch 1/2 difference images alone. The effect on photometry is limited because it is based on PSF fitting (using the PSF of the corresponding epoch).


\section{Transient Detection} \label{sec:detection}

To detect transients, we searched through the 7 NIRCam wide band Epoch2-Epoch1 difference images. We did not include the medium bands (F335M and F410M) in our transient search to reduce redundancy, as the medium-band images are shallower than the wide-band images at similar wavelength. A set of selection criteria was used to automatically generate the preliminary transient samples. We performed this selection analysis on the Epoch2-Epoch1 difference images as well as the inverted difference images to find both brightening and fading transients, respectively. The criteria aimed to eliminate noise spikes and diffraction spikes that mimicked detections while retaining as many real sources as possible to reduce the visual inspection load.  After some fine-tuning, the adopted criteria were found to produce a robust sample of transients while minimizing the loss of potentially real but marginal detections.
The effects of the selection criteria on the sample completeness will be explored in a subsequent SN rates paper based on the JADES Deep transient sample. We discuss the detection efficiency in Section \ref{sec:efficiency}, but this only analyzes the limiting magnitudes of the difference images, not the effects that the selection criteria have on the final sample.



\subsection{Source Detection} \label{subsec:source_detection}

The first source detection step involved running \texttt{DAOStarFinder} on the 7 wide-band 30mas/pixel SW and LW Epoch2-Epoch1 (Epoch1-Epoch2) difference images to search for brightening (fading) point-like sources \citep[]{Stetson1987DAOPHOT:Photometry, Bradley2024Astropy/photutils:1.12.0}. Transient type (Type\,II SN, Type\,Ia SN, variable AGN, etc.) and phase as well as physical conditions of the transient's environment (e.g., dustiness) affect the transient's wavelength of peak emission, so searching all 7 wide bands allowed us to find the widest variety of transients. Using multi-band detections with both the SW and LW detectors also helped reduce the false event rate. We performed the following selection analysis independently on each filter's difference images, merging the final recovered source catalogs at the end.

The \texttt{fwhm} (full-width half-max) parameter was set to the approximate FWHM of the detection image, which was $\sim$3 pixels (0\farcs09) for SW images and $\sim$4 pixels (0\farcs12) for resampled LW images. 
We estimated the background noise of the difference images as the standard deviation returned from \texttt{astropy} \texttt{sigma\_clipped\_stats} with \texttt{sigma}\,$=$\,3.0. The \texttt{DAOStarFinder} \texttt{threshold} parameter was set as 4 times the background standard deviation for F090W, F115W, F150W, F200W, and F277W, whereas it was set to 3 times the background standard deviation for F356W and F444W. Lower thresholds were required for F356W and F444W to reach similar completeness as with the shorter wavelength filters (see Section \ref{sec:efficiency}). We chose not to constrain the \texttt{sharpness} parameter, as bright transients that are surrounded by subtraction artifacts in the difference images can have a wide variety of \texttt{sharpness} values. The \texttt{roundness} parameter was inclusively constrained between --0.60 and 0.60 for each filter, and the image border (within one PSF size from an edge) was excluded from the search.

After generating the initial sample with \texttt{DAOStarFinder}, the next step involved eliminating sources that appeared in the difference image due to a lack of overlap between the Epoch1 and Epoch2 science images. If either the Epoch1 or Epoch2 science image contained no data within 0\farcs2 of the detected source position, the source was eliminated from the sample.


\subsection{Transient Selection Algorithm} \label{subsec:transient_selection}

After finding sources with \texttt{DAOStarFinder} and removing detections in the non-overlapping regions of Epoch1 and Epoch2, additional sample cuts were based on the sources' signal-to-noise ratios (S/N) in the difference images, with S/N defined as the ratio of measured flux to flux uncertainty. Flux and flux uncertainty were measured via aperture photometry using an $r$\,$=$\,0\farcs1 circular aperture  with a sky annulus of $r$\,$=$\,0\farcs1--0\farcs2. Aperture corrections were derived with the effective point spread functions (ePSFs) built by \citet{Ji2023JADES4.5} and are listed in Table \ref{tab_apercor}. We applied additional corrections to account for local background over-subtraction due to the source flux spilled over into the sky annulus as listed in Table~\ref{tab_apercor}. 
We determined these factors by injecting a mock PSF onto an empty array (i.e., with no background), performing aperture photometry as described above, and calculating the ratio of the measured flux to known flux of the injected mock PSF.



\begin{deluxetable*}{lccccccccc}
\tablecaption{Corrections for Photometry Aperture and Sky Over-Subtraction}
\tablehead{
\colhead{} & \colhead{F090W} & \colhead{F115W} & \colhead{F150W} & \colhead{F200W} & \colhead{F277W} & \colhead{F335M} &  \colhead{F356W} & \colhead{F410M} & \colhead{F444W} 
} 
\label{tab_apercor}
\startdata
Aperture ($r=0.1$\arcsec) & 1.356 & 1.312 & 1.318 & 1.392 & 1.656 & 1.778 & 1.822 & 1.974 & 2.067 \\
Sky over-subtraction ($r_{\rm sky}$\,$=$\,0.1--0.2\arcsec) & 1.024 & 1.025 & 1.025 & 1.028 & 1.087 & 1.095 & 1.097 & 1.093 & 1.126 \\
\enddata
\end{deluxetable*}

We adopted the following criteria to select transient sources:

\begin{enumerate}
    \item S/N\,$\geq$\,5 in at least one band
    \item S/N\,$\geq$\,3 in at least two bands
    \item 7-band S/N\,$\geq$\,14
\end{enumerate}

Transients had to satisfy all three requirements to be selected. 7-band S/N was calculated as the inverse-variance weighted flux average, \^{f}, divided by the square root of the flux variance, Var(\^{f}): 

\begin{equation} \label{ivwfa}
     \widehat{f} = \frac{\sum_{i}\cfrac{f_i}{\sigma_i ^2}}{\sum_{i}\cfrac{1}{\sigma_i ^2}}
\end{equation}

\begin{equation} \label{varf}
    Var(\widehat{f}) = \frac{1}{\sum_{i}\cfrac{1}{\sigma_i^2}}
\end{equation}

\begin{equation}
    \text{7-band S/N} = \frac{\widehat{f}}{\sqrt{Var(\widehat{f})}}
\end{equation}
where $f$ was the source flux, $\sigma$ was the flux uncertainty, and the sum was performed over the 7 NIRCam wide bands. This 7-band inverse variance weighted S/N requirement roughly translates to requiring an average S/N of $\sim$\,5 in each of the 7 bands. However, we placed a lower limit of S/N\,$=$\,1 for each element of the 7-band S/N calculation to remove unphysical S/N values. These selection criteria were carefully designed such that they would recover the majority of transients that were found with a preliminary but thorough difference image visual inspection.

These criteria reduced the potential sample to an order of hundreds of sources for each detection image. Each of these sources was visually inspected in multiple difference images and in both the Epoch1 and Epoch2 science images to determine if it should be included in the final transient sample. 

After visually inspecting the sample to remove noise spikes and subtraction artifacts, we found 38 transients brightening and 50 transients fading from Epoch1 to Epoch2.


\subsection{Supernova Selection} \label{subsec:supernova_selection}

To further isolate a sample of SNe, we applied additional criteria to remove sources whose behavior are not consistent with that of SNe.  Common contaminants are compact variable sources 
such as active galactic nuclei (AGN).  The sample could also be contaminated by other types of transients such as gamma ray burst (GRB) afterglows and tidal disruption events (TDEs).  The removal criteria included:

\begin{enumerate}
    \item If a transient that faded from Epoch1 to Epoch2 brightened again in each filter in the follow-up Epochs, it was removed from the SN sample.

    \item If a transient that brightened from Epoch1 to Epoch2 faded below its Epoch1 brightness in each filter in the follow-up epochs, it was removed from the SN sample. 

    \item AGN removal: For all the transients that are not clearly offset from the host galaxy nucleus, we have cross-matched our transient sources with previously published AGN catalogs \citep[]{Lyu2022AGNRadio, Lyu2024ActiveJWST/MIRI} and removed any AGN candidates. These previous AGN catalogs were based on an extensive search with the deepest X-ray to radio data, including the recent JWST/MIRI observations in GOODS-S. In addition, we conducted SED analysis to search for AGN evidence with the same fitting package in \citet{Lyu2024ActiveJWST/MIRI} and removed any AGN candidates. In total, we identified and removed 7 AGN candidates.

    \item If a source's host SED did not show signs of AGN activity but the source exhibited variability directly coinciding with its host's galactic nucleus such that there was no evidence of a point source appearing or disappearing, it was removed from the SN sample.
\end{enumerate}

After applying these criteria, we removed 9 transients, leaving us with 34 SN candidates that brightened from Epoch1 to Epoch2 (JADES-SN-23 sample) and 45 SN candidates that faded from Epoch1 to Epoch2 (JADES-SN-22 sample) in our SN sample, making a total of 79 SN candidates. The 9 removed variable sources include 7 likely variable AGN, one likely star, and one source of unknown type (potentially TDE). See Section \ref{subsec: other_variable_sources} and Table \ref{tab_other_variable_sources} for additional details about these non-SN variable sources. These criteria do not guarantee that our SN sample is free of contaminants. However, we will assume that these 79 transients are SNe because they appear as point-like variable sources offset from their respective host galaxies' cores. The 9 transients that are not likely to be SNe are listed in Section \ref{subsec: other_variable_sources}. 

There are 4 additional candidates which are likely real SNe, but they have 7-band S/N\,$\geq$\,12, which does not meet the S/N\,$\geq$\,14 requirement. These are not included in the statistical sample of 79 SNe (the combined JADES-SN-23 and JADES-SN-22 samples). Requiring 7-band S/N\,$\geq$\,12 rather than 14 for the selection criteria would have roughly doubled the visual inspection load for each filter, which was on the order of hundreds of sources per filter with S/N\,$\geq$\,14. We chose to implement the 7-band S/N\,$\geq$\,14 requirement to reduce visual inspection load and reduce the likelihood of contaminants in the sample. See Section \ref{subsec:marginal_detections} for additional details regarding these marginally-detected SN candidates.

We refer to each transient with its International Astronomical Union (IAU) ID issued by the Transient Name Server\footnote{https://www.wis-tns.org} (TNS).  We assign the \texttt{DAOStarFinder} position as the SN positions. Refer to Table \ref{tab_positions} for the JADES-SN-23 and JADES-SN-22 IAU IDs and positions.  The discovery of these transient sources was also reported in \citet{DeCoursey2024} and \citet{DeCoursey2023a}.

\begin{deluxetable*}{lccclcc}
\label{tab_positions}
\tablecaption{JADES-SN-22 and JADES-SN-23 positions}
\tablehead{
\colhead{JADES-SN-23 ID} & \colhead{RA} & \colhead{Dec} & \colhead{} & \colhead{JADES-SN-22 ID} & \colhead{RA} & \colhead{Dec}}
\startdata
\tr{53} (1)  & 03:32:43.9212 & -27:47:16.141 & & \tr{77} (1)   & 03:32:38.0488 & -27:43:46.465 \\
\tr{50} (2)  & 03:32:32.2073 & -27:48:59.148 & & \tr{33} (2)   & 03:32:44.8775 & -27:46:32.302 \\
\tr{88} (3)  & 03:32:33.7504 & -27:46:44.413 & & \tr{39} (3)   & 03:32:41.3980 & -27:48:49.472 \\
\tr{10} (4)  & 03:32:39.4574 & -27:50:19.666 & & \tr{93} (4)   & 03:32:43.3889 & -27:46:26.931 \\
\tr{44} (5)  & 03:32:27.3001 & -27:48:39.309 & & \tr{107} (5)  & 03:32:48.1335 & -27:47:05.745 \\
\tr{71} (6)  & 03:32:40.8009 & -27:46:05.984 & & \tr{102} (6)  & 03:32:29.4456 & -27:46:40.672 \\
\tr{27} (7)  & 03:32:32.3647 & -27:49:15.238 & & \tr{103} (7)  & 03:32:42.5164 & -27:45:51.810 \\
\tr{36} (8)  & 03:32:34.9551 & -27:48:36.682 & & \tr{13} (8)   & 03:32:36.9064 & -27:49:30.377 \\
\tr{26} (9)  & 03:32:32.4679 & -27:48:52.260 & & \tr{38} (9)   & 03:32:41.2737 & -27:50:22.290 \\
\tr{52} (10) & 03:32:40.5844 & -27:45:43.623 & & \tr{55} (10)  & 03:32:41.7927 & -27:47:39.187 \\
\tr{15} (11) & 03:32:39.6793 & -27:49:36.610 & & \tr{21} (11)  & 03:32:34.9004 & -27:48:07.155 \\
\tr{29} (12) & 03:32:31.7072 & -27:47:48.508 & & \tr{100} (12) & 03:32:30.3166 & -27:48:05.987 \\
\tr{6}  (13) & 03:32:45.1465 & -27:45:53.137 & & \tr{79} (13)  & 03:32:26.6302 & -27:48:01.875 \\
\tr{11} (14) & 03:32:38.8232 & -27:49:21.980 & & \tr{80} (14)  & 03:32:31.0676 & -27:47:58.311 \\
\tr{19} (15) & 03:32:46.3701 & -27:44:55.788 & & \tr{95} (15)  & 03:32:32.3579 & -27:46:30.298 \\
\tr{7}  (16) & 03:32:43.7610 & -27:47:00.273 & & \tr{8} (16)   & 03:32:40.1239 & -27:47:05.315 \\
\tr{28} (17) & 03:32:34.0699 & -27:48:39.324 & & \tr{34} (17)  & 03:32:28.8357 & -27:47:55.947 \\
\tr{87} (18) & 03:32:38.9635 & -27:44:20.649 & & \tr{23} (18)  & 03:32:34.9077 & -27:44:55.563 \\
\tr{60} (19) & 03:32:35.8521 & -27:47:19.044 & & \tr{66} (19)  & 03:32:44.4700 & -27:47:05.847 \\
\tr{5}  (20) & 03:32:45.1053 & -27:45:34.511 & & \tr{92} (20)  & 03:32:37.6742 & -27:45:22.492 \\
\tr{9}  (21) & 03:32:40.2246 & -27:49:49.492 & & \tr{37} (21)  & 03:32:38.9083 & -27:50:03.993 \\
\tr{83} (22) & 03:32:51.1740 & -27:45:29.209 & & \tr{20} (22)  & 03:32:35.3499 & -27:48:37.706 \\
\tr{22} (23) & 03:32:29.5180 & -27:49:16.082 & & \tr{111} (23) & 03:32:40.0216 & -27:49:06.786 \\
\tr{35} (24) & 03:32:32.1851 & -27:46:14.713 & & \tr{2} (24)   & 03:32:46.0677 & -27:47:04.425 \\
\tr{30}\tablenotemark{a} (25) & 03:32:37.1747 & -27:47:36.747 & & \tr{16} (25)  & 03:32:40.1875 & -27:47:42.854 \\
\tr{81} (26) & 03:32:38.5328 & -27:49:22.009 & & \tr{64} (26)  & 03:32:35.6305 & -27:47:49.506 \\
\tr{48} (27) & 03:32:39.2563 & -27:45:47.900 & & \tr{1} (27)   & 03:32:43.8862 & -27:46:34.074 \\
\tr{45} (28) & 03:32:37.2659 & -27:50:08.804 & & \tr{69} (28)  & 03:32:37.2123 & -27:44:59.852 \\
\tr{24}\tablenotemark{a} (29) & 03:32:32.3623 & -27:47:20.556 & & \tr{12} (29)  & 03:32:37.1281 & -27:49:41.317 \\
\tr{82} (30) & 03:32:37.6068 & -27:48:38.101 & & \tr{46} (30)  & 03:32:44.7521 & -27:48:12.124 \\
\tr{14} (31) & 03:32:43.4720 & -27:46:34.402 & & \tr{65} (31)  & 03:32:31.3916 & -27:47:14.250 \\
\tr{90} (32) & 03:32:49.0392 & -27:45:19.375 & & \tr{54} (32)  & 03:32:47.5319 & -27:47:49.588 \\
\tr{89} (33) & 03:32:48.4163 & -27:45:50.601 & & \tr{67} (33)  & 03:32:39.9484 & -27:46:06.281 \\
\tr{25} (34) & 03:32:43.4088 & -27:44:13.109 & & \tr{56} (34)  & 03:32:34.8200 & -27:48:35.918 \\
             &               &               & & \tr{68} (35)  & 03:32:52.3569 & -27:45:54.418 \\
             &               &               & & \tr{18} (36)  & 03:32:49.1404 & -27:45:24.119 \\
             &               &               & & \tr{17} (37)  & 03:32:42.3032 & -27:47:45.905 \\
             &               &               & & \tr{31} (38)  & 03:32:32.9662 & -27:45:48.419 \\
             &               &               & & \tr{61} (39)  & 03:32:27.3916 & -27:48:42.314 \\
             &               &               & & \tr{109} (40) & 03:32:35.3900 & -27:49:20.832 \\
             &               &               & & \tr{3} (41)   & 03:32:37.5384 & -27:48:39.412 \\
             &               &               & & \tr{4} (42)   & 03:32:37.5417 & -27:48:40.697 \\
             &               &               & & \tr{110} (43) & 03:32:32.3671 & -27:48:45.832 \\
             &               &               & & \tr{101} (44) & 03:32:30.4650 & -27:47:27.694 \\
             &               &               & & \tr{32} (45)  & 03:32:39.5341 & -27:45:08.041 \\
\sidehead{Marginal Detections}
\tr{84} (35) & 03:32:29.3115 & -27:48:26.467 & & \tr{96} (46)  & 03:32:38.7242 & -27:44:13.485 \\
\tr{59} (36) & 03:32:42.2813 & -27:47:46.471 & & \tr{94} (47)  & 03:32:35.5612 & -27:45:56.793 \\
\enddata
\tablecomments{ The ID numbers in parentheses indicate the Figure \ref{footprint} IDs.}
\tablenotetext{a}{Also discovered by \citet{Hayes2024GlimmersVariability} as 1402129 (\tr{30}) and 1402146 (\tr{24}).}
\end{deluxetable*}

\begin{figure*}
    \centering
    \includegraphics[width=7in]{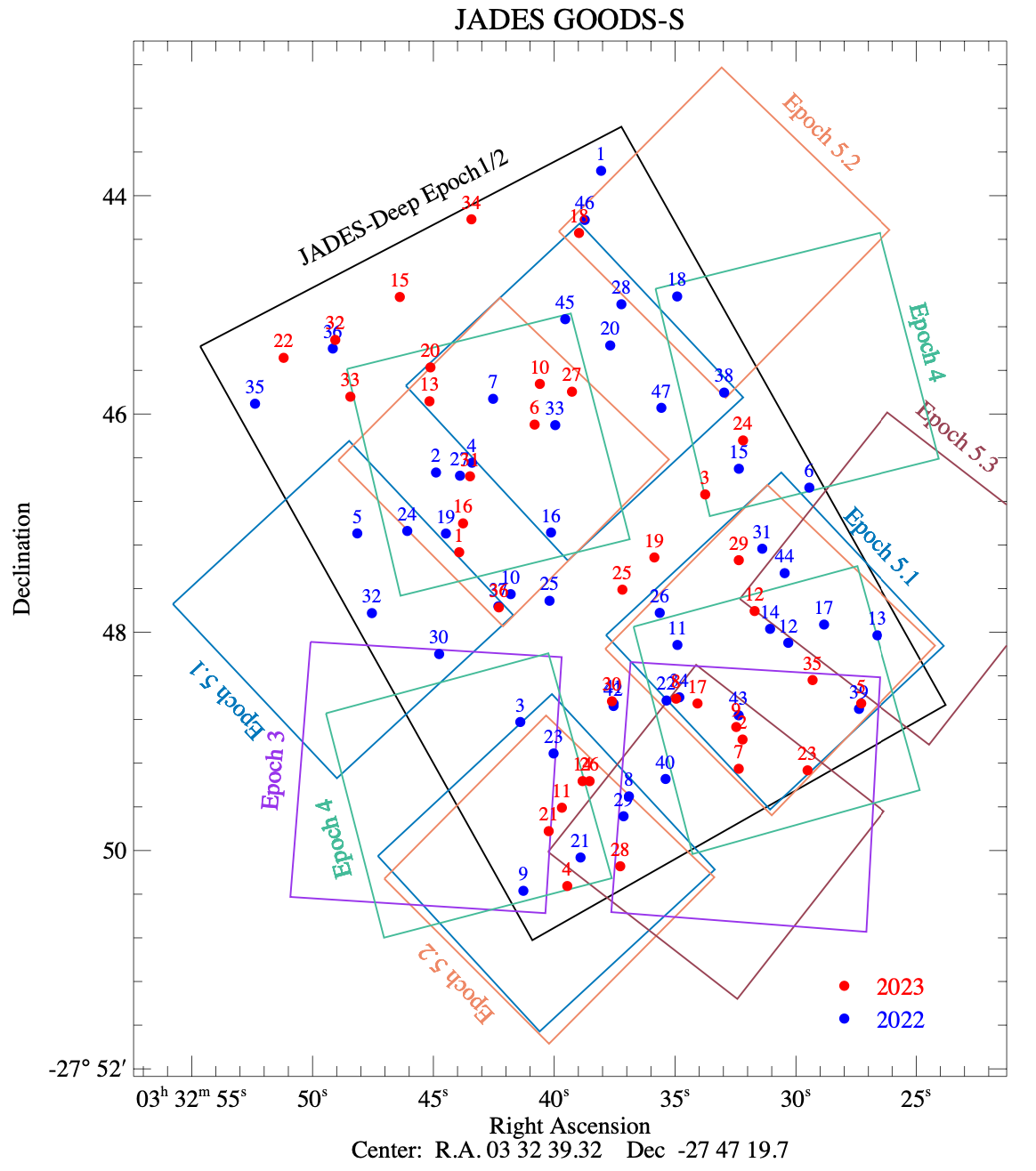}

    \medskip
    
    \caption{Footprints of the JADES Deep/Medium surveys (PID 1180) and DDT follow-up program (PID 6541) providing multi-epoch NIRCam imaging data of the GOODS-S field as listed in Table~\ref{tab_obslog}.  The locations of the SN candidates detected in the JADES-Deep 2022 and 2023 data are marked with the blue and red solid circles, respectively, with their ID numbers listed in Table~\ref{tab_positions}. The actual boundary of each footprint is more irregular due to dithering, so the coverage of sources close to the footprint boundaries needs to be checked against the real data. }
    \label{footprint}
\end{figure*}


\subsection{Detection Limit and Completeness} \label{sec:efficiency}
While we defer the full analysis of the effects that our selection criteria have on the sample completeness to a forthcoming supernova rates paper, we determined the limiting magnitude to which we could recover sources in the JADES Deep Field via mock source injection and recovery in the 7 wide-band Epoch2-Epoch1 difference images. We artificially inserted mock PSFs of known magnitude into a representative 1000x1000 pixel cutout from the Epoch2 science image at random locations, subtracted the corresponding Epoch1 science image cutout, ran \texttt{DAOStarFinder} on the difference image with the parameters specified in Section \ref{subsec:source_detection}, and crossmatched our recovered source catalog to our input source catalog. The representative image cutout contained large spiral galaxies to demonstrate that we may not achieve 100\% completeness even at bright magnitudes if an SN event occurs near the galaxy center. The mock PSFs were generated using the model PSFs (mPSFs) from \citet{Ji2023JADES4.5}, which were constructed specifically for JADES NIRCam products with WebbPSF, with the \texttt{EPSFModel} normalization radius set to 0\farcs3. 

\begin{figure}
    \centering
 {\includegraphics[width=9cm]{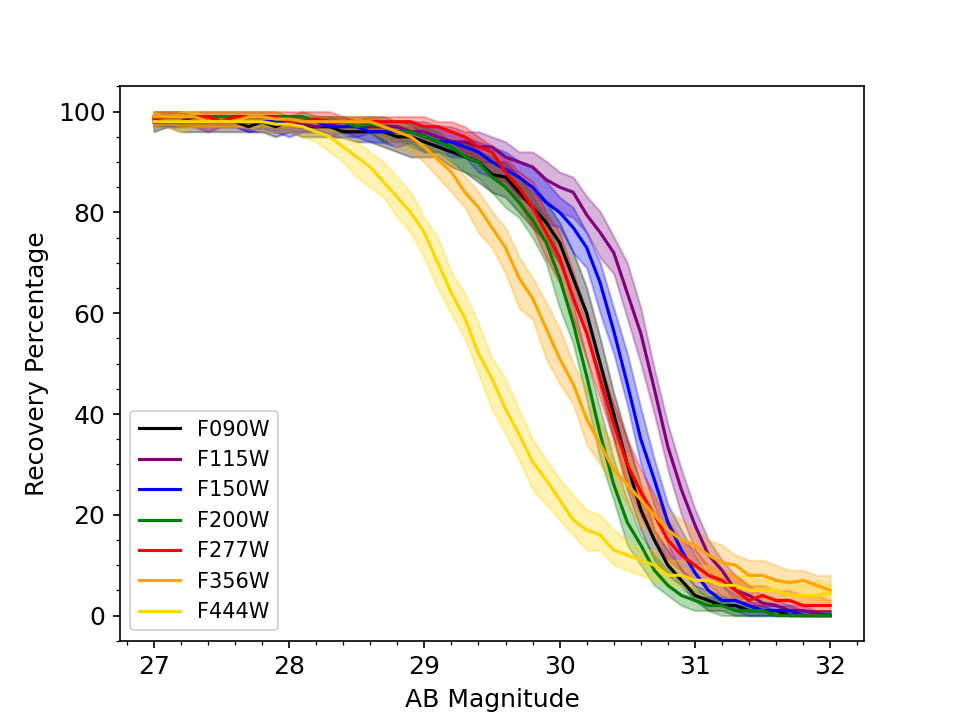}}
    \caption{Maximum possible recovery percentage (i.e., without the full transient selection criteria applied) as a function of mock PSF input magnitude for the 7 wide-band  Epoch2-Epoch1 difference images. This is based on an iterative PSF injection and recovery scheme ranging from 27 to 32 magnitude. The detection threshold was set to 3\,$\sigma$ for F356W/F444W and 4\,$\sigma$ for the rest as described in Section~\ref{subsec:source_detection}.}
    \label{deep_detection_efficiency}
\end{figure}

This was an iterative analysis, with 100 mock PSFs of the same magnitude injected in each iteration. The injection magnitude increased by 0.1 magnitude per iteration from 27 to 32 AB magnitudes. The mock source injection and recovery was performed 100 times for each magnitude. Figure \ref{deep_detection_efficiency} shows the median recovery percentage as a function of input PSF magnitude for each wide-band filter, with the 16\% and 84\% confidence intervals shaded in. We emphasize that the recovery percentages shown in Figure \ref{deep_detection_efficiency} are upper limits, as we did not apply the full transient selection criteria to the recovered sources. We simply recorded how many mock sources were recovered by \texttt{DAOStarFinder} using the parameters described in Section \ref{subsec:transient_selection}. 

F115W (purple) and F150W (blue) probe the deepest, with 50\% recovery rate around 30.65 and 30.45, respectively.  
The SW bands and F277W recovery rate curves follow similar shapes, dropping off sharply below $\sim$\,80\% recovery around 29.7-30.2 AB magnitude and  reaching $\sim$\,0\% recovery around 31-31.7 AB magnitude. At brighter magnitudes than $\sim$\,30.2, F356W (orange) and F444W (yellow) recover fewer sources than the other filters, with F444W consistently recovering fewer sources than F356W until it reaches a $\sim$\,100\% recovery rate above $\sim$\,28 AB magnitude. 


\subsection{PSF Photometry} \label{subsec:psf_photometry}

We measured PSF photometry from the 9-band NIRCam Epoch2-Epoch1/Epoch1-Epoch2 native pixel scale difference images. We adopt the PSF fitting method developed in \citet{Pierel2024JWST1.78} for measuring photometry on Level 3 (i.e., drizzled mosaic) JWST images, because of the need to create difference images on drizzled data for maximum S/N and cosmic ray rejection. Drizzled images are created from individual dithered ``CAL'' exposures, and have been bias-subtracted, dark-subtracted, flat-fielded, and corrected for geometric distortions. Unlike \citet{Pierel2024JWST1.78}, we have SN-free template images for all of our observations and so we replace the generic drizzled images with the difference images in all filters produced in Section \ref{subsec:host_subtraction}. We then implement the Level 3 PSF fitting routine from \citet{Pierel2024JWST1.78} using $5\times5$ pixel cutouts. The routine uses Level 2 PSF models from {\tt webbpsf}\footnote{\url{https://webbpsf.readthedocs.io}} that are temporally and spatially dependent and include a correction to the infinite aperture flux, which are then drizzled together to create a Level 3 PSF model. These total fluxes, which are in units of MJy/sr, are converted to AB magnitudes using the native pixel scale of each image ($0.03\arcsec/$pix for SW, $0.06\arcsec/$pix for LW). 

Refer to Tables \ref{tab_JD23_photometry} and \ref{tab_JD22_photometry}, respectively, for the JADES-SN-23 Epoch2-Epoch1 and JADES-SN-22 Epoch1-Epoch2 photometry. We display photometry for the subset of SNe within Epochs 3, 4, 5.1, 5.2, and 5.3 in Tables \ref{tab_E3_photometry}, \ref{tab_E4_photometry}, \ref{tab_E5p1_photometry}, \ref{tab_E5p2_photometry}, and \ref{tab_E5p3_photometry}, respectively. To account for the systematic zeropoint uncertainty, we added 0.01 magnitude in quadrature with the statistical uncertainty from the PSF photometry measurement, and we added a 0.01 magnitude uncertainty floor.

For non-detected sources, we report a 2$\sigma$ upper limit.  To measure the detection limits with the difference images, we selected a representative sky area (e.g., free from significant residuals due to the subtraction of large bright galaxies) and placed 2401 $r$\,$=$\,0\farcs1 circular apertures in a square grid pattern of 49\,$\times$\,49 with 20-pixel steps.  We then performed aperture photometry with a sky annulus of $r$\,$=$\,0\farcs1--0\farcs2 with the aperture corrections listed in Table~\ref{tab_apercor}, and calculated the standard deviation by fitting a Gaussian to the histogram of the measured sky signals.  Note that since the image differencing removes non-transient sources, the resultant histogram is symmetric toward the positive/negative directions.

There are a few cases where a source has disagreeing Epoch5.1/5.2/5.3 photometry for the same filter, which is unlikely to reflect real changes in brightness because the Epoch5.1/5.2/5.3 images were all taken on the same day. Rather, these disagreements arose from subtraction residuals contaminating one of the epoch's difference images or because the source fell very close to the image edge in one of the epochs. In these cases, we drop the data from the contaminated epoch rather than attempt to improve the subtraction (as described in Section \ref{subsec:host_subtraction}) because there was non-contaminated data from the same day with the same filters available. We denote these cases with superscript ``a" in Tables \ref{tab_E5p1_photometry}-\ref{tab_E5p3_photometry}.

Additionally, note that \tr{20}'s photometry in Table \ref{tab_JD22_photometry} is measured from the science image, as \tr{20} exploded prior to the Epoch1 imaging and remains bright in the Epoch2 and follow-up imaging. Photometry measured from the difference images would underestimate \tr{20}'s brightness. \tr{20}'s host is faint (see images in Appendix \ref{sec:deep_2022}) and \tr{20} is offset from its host, so the host light is not strongly contaminating the science image photometry. See Section \ref{subsec:classification} for additional details about \tr{20}'s likely nature.




\subsection{Host-Galaxy Subtraction} \label{subsec:host_subtraction}

Visual inspection of individual sources in the difference images showed that the subtraction of SN host galaxies sometimes left a significant residual, affecting the resultant photometry.  A closer examination indicates that the registration of the 2023/2022 JADES Deep images can be off by up to $\sim$\,0.5 pixels ($=$\,0\farcs015) at places, possibly suggesting insufficient accuracy of astrometric calibration for part of the data set.  

For those sources that suffer from poor host-galaxy subtraction, we improved the subtraction as follows: (1) cut out a section of 601$\times$601 pixels (i.e., $\pm$300 pixels; 0.03"/pixel) around the SN from the 2022 and 2023 images of each filter, (2) resample each pixel by 8x8 rebinning with bilinear interpolation, (3) apply integer shifts to the reference (i.e., SN-free) images in a 9$\times$9 grid of the resampled 1/8-scale pixels (i.e., up to $\pm$0.5 original pixel $=$\,0\farcs015 in both x and y directions), (4) subtract the shifted reference images from the target image, and (5) box-average 8x8 pixels in the resultant difference images, bringing them back to the original pixel scale of 0.03"/pixel.  We visually inspected the residual for the 9$\times$9 host-subtracted images and selected the one with the cleanest subtraction.  The quality of the subtraction is fairly stable among the groups of SW and LW images, respectively, but not necessarily between SW and LW images.  Therefore, we identified the best offsets for the SW and LW images separately, using F200W and F356W images.
We then measured 9-band PSF photometry in these local difference images with the cleanest subtraction.  

Note that \tr{9}, \tr{14}, and \tr{59} are positioned near the centers of large spiral galaxies. Despite performing refined host subtractions, they still suffer from residual host contamination in the follow-up epoch difference images because of the change in PA.
(In comparison, the Epoch2$-$Epoch1 difference images are free of contamination.) Their follow-up epoch PSF photometry, shown in Tables \ref{tab_E3_photometry}, \ref{tab_E4_photometry}, \ref{tab_E5p1_photometry}, \ref{tab_E5p2_photometry}, and \ref{tab_E5p3_photometry}, is likely overestimated. 


\section{Results} \label{sec:results}


\subsection{The JADES Supernova Sample} \label{sec:sample}

We have identified a total of 79 SNe in the JADES Deep Survey data, consisting of 34 SNe that brightened from Epoch1 to Epoch2 (JADES-SN-23 sample) and 45 SNe that faded from Epoch1 to Epoch2 (JADES-SN-22 sample). There are 2 additional marginally-detected SNe in each sample that do not meet the 7-band S/N\,$\geq$\,14 requirement and are not included in the statistical JADES-SN-22 or JADES-SN-23 samples (see Section \ref{subsec:marginal_detections}). The IAU IDs and positions of the JADES-SN-23 and JADES-SN-22 sources are listed in Table \ref{tab_positions}, and Figure \ref{footprint} shows each SN's position in the JADES Deep footprint.


In Appendices~\ref{sec:deep_2023} and \ref{sec:deep_2022}, we show 30mas scale stamp images in a variety of NIRCam filters for each source in the JADES-SN-23 and JADES-SN-22 samples, respectively. For each source, we show three single-filter images and the respective 3-color image for Epoch1, Epoch2, and the Epoch2-Epoch1 (or Epoch1-Epoch2) difference images. The checker-pattern residuals that appear in some of the LW difference images arise from the subtraction of resampled science images.

\begin{figure*}
    \centering
 {\includegraphics[width=11cm]{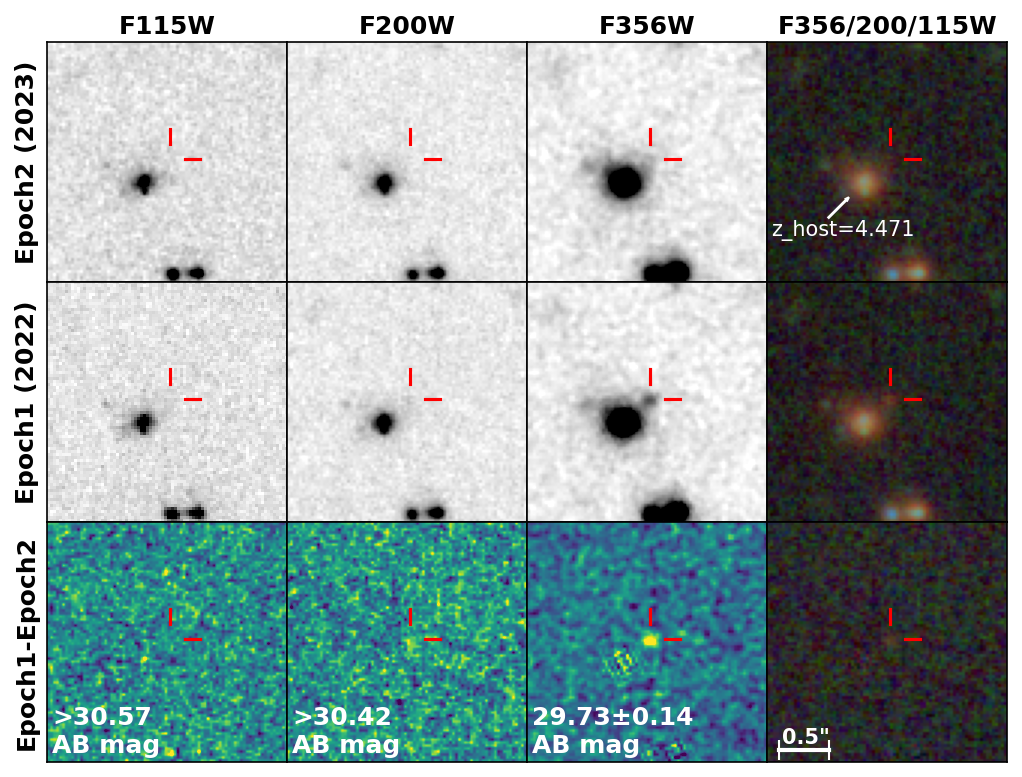}}
    \caption{Images of \tr{93}, which belongs to host galaxy JADES-GS+53.18086-27.77420. \textit{Top}: The Epoch2 F115W, F200W, and F356W NIRCam images that show the lack of emission from \tr{93} to the upper right of its host because it has faded from Epoch1 to Epoch2. The right-most panel shows the F356W/F200W/F115W Epoch2 RGB image. \textit{Middle}: The Epoch1 F115W, F200W, and F356W NIRCam images (and F356W/F200W/F115W RGB image) of \tr{93}. The F356W image shows the emission from \tr{93} to the upper right of its $z_{\mathrm{spec}}$\,$=$\,4.471 host \citep{Inami2017TheGalaxies}. \textit{Bottom}: The Epoch1-Epoch2 F115W, F200W, and F356W difference images (and F356W/F200W/F115W RGB difference image) showing \tr{93}'s change in brightness. It is clear that there is emission in the F356W difference image, but it is incredibly faint. This is a striking example of JWST's power to detect faint, high-redshift SNe.}
    \label{tr93_in_paper}
\end{figure*}

In this section, we highlight one SN that demonstrates the power of JWST as a high-redshift supernova survey tool. \tr{93}, which belongs to host galaxy JADES-GS+53.18086-27.77420, is one of the highest-redshift sources in the JADES-SN-22 sample, with $z_{\mathrm{spec}}$\,$=$\,4.471 \citep{Inami2017TheGalaxies}. Figure \ref{tr93_in_paper} shows the F115W, F200W, and F356W Epoch1 and Epoch2 NIRCam images of \tr{93} as well as their respective difference images and RGB images. There is clearly point-like emission in the Epoch1 F356W image at the location of \tr{93} that has faded entirely in the F356W Epoch2 image. Positive emission (yellow) is clearly seen in the F356W difference image at \tr{93}'s position, although the signal is as faint as $m_{F356W}$\,$=$\,29.73\,$\pm$\,0.14, which only JWST has the power to detect. There is no robust emission in the difference images at wavelengths shorter than F277W (see Table~\ref{tab_JD22_photometry}), which is an additional indicator of its high-redshift nature because the SN emission has been redshifted away from shorter wavelengths.

Both the JADES-SN-22 and JADES-SN-23 samples contain multiple $z$\,$>$\,4 SNe, representing a new frontier for supernova science.

We note that \citet{Hayes2024GlimmersVariability} previously published the discovery of two of the JADES-SN-23 SNe. 
As marked in Table~\ref{tab_positions}, \tr{30} appears in their paper as 1402129 and \tr{24} appears as 1402146.


\subsection{Marginal Detections} \label{subsec:marginal_detections}

Two marginally-detected SN candidates were found in the Epoch2-Epoch1 difference images (\tr{84} and \tr{59}) and two more were found in the Epoch1-Epoch2 difference images (\tr{96} and \tr{94}). See Section \ref{subsec:supernova_selection} for a description of why these candidates are not included in the JADES-SN-22 and JADES-SN-23 samples. These sources are listed in the position table (Table \ref{tab_positions}), redshift tables (Tables \ref{tab_JD23_redshifts} and \ref{tab_JD22_redshifts}) and photometry tables (Tables \ref{tab_JD23_photometry}, \ref{tab_JD22_photometry}, \ref{tab_E3_photometry}, \ref{tab_E4_photometry}, \ref{tab_E5p1_photometry}, \ref{tab_E5p2_photometry}, and \ref{tab_E5p3_photometry}) but are separated from the JADES-SN-22 and JADES-SN-23 samples. 

\begin{figure*}
    \centering
 {\includegraphics[width=11cm]{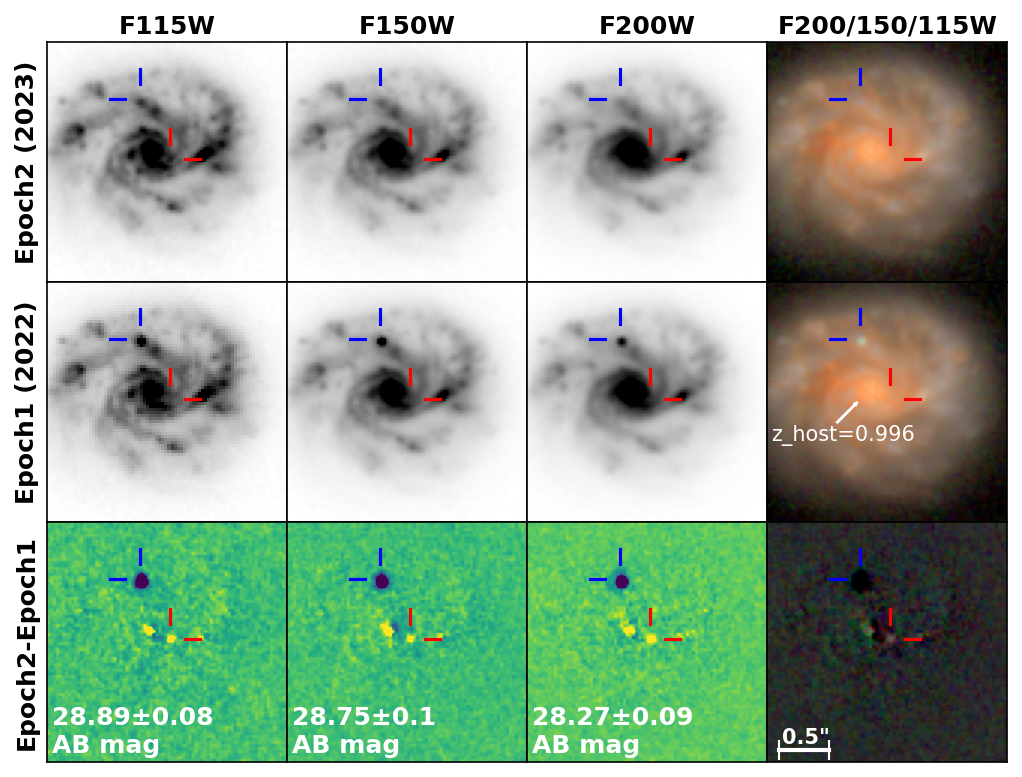}}
    \caption{Short-wavelength images of \tr{59} (indicated by red crosshairs) and \tr{17} (indicated by blue crosshairs), which both belong in host galaxy JADES-GS+53.17622-27.79620. \textit{Top}: The Epoch2 F115W, F150W, and F200W NIRCam images that show the emission from \tr{59} near the center of the galaxy and faint emission from fading \tr{17} to the upper left of the galactic center. \textit{Middle}: The Epoch1 F115W, F150W, and F200W NIRCam images that show an absence of emission from \tr{59} because it had not yet exploded, and that show bright emission from \tr{17} north of the galaxy center. \textit{Bottom}: The Epoch2-Epoch1 F115W, F150W, and F200W difference images showing positive emission (yellow) in the center from \tr{59} and negative emission (blue) north of the center from \tr{17}. Note that the positive emission to the left of \tr{59} is a subtraction artifact from the galactic center. We know that \tr{59} is not a subtraction artifact because there is point-like emission in the Epoch2 images at \tr{59}'s position that is not present in the Epoch1 images.}
    \label{tr59_stamp}
\end{figure*}

We highlight one of the interesting marginal detections, \tr{59}, to show JWST's ability to find SNe relative to HST. \tr{59}, shown by the red crosshairs in Figure \ref{tr59_stamp}, is located at (03:32:42.2813, -27:47:46.471) and its host galaxy JADES-GS+53.17622-27.79620 has $z_{\mathrm{spec}}$\,$=$\,0.9961 \citep{LeFevre2013The24.75}. While its redshift is relatively modest, \tr{59} is interesting because it lies near the center of a dusty luminous infrared galaxy (LIRG) with log(L$_\odot$)\,$\approx$\,11.35 (M. Florian et al. in preparation). There is a population of dust-obscured SNe that creates large uncertainties in core-collapse SN rate estimates because dust obscuration makes them difficult to discover \citep{Kool2018FirstDetection}. \tr{59} lies in such a crowded, dusty region and is relatively faint, peaking at just $m_{F200W}$\,$=$\,28.27\,$\pm$\,0.09, making it inaccessible to telescopes other than JWST.

In the difference images in the bottom panel of Figure \ref{tr59_stamp}, \tr{59} appears as the central yellow dot (the other yellow dot to its left is a subtraction artifact from the galaxy center). We know that \tr{59}'s emission is not a subtraction artifact because there is clearly a point source present in the Epoch2 F115W, F150W, and F200W images (Figure \ref{tr59_stamp} top panel) that is not present in the Epoch1 F115W, F150W, and F200W images (Figure \ref{tr59_stamp} middle panel). Note, however, that \tr{59}'s Epoch5.1 and Epoch5.2 LW photometry measurements are dropped from Tables \ref{tab_E5p1_photometry} and \ref{tab_E5p2_photometry}, respectively, due to contaminating subtraction residuals. 

 The large blue dot to the upper left of \tr{59}'s emission in the Figure \ref{tr59_stamp} difference images is the negative difference emission from \tr{17} (indicated by blue crosshairs), a SN that exploded prior to Epoch1 imaging and faded significantly in the year between the Epoch1 and Epoch2 imaging. In the Epoch1-Epoch2 difference image, \tr{17} peaks in F150W at 26.13 $\pm$ 0.02 mag, making it bright enough for HST detection. This transient was reported by the JADES team in 2023 on TNS \citep{DeCoursey2023a}. 

 We compare \tr{59} and \tr{17} because, within one galaxy, they highlight the vast improvement in detection threshold that JWST has over HST.  HST was limited to relatively bright SNe that exploded away from the galactic centers, whereas JWST has the capacity to detect SNe that are multiple magnitudes fainter and in dustier regions closer to centers of galaxies.

\subsection{SN SEDs}

We constructed SEDs for each SN, with multi-epoch SEDs shown for the JADES-SN-2023 sources that fell within the follow-up observations. Refer to Appendices \ref{sec:deep_2023} and \ref{sec:deep_2022} for the SEDs for each of the JADES-SN-23 and JADES-SN-22 sources, respectively. 

\begin{figure}
    \centering
 {\includegraphics[width=9cm]{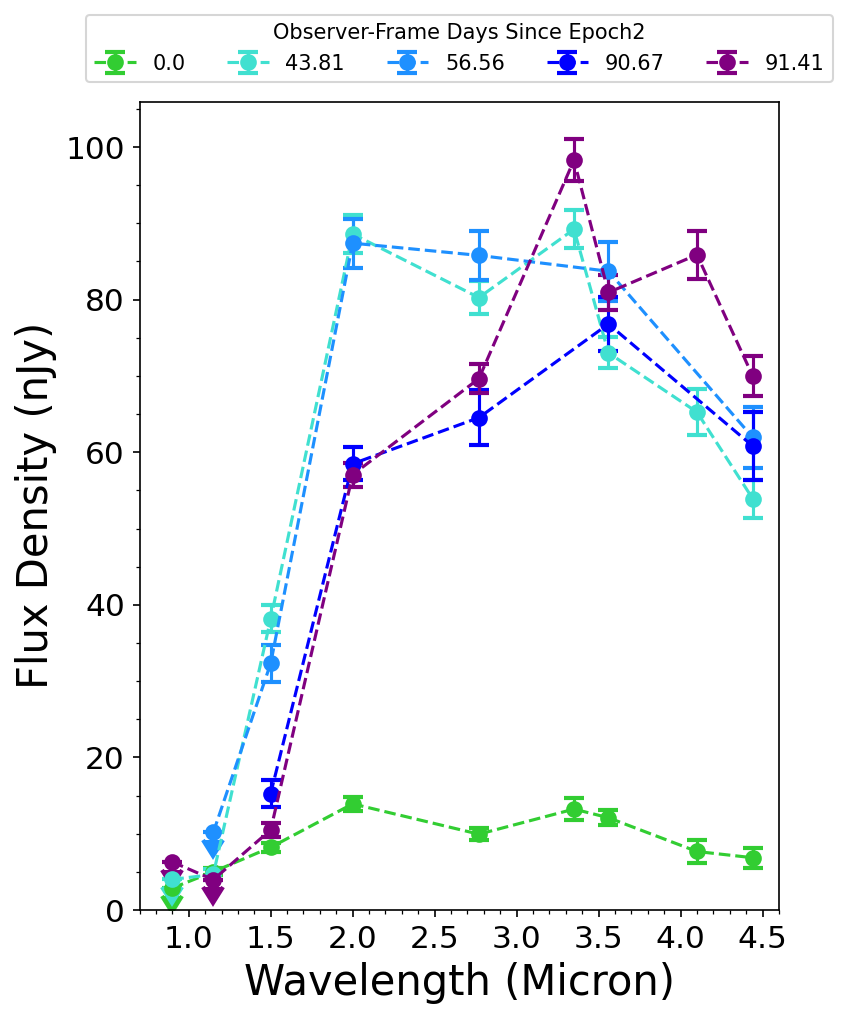}}
    \caption{Multi-epoch SED of \tr{26}, a SN Ic-BL at $z$\,$=$\,2.83 \citep{Siebert2024} in host galaxy JADES-GS+53.13533-27.81457. The x-axis is observer-frame wavelength, the y-axis is the difference image (EpochX-Epoch1) flux density in nJy, and SEDs are colored by observer-frame days since Epoch2 observations. \tr{26} is relatively faint in Epoch2 (green) but reaches near peak in Epoch3 (cyan; 43.81 days after Epoch2). It remains near peak in Epoch4 (light blue; 56.56 days after Epoch2), and is post peak in Epoch5.2 (dark blue; 90.67 days after Epoch2) and Epoch5.3 (purple; 91.41 days after Epoch2). The SED peak shifts to longer wavelength with time.}
    \label{tr26_sed}
\end{figure}

Here, we highlight the multi-epoch SED of \tr{26} (host $z_{\mathrm{spec}}$\,$=$\,2.83), which exhibited significant brightening between Epoch2 and Epoch3 (Figure \ref{tr26_sed}). \tr{26} was relatively faint in Epoch2 but brightened significantly in Epoch3, indicating that it was near its peak 43.81 observer-frame days after Epoch2. The Epoch3 SED peaks around 2$\mu$m (observer frame), which would be $\sim$\,0.5$\mu$m in the rest-frame assuming $z$\,$=$\,2.83. In Epoch4, 56.56 days after Epoch2, \tr{26} still appears to be near peak. \tr{26} starts to fade in Epochs 5.2 (90.67 days after Epoch2) and 5.3 (91.41 days after Epoch2), and the SED peak shifts towards longer wavelength. We do not display Epoch5.1 because it only covers F200W and F277W. See Section \ref{at2023adta} for details regarding the light curve fitting and classification of \tr{26}, and refer to  \citet{Siebert2024} for a complete analysis of \tr{26}.

\subsection{Host Galaxy Assignment} \label{subsec:redshifts}

In estimating SN redshifts, we assigned the host galaxy based on a ``directional light radius" (DLR) analysis \citep{Gupta2016HostSurveys}. We calculated the ratio of the galaxy light radius to the distance between host center and supernova for each galaxy within 3\arcsec\ of the supernova. The galaxy with the lowest ratio was assigned as the host. This method is more robust than simply assigning the closest galaxy as the host, as a supernova may be within the light radius of a large spiral galaxy (the likely host) but still be physically closer to a much more compact galaxy that is not likely the host. In this case, the DLR method would assign the large spiral galaxy as the host.

We used two galaxy catalogs internal to the JADES collaboration to assign hosts, one of which was produced in March 2023 and the other produced in August 2023. The main differences between the two catalogs are that the August 2023 catalog has a more complex detection algorithm designed to recover fainter and more compact sources, and it further deblends large objects with high-pass filtering to search for satellite objects in their exteriors. Host galaxy positions were obtained primarily from the August 2023 catalog, but we also referred to the March 2023 catalog if the segmentation mapping in the August 2023 catalog had not properly de-blended the host galaxy from neighboring galaxies. 

We list the JADES IDs of the assigned hosts for the JADES-SN-23 and JADES-SN-22 sources in Tables \ref{tab_JD23_redshifts} and \ref{tab_JD22_redshifts}, respectively. We also list the JADES FitsMap IDs, which come from the public JADES data-release 2 (DR2) catalog available on MAST\footnote{https://archive.stsci.edu/hlsp/jades} and can be searched using the JADES FitsMap viewer\footnote{https://jades-survey.github.io/viewer/}. Some sources are missing FitsMap IDs because the public DR2 version of the JADES catalog either does not include the assigned host because it is too close to a diffraction spike or because the segmentation scheme for the public catalog blends the assigned host with another nearby galaxy.

\subsection{Host Galaxy Redshifts}

We show the redshifts for the JADES-SN-23 and JADES-SN-22 samples in Tables \ref{tab_JD23_redshifts} and \ref{tab_JD22_redshifts}, respectively.  Out of our sample of 79 SNe, 47 have hosts with spectroscopic redshifts ($\sim$\,59\% of sample).  For the SN hosts without a spectroscopic redshift, we have adopted the 
latest JADES photometric redshifts available within the JADES collaboration (K.\ Hainline et al., in private communication), which were derived using the galaxy SED fitting code 
\texttt{EAZY} \citep{Brammer2008EAZY:Code,Hainline2023TheGOODS-N} with photometry measured from the Epoch1 mosaic.
We have also calculated the 16\% and 84\% redshift error bounds to indicate the reliability of the photometric redshift (see Section~\ref{photo-z} for more detail).

Note that three supernovae could not be assigned redshifts: \tr{25}, \tr{32}, and \tr{101}. \tr{32} has no obvious host, so it cannot be assigned a redshift. There is a nearby star which may be obscuring its host. \tr{25} and \tr{101} cannot be assigned redshifts because their hosts are too faint to yield well-constrained photometric redshifts. See Section \ref{subsec:highz_sne} for more details on \tr{101}'s potential host and redshift. \tr{68}'s redshift is also unclear because its host galaxy is unclear (although it was cautiously assigned a host with $z_{\mathrm{spec}}$\,$=$\,1.114). See Section \ref{subsec:highz_sne} for details.

In Tables~\ref{tab_JD23_redshifts} and \ref{tab_JD22_redshifts}, there is also a Redshift Rank column, where 1 indicates spectroscopic redshift, 2 indicates JADES \texttt{EAZY} photometric redshift with little to no SN contamination in the Epoch1 SED, and 3 indicates that the SED was re-fit with \texttt{eazy-py} using the Epoch2 data only because SN emission contaminated the Epoch1 SED. In some of the z-rank\,$=$\,3 cases, there still may be SN emission contaminating the Epoch2 SED, making it difficult to derive reliable photometric redshifts.

\subsubsection{JADES Photometric Redshifts (Epoch1)}
\label{photo-z}

The JADES photometric redshifts were calculated from the Epoch1 NIRCam photometry using  \texttt{EAZY} \citep{Brammer2008EAZY:Code}, following a similar procedure to what was done in \citet{Hainline2023TheGOODS-N}. Strong features in galaxy SEDs, like the 912\,\AA \: Lyman break, 1216\,\AA\ Ly$\alpha$ break, 4000\,\AA/Balmer breaks, and nebular line emission, can be observed photometrically, and we estimated the redshifts of these host galaxies by fitting with templates. \texttt{EAZY} linearly combines galaxy SED templates and iterates over a defined redshift grid to determine which redshift yields the best fit to the observed SED. Refer to \citet{Hainline2023TheGOODS-N} for the specific template sets used to generate the fits.

The input redshift grid covers from $z_{\mathrm{min}}$\,$=$\,0.01 to $z_{\mathrm{max}}$\,$=$\,21.0 with $\delta z$\,$=$\,0.01.  For each SN host without a spectroscopic redshift, the template SEDs were fit to the 14-band 0\farcs1 circular aperture photometry from HST/ACS (F435W, F606W, F775W, F814W, F850LP) and JWST/NIRCam (F090W, F115W, F150W, F200W, F277W, F335M, F356W, F410M, F444W). 

\texttt{EAZY} returns a $\chi^2(z)$ curve based on its ability to fit the galaxy photometry at each redshift. We convert this $\chi ^2$ curve to a probability distribution $\it{P(z)}$ assuming a uniform redshift prior: $P(z)$\,$=$\,$\exp[-\chi^2(z)/2]$, which we normalize such that $\int P(z) dz$\,$=$\,1. 

To estimate the most probable redshift from this $P(z)$ distribution, we convert the $P(z)$ distribution into a cumulative $P(z)$ distribution and then adopt the 50\% cumulative $P(z)$ value as the host galaxy's redshift (and SN's redshift), with the 16\% and 84\%  cumulative $P(z)$ values as the error bounds. Figure 13 of \citet{Rieke2023JADESImaging} compares photometric redshifts derived using a similar method with \texttt{EAZY} to spectroscopic redshifts, which has an overall outlier fraction of only 5\%, an average offset between spectroscopic and photometric redshifts of 0.05, and a scatter around the relation of $\sigma_{\mathrm{NMAD}}$\,$=$\,0.024 (where ``NMAD" stands for normalized median absolute deviation).

\subsubsection{Photometric Redshifts with SN-free Epoch2 Data}

The JADES \texttt{EAZY} fits were run with photometry measured from the Epoch1 images resulting in 11 of the JADES-SN-22 host galaxy SEDs being contaminated with SN emission. This was not a problem for extended galaxies where the SN emission was $\gtrsim$0\farcs2 from the galaxy center, as the point-like SN emission did not overlap with the 0\farcs1 radius aperture and 0\farcs1-0\farcs2 radius annulus used to measure the host photometry. However, this resulted in inaccurate photometric redshifts for compact galaxies whose 0\farcs1 radius aperture was contaminated by the SN emission. To obtain more accurate photometric redshifts for these sources (z-rank\,$=$\,3), we re-fit the SEDs using photometry from the Epoch2 images with \texttt{eazy-py} \footnote{https://github.com/gbrammer/eazy-py/}, using the \texttt{tweak\_fsps\_QSF\_12\_v3.param} templates created using the code \texttt{fsps} \citep{Conroy2009TheGalaxies}. 

The SN emission in these galaxies may not have entirely faded out by the time the Epoch2 NIRCam images were taken, so there may still be SN contamination in the host galaxy SEDs. However, these SNe were an additional year (in the observer frame) post-peak in the Epoch2 images than in the Epoch1 images, so the potential contamination was less severe in the Epoch2 SEDs. For a subset of these 11 host galaxies, we re-measured the host centroid with \texttt{photutils.segmentation} in the Epoch2 images because the SN emission in the Epoch1 images biased the host centroid. We re-measured the host centroid if $\sim$\,25\% or more of the 0\farcs1 radius aperture area was not overlapping with host emission in the Epoch2 F200W image. 

\begin{deluxetable*}{cccccccc}
\label{tab_JD23_redshifts}
\tablecaption{JADES-SN-23 host IDs and redshifts}
\tablehead{
\colhead{ID} & \colhead{JADES Host ID} & \colhead{FitsMap ID} & \colhead{z} & \colhead{z-low} & \colhead{z-high} & \colhead{z Rank} & \colhead{Spec-z Reference}
}
\startdata
\tr{53}  & JADES-GS+53.18299-27.78781 & 205579  & 4.35  & 4.31 & 4.38 & 2 &                                    \\
\tr{50}  & JADES-GS+53.13420-27.81642 & 197732  & 4.117 &      &      & 1 & \citet{Garilli2021TheMeasurements} \\
\tr{88}  & JADES-GS+53.14065-27.77896 & 127158  & 3.74  & 3.57 & 3.81 & 2 &                                    \\
\tr{10}  & JADES-GS+53.16439-27.83877 & 82314   & 3.61  &      &      & 1 & DDT-6541                           \\
\tr{44}  & JADES-GS+53.11368-27.81093 & 199353  & 3.21  & 3.09 & 3.79 & 2 &                                    \\
\tr{71}  & JADES-GS+53.16995-27.76844 & 211968  & 3.090 &      &	  & 1 & \citet{Inami2017TheGalaxies}       \\
\tr{27}  & JADES-GS+53.13485-27.82088 & 96906   & 2.90  &      &      & 1 & DDT-6541                           \\
\tr{36}  & JADES-GS+53.14564-27.81019 & 199488  & 2.86  & 2.76 & 2.91 & 2 &                                    \\
\tr{26}  & JADES-GS+53.13533-27.81457 & 198373  & 2.83  &      &	  & 1 & DDT-6541                           \\
\tr{52}  & JADES-GS+53.16908-27.76216 & 213818  & 2.78  & 2.66 & 2.90 & 2 &                                    \\
\tr{15}  & JADES-GS+53.16529-27.82678 & 91430   & 2.77  & 1.91 & 2.89 & 2 &                                    \\
\tr{29}  & JADES-GS+53.13210-27.79678 & \nodata & 2.73  &      &      & 1 & DDT-6541                           \\
\tr{6}   & JADES-GS+53.18807-27.76475 & 213100  & 2.623 &      &      & 1 & \citet{DEugenio2024}               \\
\tr{11}  & JADES-GS+53.16175-27.82281 & 95007   & 2.344 &      &      & 1 & \citet{Bunker2023}                 \\
\tr{19}  & JADES-GS+53.19326-27.74884 & 217127  & 2.24  & 2.11 & 2.35 & 2 &                                    \\
\tr{7}   & JADES-GS+53.18230-27.78338 & 207072  & 2.06  &      &      & 1 & DDT-6541                           \\
\tr{28}  & JADES-GS+53.14194-27.81095 & 104790  & 1.94  & 1.82 & 2.01 & 2 &                                    \\
\tr{87}  & JADES-GS+53.16266-27.73909 & 218643  & 1.932 &      &	  & 1 & \citet{Momcheva2016TheGalaxies}    \\
\tr{60}  & JADES-GS+53.14929-27.78859 & 205311  & 1.912 &      & 	  & 1 & \citet{Momcheva2016TheGalaxies}    \\
\tr{5}   & JADES-GS+53.18792-27.75959 & 214338  & 1.86  & 1.77 & 1.96 & 2 &                                    \\
\tr{9}   & JADES-GS+53.16765-27.83040 & 193080  & 1.854 &      &      & 1 & \citet{Momcheva2016TheGalaxies}    \\
\tr{83}  & JADES-GS+53.21332-27.75812 & 214628  & 1.748 &      &      & 1 & \citet{DEugenio2024}               \\
\tr{22}  & JADES-GS+53.12301-27.82107 & 196414  & 1.62  &      &      & 1 & DDT-6541                           \\
\tr{35}  & JADES-GS+53.13450-27.77101 & 210860  & 1.500 &      & 	  & 1 & \citet{Momcheva2016TheGalaxies}    \\
\tr{30}  & JADES-GS+53.15491-27.79346 & 204024  & 1.19  & 1.08 & 1.29 & 2 &                                    \\
\tr{81}  & JADES-GS+53.16056-27.82275 & 94908   & 1.171 &      &	  & 1 & \citet{Urrutia2019TheRelease}      \\
\tr{48}  & JADES-GS+53.16358-27.76329 & 213475  & 1.16  & 1.12 & 1.21 & 2 &                                    \\
\tr{45}  & JADES-GS+53.15543-27.83588 & 191322  & 1.139 &      &	  & 1 & \citet{Vanzella2008TheIII}         \\
\tr{24}  & JADES-GS+53.13491-27.78894 & \nodata & 1.01  &      &      & 1 & DDT-6541                           \\
\tr{82}  & JADES-GS+53.15647-27.81084 & 199156  & 0.665 &      &	  & 1 & \citet{Mignoli2005ThePopulation}   \\
\tr{14}  & JADES-GS+53.18106-27.77624 & 209108  & 0.657 &	  &       & 1 & \citet{Momcheva2016TheGalaxies}    \\
\tr{90}  & JADES-GS+53.20452-27.75542 & 215360  & 0.533 &      &	  & 1 & \citet{Momcheva2016TheGalaxies}    \\
\tr{89}  & JADES-GS+53.20180-27.76414 & 212911  & 0.210 &      &      & 1 & \citet{Momcheva2016TheGalaxies}    \\
\tr{25}  & JADES-GS+53.18087-27.73695 & 155113  & ?     &      &      &   &                                    \\
\sidehead{Marginal Detections} 
\tr{84}  & JADES-GS+53.12212-27.80734 & 200342  & 1.86  & 1.75 & 1.95 & 2 &                                    \\
\tr{59}  & JADES-GS+53.17622-27.79620 & 203278  & 0.996 &      &	  & 1 & \citet{LeFevre2013The24.75}        \\
\enddata
\end{deluxetable*}
\begin{deluxetable*}{cccccccc}
\label{tab_JD22_redshifts}
\tablecaption{JADES-SN-22 host IDs and redshifts}
\tablehead{
\colhead{ID} & \colhead{JADES Host ID} & \colhead{FitsMap ID} & \colhead{z} & \colhead{z-low} & \colhead{z-high} & \colhead{z Rank} & \colhead{Spec-z Reference} 
}
\startdata
\tr{77}  & JADES-GS+53.15852-27.72956 & 286671  & 4.82   & 4.77 & 4.86 & 3 &                                     \\
\tr{33}  & JADES-GS+53.18698-27.77563 & 129396  & 4.82   & 4.38 & 5.31 & 3 &                                     \\
\tr{39}  & JADES-GS+53.17257-27.81377 & 198564  & 4.504  &      &      & 1 & \citet{DEugenio2024}                       \\
\tr{93}  & JADES-GS+53.18086-27.77420 & 209839  & 4.471  &      &      & 1 & \citet{Inami2017TheGalaxies}        \\
\tr{107} & JADES-GS+53.20046-27.78476 & 123019  & 4.24   & 4.15 & 4.32 & 2 &                                     \\
\tr{102} & JADES-GS+53.12267-27.77795 & 127777  & 3.96   & 3.87 & 4.10 & 3 &                                     \\
\tr{103} & JADES-GS+53.17714-27.76441 & \nodata & 3.605  &      &      & 1 & \citet{LeFevre2015The6}             \\
\tr{13}  & JADES-GS+53.15375-27.82513 & 280649  & 3.58   & 3.46 & 3.72 & 2 &                                     \\ 
\tr{38}  & JADES-GS+53.17203-27.83956 & 190256  & 3.166  &      &      & 1 & \citet{Oesch2023TheFields}          \\
\tr{55}  & JADES-GS+53.17415-27.79405 & 203861  & 2.79   & 2.76 & 2.90 & 2 &                                     \\
\tr{21}  & JADES-GS+53.14541-27.80197 & 201806  & 2.73   & 2.34 & 2.88 & 3 &                                     \\
\tr{100} & JADES-GS+53.12631-27.80165 & 111073  & 2.62   & 2.51 & 4.16 & 3 &                                     \\
\tr{79}  & JADES-GS+53.11090-27.80056 & 202086  & 2.617  &      &	   & 1 & \citet{Bunker2023}                    \\
\tr{80}  & JADES-GS+53.12953-27.79964 & 202378  & 2.617  &      &	   & 1 & \citet{LeFevre2015The6}          \\
\tr{95}  & JADES-GS+53.13481-27.77504 & \nodata & 2.56   & 2.17 & 2.85 & 3 &                                     \\  
\tr{8}   & JADES-GS+53.16716-27.78481 & \nodata & 2.48   & 1.80 & 2.87 & 3 &                                     \\ 
\tr{34}  & JADES-GS+53.12024-27.79892 & 202484  & 2.323  &      &	   & 1 & \citet{Momcheva2016TheGalaxies}  \\
\tr{23}  & JADES-GS+53.14550-27.74871 & 217137  & 2.315  &      &      & 1 & \citet{DEugenio2024}                       \\
\tr{66}  & JADES-GS+53.18522-27.78504 & 206641  & 2.29   & 2.16 & 2.52 & 2 &                                     \\
\tr{92}  & JADES-GS+53.15696-27.75625 & 286357  & 2.02   & 1.77 & 2.31 & 3 &                                     \\
\tr{37}  & JADES-GS+53.16208-27.83453 & 85705   & 2.01   & 1.85 & 2.05 & 2 &                                     \\
\tr{20}  & JADES-GS+53.14729-27.81047 & 105068  & 2.00   & 1.63 & 2.35 & 3 &                                     \\
\tr{111} & JADES-GS+53.16681-27.81856 & 197183  & 1.92   &      &	   & 1 & DDT-6541                         \\
\tr{2}   & JADES-GS+53.19194-27.78455 & 206738  & 1.79   & 1.53 & 2.04 & 3 &                                     \\
\tr{16}  & JADES-GS+53.16747-27.79525 & 203597  & 1.771  &      &	   & 1 & \citet{Momcheva2016TheGalaxies}  \\
\tr{64}  & JADES-GS+53.14841-27.79697 & 203195  & 1.766  &      &	   & 1 & \citet{DEugenio2024}                    \\
\tr{1}   & JADES-GS+53.18283-27.77612 & 128996  & 1.688  &      &      & 1 & \citet{DEugenio2024}                       \\
\tr{69}  & JADES-GS+53.15503-27.74995 & 216898  & 1.62   & 1.57 & 1.69 & 2 &                                     \\
\tr{12}  & JADES-GS+53.15461-27.82810 & 193831  & 1.567  &      &	   & 1 & \citet{Momcheva2016TheGalaxies}  \\
\tr{46}  & JADES-GS+53.18645-27.80334 & 109974  & 1.42   & 1.36 & 1.53 & 3 &                                     \\
\tr{65}  & JADES-GS+53.13085-27.78716 & 205746  & 1.415  &      &      & 1 & \citet{Inami2017TheGalaxies}        \\
\tr{54}  & JADES-GS+53.19795-27.79706 & 203217  & 1.36   & 1.23 & 1.57 & 2 &                                     \\
\tr{67}  & JADES-GS+53.16633-27.76866 & 211770  & 1.294  &      &	   & 1 & \citet{Inami2017TheGalaxies}     \\
\tr{56}  & JADES-GS+53.14511-27.80991 & 199527  & 1.244  &      &	   & 1 & \citet{Urrutia2019TheRelease}    \\
\tr{68}  & JADES-GS+53.21813-27.76576 & 208546  & 1.114? &      &	   & 1 & \citet{Cooper2012TheField-South} \\
\tr{18}  & JADES-GS+53.20463-27.75680 & 214806  & 1.094  &      &	   & 1 & \citet{Vanzella2008TheIII}       \\
\tr{17}  & JADES-GS+53.17622-27.79620 & 203278  & 0.996  &      &	   & 1 & \citet{LeFevre2013The24.75}      \\
\tr{31}  & JADES-GS+53.13759-27.76325 & 213336  & 0.953  &      &	   & 1 & \citet{Urrutia2019TheRelease}    \\
\tr{61}  & JADES-GS+53.11407-27.81176 & 104035  & 0.669  &      &	   & 1 & \citet{Urrutia2019TheRelease}    \\
\tr{109} & JADES-GS+53.14735-27.82249 & 195998  & 0.669  &      &      & 1 & \citet{Urrutia2019TheRelease}       \\
\tr{3}   & JADES-GS+53.15647-27.81084 & 199156  & 0.665  &      &	   & 1 & \citet{Mignoli2005ThePopulation} \\
\tr{4}   & JADES-GS+53.15647-27.81084 & 199156  & 0.665  &      &	   & 1 & \citet{Mignoli2005ThePopulation} \\
\tr{110} & JADES-GS+53.13430-27.81269 & 198750  & 0.540  &      &	   & 1 & \citet{Xue2011TheCatalogs}       \\
\tr{101} & \nodata                    & \nodata & ?      &      &      &   &                                     \\
\tr{32}  & \nodata                    & \nodata & ?      &      &      &   &                                     \\
\sidehead{Marginal Detections} 
\tr{96}  & JADES-GS+53.16139-27.73710 & 219050  & 3.913  &      &	   & 1 & \citet{LeFevre2015The6}          \\
\tr{94}  & JADES-GS+53.14817-27.76576 & 212759  & 2.67   & 2.61 & 2.87 & 2 &                                     \\ 
\enddata
\end{deluxetable*}


\subsection{$z$\,$\geq$\,4 Supernova Candidates} \label{subsec:highz_sne}

The JADES-SN-22 and JADES-SN-23 samples contain 7 SNe associated with hosts at $z$\,$\geq$\,4. Out of these 7 supernovae, 3 have hosts with spectroscopic redshifts (\tr{50}, \tr{39}, and \tr{93}), and 2 have hosts with robust z-rank\,$=$\,2 photometric redshifts (\tr{53} and \tr{107}). The other two, \tr{77} and \tr{33}, have z-rank\,$=$\,3 host photometric redshifts.

\subsubsection{\tr{77} and \tr{33}: SNe at $z_{phot}$\,$=$\,4.82}

\begin{figure*}
    \centering
 {\includegraphics[width=14cm]{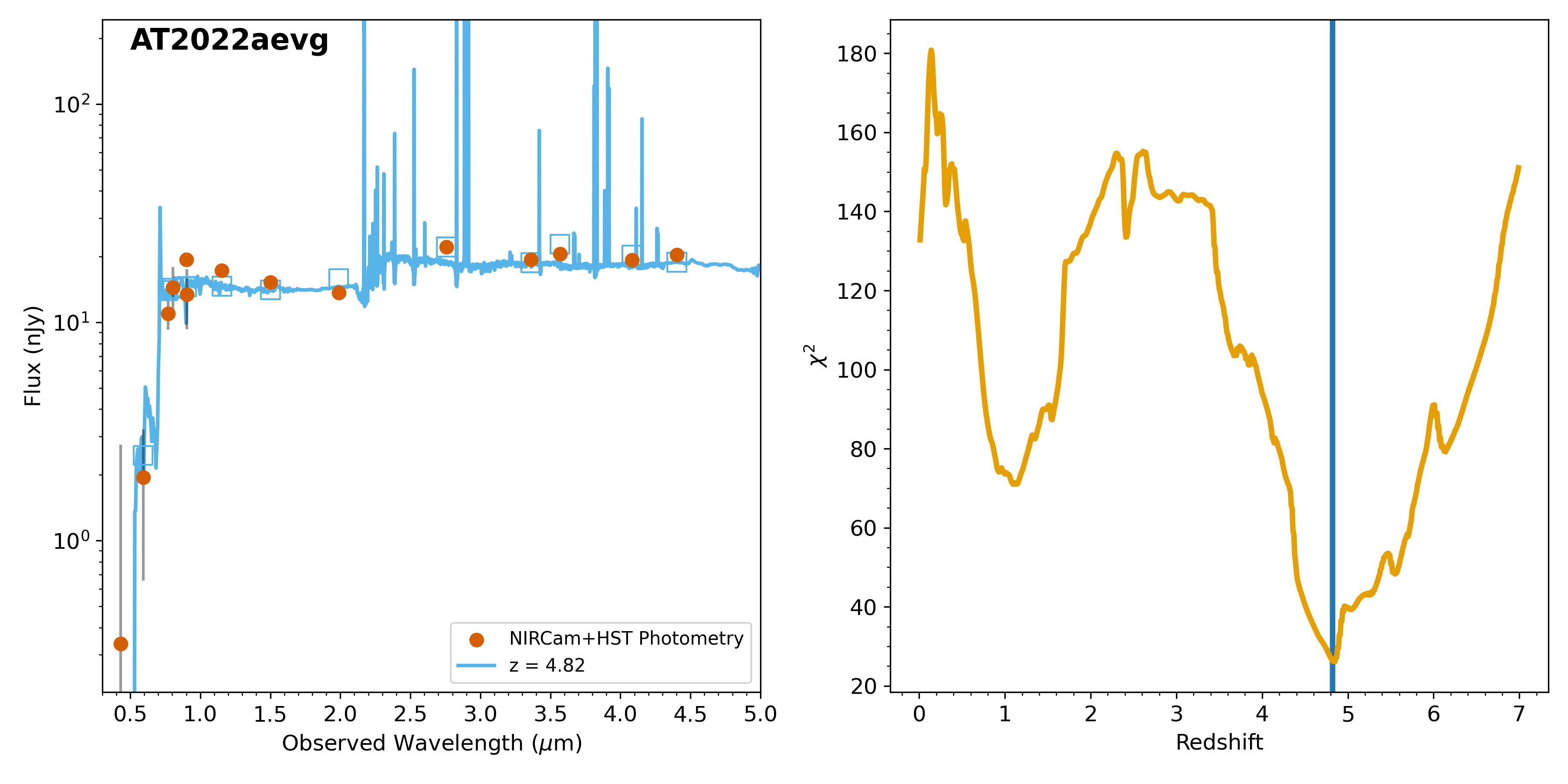}}
 \qquad
 {\includegraphics[width=14cm]{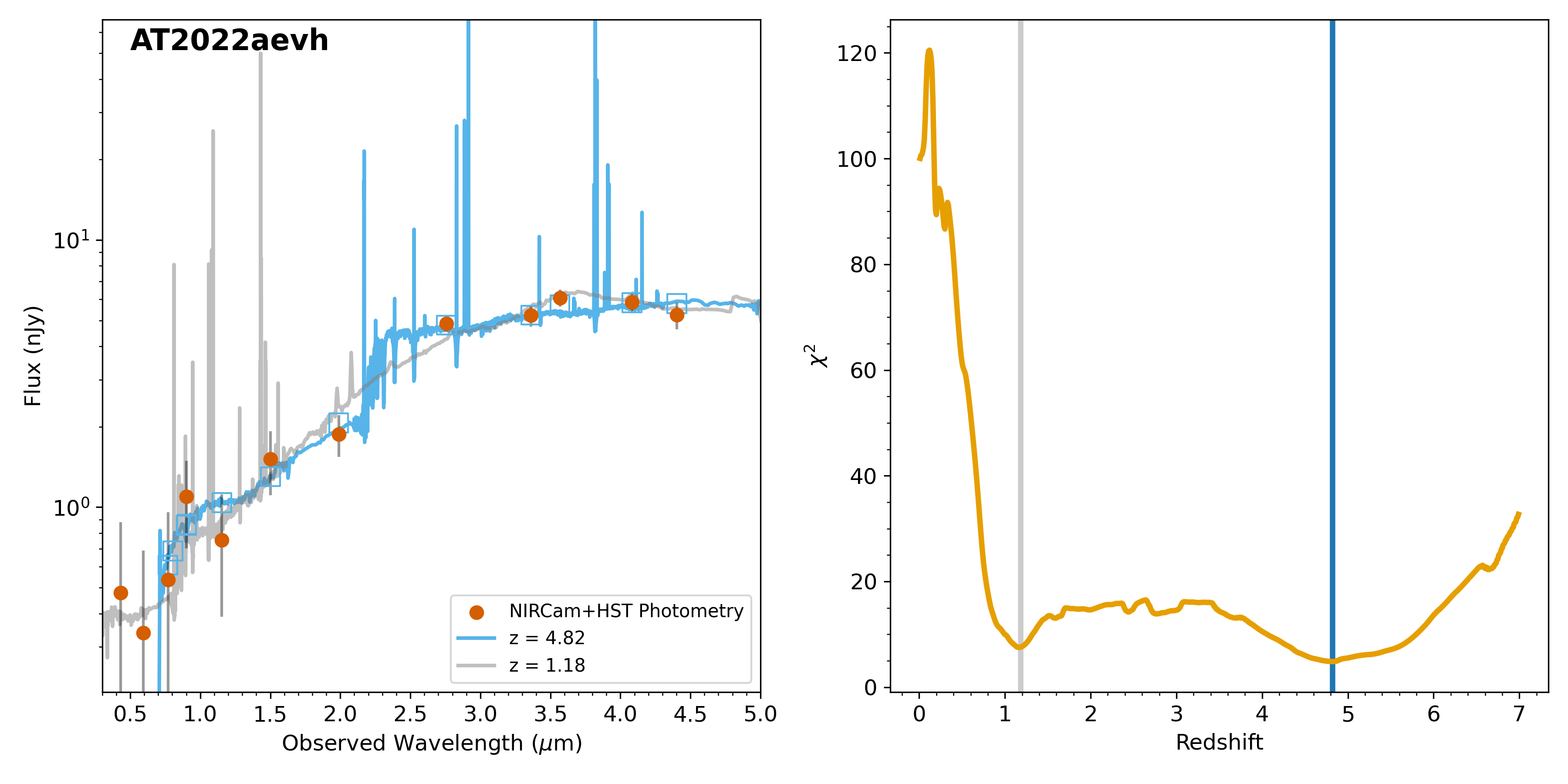}}
    \caption{$\it{Left}$:  Host galaxy SED fits for \tr{77} (top; host is JADES-GS+53.15852-27.72956) and \tr{33} (bottom; host is JADES-GS+53.18698-27.77563). HST/ACS and JWST/NIRCam measured host photometry is shown as dark orange circles with gray error bars, and the assigned host SED is shown in blue. \tr{77}'s $z$\,$=$\,4.82 SED fits the data very well, providing confidence in the $z$\,$=$\,4.82 redshift solution. \tr{33}'s $z$\,$=$\,4.82 SED (blue) also fits the host photometry well, although the fit is less certain at SW. We also show the $z$\,$=$\,1.18 SED (gray) for \tr{33}'s host, but it does not fit the host photometry as well as the $z$\,$=$\,4.82 SED at 2$\mu$m and 2.77$\mu$m. $\it{Right}$: $\chi^2$ vs. SED fit redshift for \tr{77} (top) and \tr{33} (bottom). The $\chi^2$ minima indicate the redshifts at which the SED best fits the measured photometry. \tr{77} has one clear $\chi^2$ minimum around $z$\,$=$\,4.82, providing further confidence in this redshift. \tr{33}'s $\chi^2$ surface shows two local minima:  $z$\,$=$\,4.82 and $z$\,$=$\,1.18. We assign $z$\,$=$\,4.82 as the host redshift, which has the global $\chi^2$ minimum.}
    \label{galaxy_seds}
\end{figure*}

In the left panel of Figure \ref{galaxy_seds}, we show the host SED fits generated for \tr{77} (top left; host is JADES-GS+53.15852-27.72956) and \tr{33} (bottom left; host is JADES-GS+53.18698-27.77563), the two $z$\,$\geq$\,4 SNe candidates whose assigned hosts have z-rank\,$=$\,3 redshifts. The right panels of Figure \ref{galaxy_seds} shows the $\chi^2$ vs. redshift plots associated with \tr{77} and \tr{33}'s host SED fits. $\chi^2$ minima indicate the most likely redshifts based on how well the SED fits the galaxy's photometry at each redshift.

The $z$\,$=$\,4.82 SED fit for \tr{77}'s host is very robust, showing a clear Lyman-alpha break at observer-frame $\sim$\,0.7$\mu$m. The associated $\chi^2$ surface shows a clear minimum at $z$\,$=$\,4.82, adding additional confidence to the $z$\,$=$\,4.82 redshift solution. 

\tr{33}'s $z$\,$=$\,4.82 host SED fits the measured host photometry fairly well, although the fit is less certain at SW. The $\chi^2$ surface reflects this uncertainty, as it does not have one clear minimum. Rather, it shows two local minima at $z$\,$\sim$\,4.82 (blue vertical line) and $z$\,$\sim$\,1.18 (gray vertical line). The $z$\,$=$\,1.18 galaxy SED (gray) does not fit the 2$\mu$m and 2.77$\mu$m host photometry as well as the Balmer break in the $z$\,$=$\,4.82 galaxy SED (blue), and the $z$\,$=$\,4.82 SED has a slightly lower $\chi^2$.  Additionally, the $z$\,$=$\,1.18 SED solution is partially driven by the HST/ACS photometry, which are non-detections. We therefore assign $z$\,$=$\,4.82 as the host redshift.

It is important to note that the Epoch2 photometry that was used to generate \tr{33}'s host SED fit could be contaminated with leftover SN emission (z-rank\,$=$\,3), which would invalidate this redshift solution since the SED-fitting template does not account for SN emission. \tr{33} is captured by Epoch1, Epoch2, Epoch4, and Epoch5.2. We measured PSF photometry at \tr{33}'s position in each of these epoch's science images to try to discern if \tr{33} was still present in Epoch2 and continued to fade in Epochs 4 and 5.2. However, the emission in Epoch2, from either \tr{33} or its host, was below the detection threshold of the shallower Epoch4 and Epoch5.2 images. Thus, we could not discern if the Epoch2 emission is from \tr{33} or its host. There is another potential host to the left of \tr{33} (see image in Appendix \ref{sec:deep_2022}). The z-rank\,$=$\,2 SED solution of this galaxy provides a $z$\,$>$\,4 solution, but this SED could be contaminated by \tr{33} since the galaxy photometry was measured from the Epoch1 science image, when \tr{33} was clearly present. We measured photometry in the Epoch2 science image at this potential host's position and performed SED fitting with \texttt{eazy-py} (z-rank\,$=$\,3), but the emission was too faint to yield a well-constrained SED fit. 

\subsubsection{\tr{39}: SN at $z_{spec}$\,$=$\,4.504}

We note that \tr{39} appears in a small island of emission to the upper right of its assigned host galaxy (JADES-GS+53.17257-27.81377), which has $z_{\mathrm{spec}}$\,$=$\,4.504 (see Appendix \ref{sec:deep_2022} for image).  The segmentation mapping of both the March 2023 and August 2023 versions of the JADES galaxy catalogs associate this island with the $z_{\mathrm{spec}}$\,$=$\,4.504 host galaxy. To ensure that the island whose position coincides with \tr{39} is also at $z$\,$\sim$\,4.5, we performed galaxy SED fitting at this location using photometry from the Epoch2 science image, in which \tr{39} had faded. The SED fitting yielded a redshift solution consistent with $z$\,$\sim$\,4.5, so we associate the island and thus \tr{39} with the $z_{\mathrm{spec}}$\,$=$\,4.504 galaxy.

\subsubsection{\tr{53}: SN at $z_{phot}$=4.35}

\tr{53}'s assigned host galaxy is JADES-GS+53.18299-27.78781, which has $z_{\mathrm{phot}}$\,$=$\,4.35. We note that \citet{Inami2017TheGalaxies} assign $z_{\mathrm{spec}}$\,$=$\,3.096 to a galaxy with coordinates that are close to \tr{53}. However, there are multiple galaxies near \tr{53} (see image in Appendix \ref{sec:deep_2023}), and it is unclear if the galaxy listed in \citet{Inami2017TheGalaxies} is the same galaxy as \tr{53}'s assigned host. Additionally, the spec-$z$ is based on only one spectral feature, so it is not a high-confidence spec-$z$. Thus, we report $z_{\mathrm{phot}}$\,$=$\,4.35 for \tr{53}'s host's redshift.

\subsubsection{\tr{68}: SN with Potential $z$\,$>$\,5 Host?}
\begin{figure}
    \centering
 {\includegraphics[width=8cm]{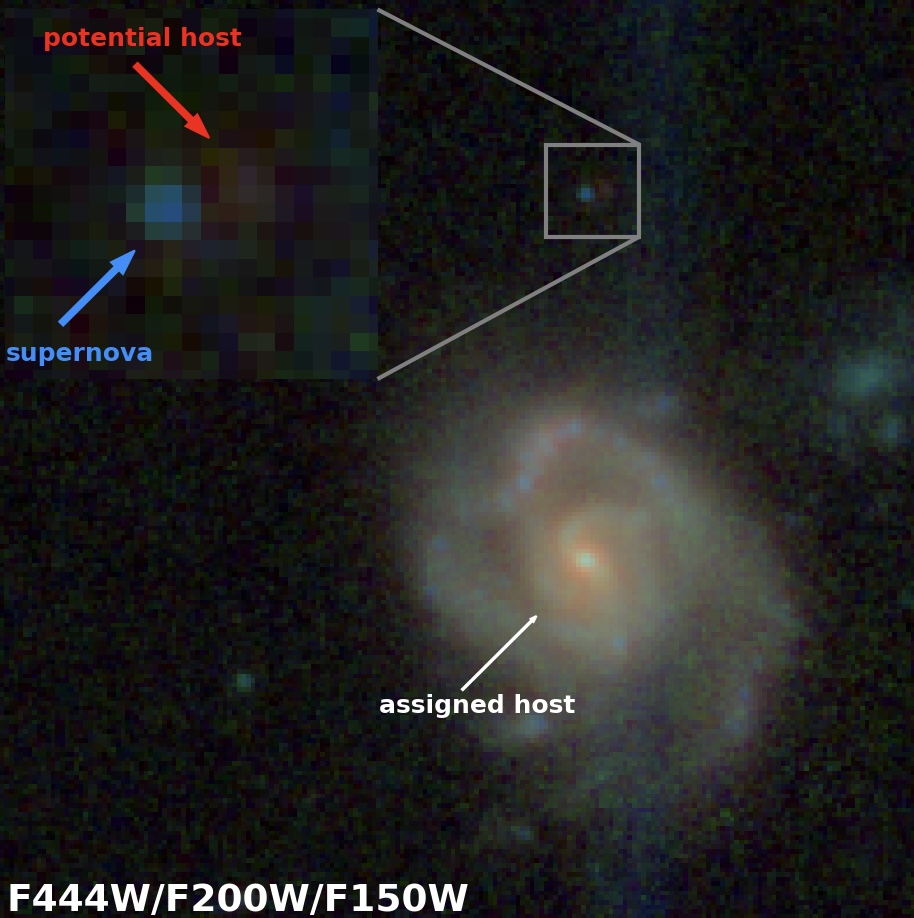}}
    \caption{\tr{68} appears as the blue dot above the spiral galaxy, which is \tr{68}'s assigned host ($z_{\mathrm{spec}}$\,$=$\,1.114). To the upper left of the spiral galaxy is a zoomed-in image of \tr{68}, which shows a faint red blob to its right. This faint red blob is another potential host galaxy for \tr{68}. Galaxy SED fitting of the red blob in the Epoch2 imaging, when the supernova had faded away, yielded a photometric redshift solution of $z$\,$\sim$\,5.8. \tr{68}'s host is unclear, but SN light curve fitting yielded a better solution at $z$\,$\sim$\,1 than at $z$\,$\sim$\,5.8, so we assign the spiral galaxy as \tr{68}'s likely host.}
    \label{tr68}
\end{figure}

\tr{68} is a high-redshift SN candidate because it exploded right next to a faint blob of LW emission that is not listed in the JADES galaxy catalogs (see Figure \ref{tr68}). \tr{68} appears as the blue compact emission and the potential host is the red blob of emission to \tr{68}'s right. Galaxy SED fitting of the Epoch2 images (when the SN was gone) yields a $z$\,$\sim$\,5.8 solution, although the potential host is quite faint so the fit is not secure. \tr{68} is also positioned in the outskirts of a $z_{\mathrm{spec}}$\,$=$\,1.114 large spiral galaxy, which we assign as the host (JADES-GS+53.21813-27.76576). We assign the lower redshift galaxy as the host because the SN light curve fitting yielded a much better fit at $z$\,$\sim$\,1 than at $z$\,$\sim$\,5.8. \tr{68} is also quite blue, which is more consistent with a lower redshift normal SN solution. If \tr{68} were at $z$\,$\sim$\,5.8, it would be brighter in rest-frame 0.2$\mu$m than in rest-frame 0.3$\mu$m or 0.7$\mu$m, which is unlikely for a typical SN. For theoretical pair-instability, low-metallicity superluminous, or population III SNe, the expectation is that the SN is very blue at early times. However, with only one observation, we cannot determine if \tr{68} is an exotic SN.

We still mention this high-redshift solution for \tr{68} because we consider it unlikely for a SN on the outskirts of a large spiral galaxy to explode so close to another potential host (see DLR host-assignment method described in Section \ref{subsec:redshifts}).  Also, the $z$\,$\sim$\,5.8 light curve fitting may not have yielded a good solution because the SN models at $z$\,$\sim$\,5.8 may not be accurate. Important future work includes developing more sophisticated SN models at high-redshift.

\subsubsection{\tr{101}: Potential High-z SN with Undetermined Redshift}

\tr{101} is another high-redshift SN candidate shown in the left panel of Figure \ref{z16_candidates}. At its position, there was LW emission in Epoch1 that faded in Epoch2, but there remained some LW emission in Epoch2. It is unclear if this remaining emission is leftover SN emission or host galaxy emission. We performed galaxy SED fitting with the Epoch2 science image photometry assuming that the Epoch2 emission was from a host, but the resulting fit was inconclusive because the Epoch2 emission is so faint. The SED fitting did, however, provide a $z$\,$\sim$\,6 solution, making \tr{101} a high-redshift SN candidate. We are not confident in this solution, so we choose not to assign a redshift to \tr{101}. It should be noted in general that SNe discovered with JWST tend to give a $z$\,$\sim$\,6--7 solution when their JWST/NIRCam and HST/ACS photometry is fit with galaxy SED templates because they appear to ``drop out" in the HST/ACS bands (because they did not exist at the time of HST imaging), making them F814W dropouts.


\subsection{Properties of the JADES SN Sample} \label{subsec: properties}

\begin{figure*}
    \centering
 {\includegraphics[width=0.49\linewidth]{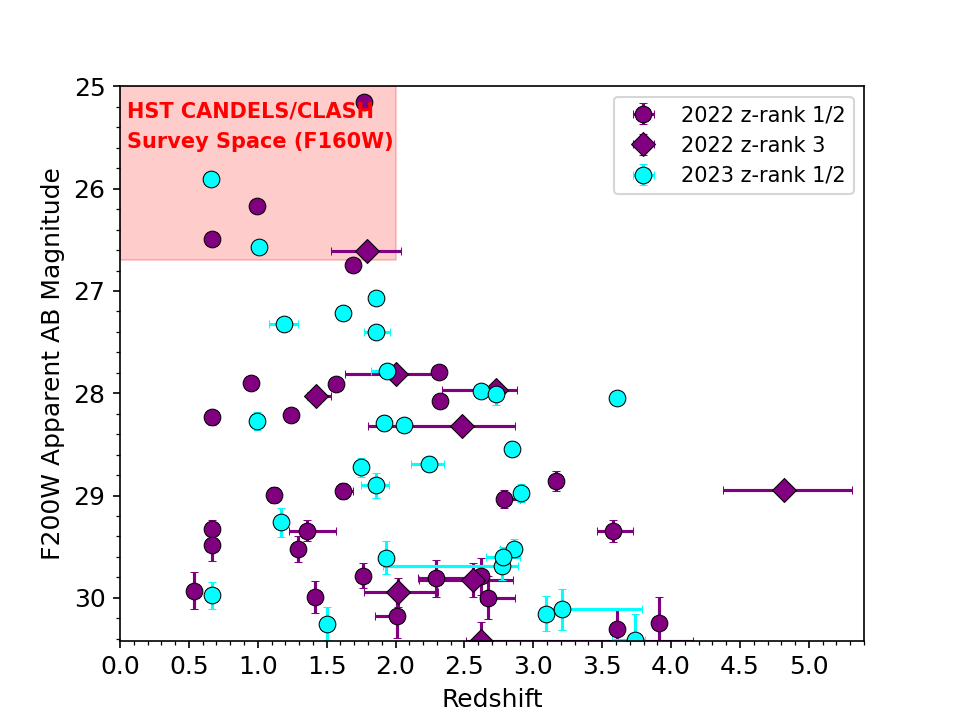}}
{\includegraphics[width=0.49\linewidth]{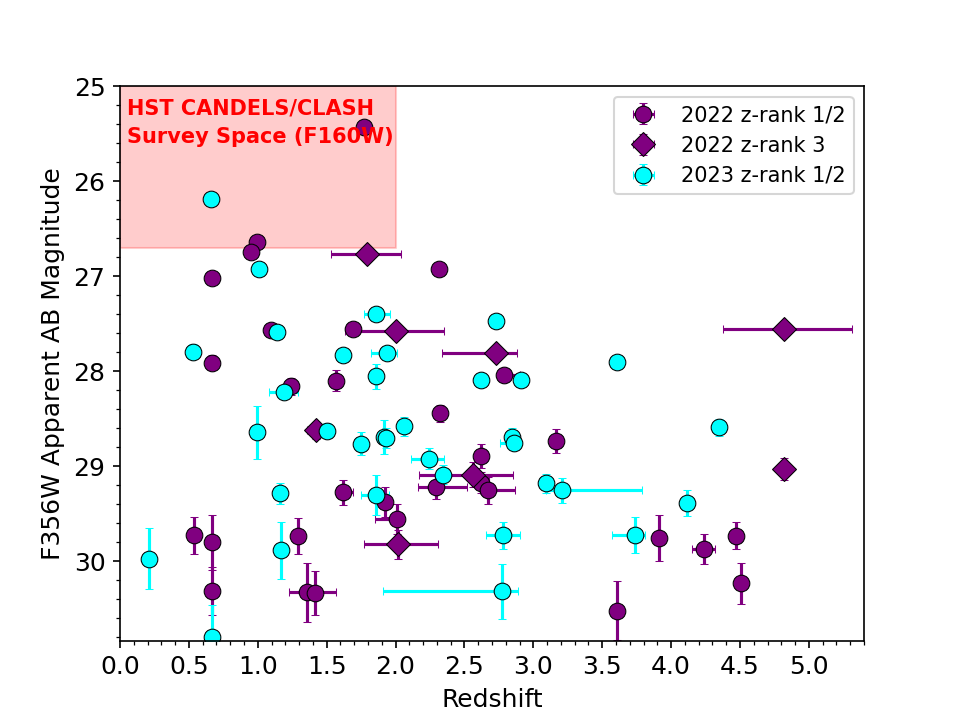} }
    \caption{Apparent magnitude vs. redshift for SNe detected in F200W (left) and F356W (right). The JADES-SN-22 sample is shown in purple and the JADES-SN-23 sample is shown in cyan. SNe that are below the Epoch1/Epoch2 F200W and/or F356W detection thresholds are not shown. Circles indicate robust redshifts (rank 1 or 2, meaning spectroscopic or robust photometric redshift), and diamonds indicate less robust redshifts (rank 3, meaning that the host galaxy SED used to generate the photometric redshift may have been contaminated with SN emission). The HST CANDELS/CLASH survey parameter space is indicated by the red shaded area.}
    \label{magnitude_redshift}
\end{figure*}

The JADES-SN-22 and JADES-SN-23 samples have greatly increased the number of high-redshift SN detections. We plot F200W and F356W apparent AB magnitude vs redshift and compare to the parameter space covered by F160W in the HST CANDELS/CLASH survey in Figure \ref{magnitude_redshift}. 

JWST more than doubles the supernova redshift regime and expands the AB magnitude parameter space by multiple magnitudes beyond what was possible with HST. Comparing the plots at $z$\,$>$\,2 shows that longer wavelength images are more optimal for detecting high-redshift SNe, as high-redshift SNe generally peak at longer wavelengths. There is a stronger trend of brightness decreasing with redshift in the F200W plot than in the F356W plot. 

\begin{figure*}
    \centering
 {\includegraphics[width=11cm]{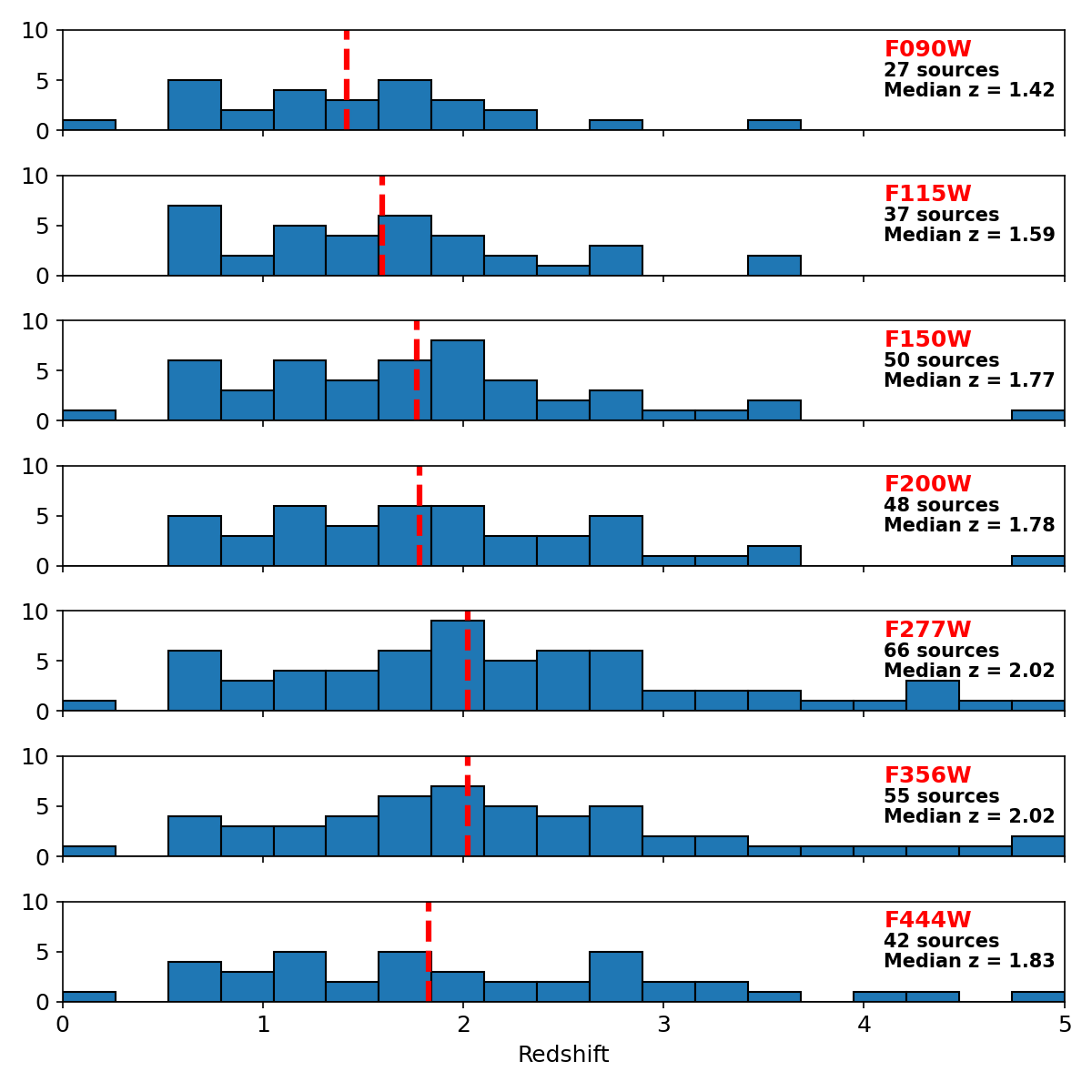}}
    \caption{Redshift distribution of SNe recovered by each of the NIRCam wide bands. For each filter, a ``recovery'' means that \texttt{DAOStarFinder} detected the source in that filter's Epoch2-Epoch1 (or Epoch1-Epoch2) difference image, and the source passed all of the selection criteria detailed in Section \ref{sec:detection}. The red dashed line in each panel indicates the median redshift in the distribution. The F277W and F356W bands result in the highest fraction of candidates detected (66 and 55 sources, respectively), and these bands have the highest median redshift distribution ($z_{\mathrm{med}}$\,$=$\,2.02).}
    \label{redshift_histogram}
\end{figure*}

In Figure \ref{redshift_histogram}, we show a redshift histogram for each of the 7 NIRCam wide bands that we used to conduct the SN search. Note again that we searched each band separately, combining the full sample only after finalizing the SN sample for each individual band. While the finalized SN sample for each band depends on the SN's S/N in the 6 other bands (see Section \ref{subsec:transient_selection}), the histogram shows the redshift distribution of sources that \texttt{DAOStarFinder} was capable of detecting in each band. 

The median recovered SN redshift for each filter, indicated by the vertical dashed red lines in Figure \ref{redshift_histogram}, shows that the longer wavelength filters detect higher-redshift supernovae more effectively (except for F444W, which is less sensitive due to a higher sky background). F277W and F356W each recover SNe with median redshift $z$\,$=$\,2.02, whereas F200W and F150W  recover SNe with median redshifts $z$\,$=$\,1.78 and $z$\,$=$\,1.77, respectively.  Although the median redshifts are not so different since the majority of detected SNe are at $z$\,$\leq$\,2 (where the SW and LW data are equally effective), the slight shift toward higher redshift with F277W and F356W reflects the existence of a significant high-redshift tail in the SN samples with these LW filters as seen in Figure~\ref{redshift_histogram}. These additional high-redshift detections cause F277W and F356W to recover the highest fraction of candidates detected (66 and 55 sources, respectively). 

NIRCam LW filters such as F277W and F356W are especially powerful for detecting high-redshift SNe.  This is because these mid-infrared filters sample the  Rayleigh-Jeans part of a SN spectrum up to a high redshift, meaning that the source would brighten at higher redshift as these filters sample the SN SED closer and closer to its peak.  This partially compensates the dimming due to the increased distance, causing the F277W/F356W apparent magnitudes of high-redshift SNe to drop much more slowly compared to those at shorter-wavelengths \citep{Escude1997}.



\subsection{Any Exotic SNe such as SLSNe and PISNe?}

\begin{figure*}
    \centering
 {\includegraphics[width=15cm]{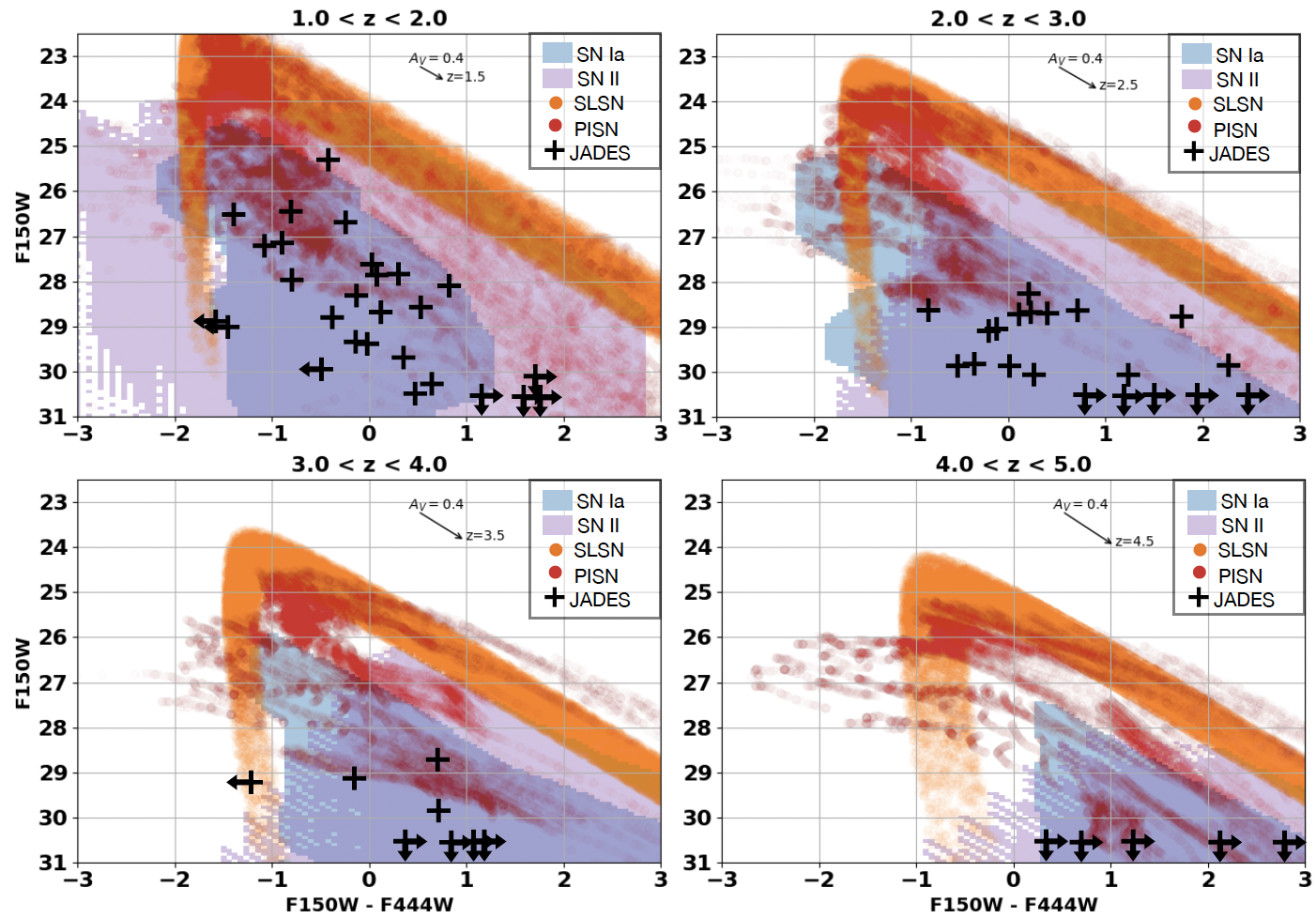}}
    \caption{F150W vs. F150W-F444W color-magnitude diagram binned by redshift (1\,$<$\,$z$\,$<$\,2 in upper left, 2\,$<$\,$z$\,$<$\,3 in upper right, 3\,$<$\,$z$\,$<$\,4 in lower left, 4\,$<$\,$z$\,$<$\,5 in lower right). The black plus signs show the JADES sample, with arrows indicating upper limits. The underlying shaded regions indicate the expected parameter space populated by fiducial SNe Ia (blue), SNe II (purple), SLSNe (orange), and PISNe (red). The SN~Ia template is from \citet{Hsiao2007K-CorrectionsSupernovae} with the peak $r$ band magnitude of $-19.3~\mathrm{mag}$. For SNe~II, we take the Nugent template  (\url{https://c3.lbl.gov/nugent/nugent_templates.html})
    with the peak $r$ band magnitude of $-18.5~\mathrm{mag}$. The SLSN template is from \citet{Moriya2022DiscoveringTelescope} and the PISN models are from \citet{Kasen2011PairBreakout}. For all the models, we adopt $A_V$\,$=$\,0, 1, and 2~mag to determine the populating regions. All of the JADES SNe fall within the expectations for normal SNe except for \tr{13}, which is the bluest point in the 3\,$<$\,$z$\,$<$\,4 panel. \tr{13} falls in an area populated by SLSNe, making it a SLSN candidate. However, see Section \ref{tr13} for an explanation of why we do not consider \tr{13} to be a SLSN.}
    \label{f150w_vs_f150w-f444w}
\end{figure*}

To determine if the sample contains any superluminous supernovae (SLSNe) or pair-instability supernovae (PISNe), we plot the F150W vs. F150W-F444W color-magnitude diagram (CMD) in four redshift bins (1\,$<$\,$z$\,$<$\,2, 2\,$<$\,$z$\,$<$\,3, 3\,$<$\,$z$\,$<$\,4, and 4\,$<$\,$z$\,$<$\,5) in Figure \ref{f150w_vs_f150w-f444w}. The JADES-SN-22 and JADES-SN-23 data points are overlaying fiducial populations of SNe Ia, SN II, SLSNe, and PISNe. All of the JADES SNe fall within expectations for normal SNe except for \tr{13}, which is the bluest point in the 3\,$<$\,$z$\,$<$\,4 (bottom left) panel. It falls in a region populated by the fiducial SLSNe, making it a SLSN candidate. After analyzing its light curve and Epoch2 photometry, however, we do not believe that \tr{13} is a SLSN (see Section \ref{tr13}). None of the JADES SNe fall within the PISNe regions, so there are no PISNe candidates in the sample.

\subsection{Supernovae Interloping as $z$\,$\sim$\,16 Galaxy Candidates} \label{subsec:z16_interlopers}

\begin{figure*}
    \centering
 {\includegraphics[width=8.5cm]{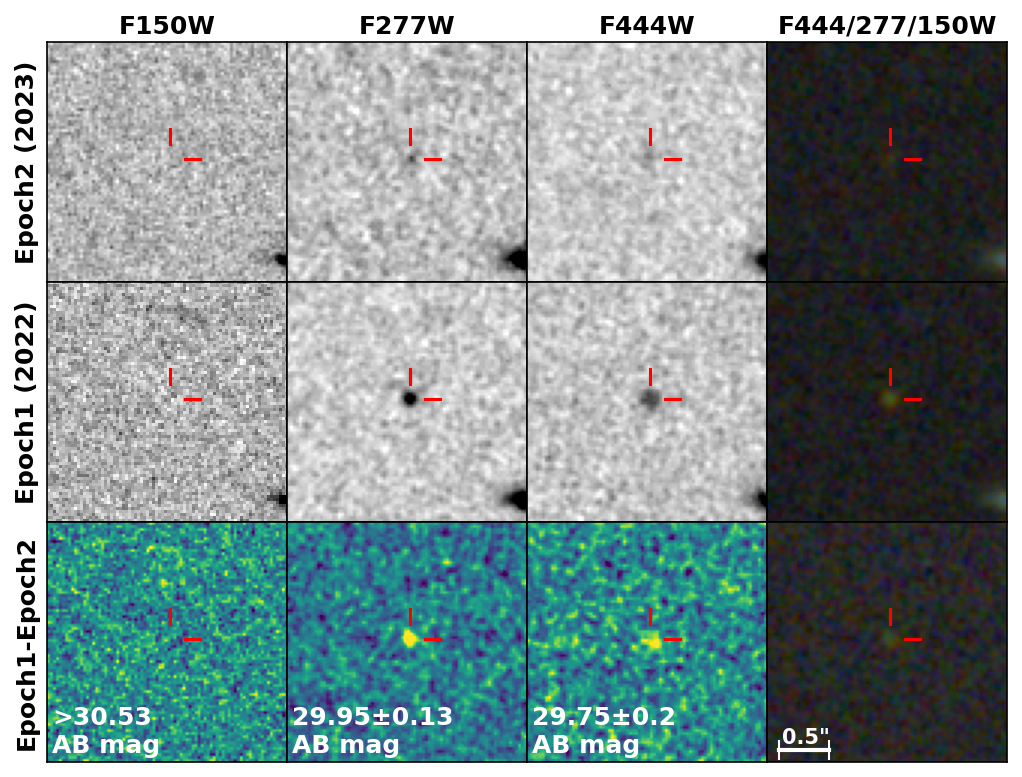}}
 \qquad
 {\includegraphics[width=8.5cm]{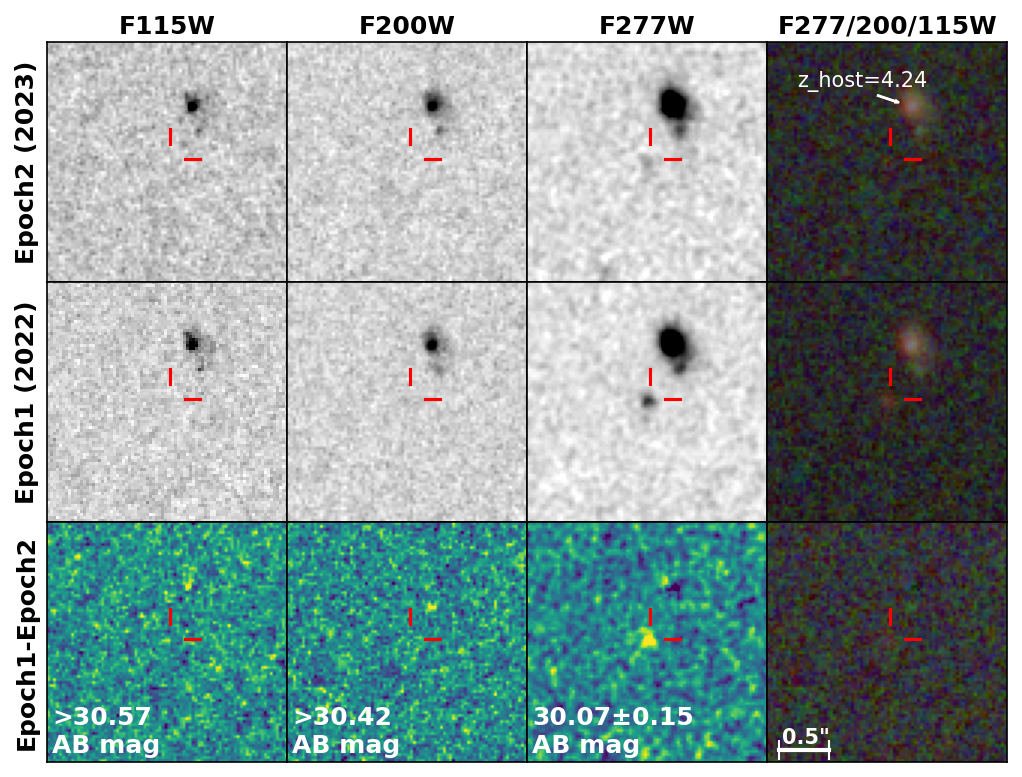}}
  \caption{Epoch1, Epoch2, and difference images of \tr{101} (left panel; F444W/F277W/F150W) and \tr{107} (right panel; F277W/F200W/F115W). \textit{Top row}: The Epoch2 NIRCam images that show very faint LW emission from \tr{101} and \tr{107} (or emission from \tr{101}'s potential host galaxy) because they faded between Epoch1 and Epoch2. Neither source shows SW emission in Epoch2. \textit{Middle row}: Epoch1 NIRCam images showing LW emission from \tr{101} and \tr{107}. Prior to the Epoch2 imaging, \tr{101} was considered a $z_{\mathrm{phot}}$\,$=$\,15.77 galaxy candidate \citep{Hainline2023TheGOODS-N} because of its compact morphology and lack of emission at wavelengths short of F200W, mimicking a Lyman break. \tr{107} was a $z_{\mathrm{phot}}$\,$=$\,15.89 galaxy candidate in the JADES March 2023 galaxy catalog for similar reasons. \textit{Bottom row}: The Epoch1-Epoch2 difference images showing fading LW brightness for \tr{101} and \tr{107}. While the difference image emission is faint, the sources clearly faded at LW between Epoch1 and Epoch2, indicating that they are transients (likely SNe) rather than $z$\,$\sim$\,16 galaxy candidates.}
  \label{z16_candidates}
\end{figure*}

Two of the sources that faded at LW between the Epoch1 and Epoch2 imaging were $z$\,$\sim$\,16 galaxy candidates prior to the Epoch2 imaging. These sources are \tr{101} and \tr{107}, shown in Figure \ref{z16_candidates}.

\tr{101} (left side of Figure \ref{z16_candidates}) is the z$_{\mathrm{phot}}$\,$=$\,15.77 galaxy candidate JADES-GS-53.12692-27.79102 in \citet{Hainline2023TheGOODS-N}, and \tr{107} (right side of Figure \ref{z16_candidates}) is a z$_{\mathrm{phot}}$\,$=$\,15.89 galaxy candidate from the JADES March 2023 galaxy catalog (although this source was eventually rejected as a high-redshift galaxy due to 4$\sigma$ emission in F090W; see JADES-GS-53.20055-27.78493 in \citealt{Hainline2023TheGOODS-N}). These redshift solutions resulted from $\tt{EAZY}$ SED fitting, with the exact methodology described in \citet{Hainline2023TheGOODS-N}. These sources had such high redshift solutions because they lacked SW emission in the Epoch1 NIRCam imaging and were below the HST detection threshold, showing up as F150W dropouts.

With two epochs of imaging data (Epoch1 and Epoch2), we determined that \tr{101} and \tr{107} are not $z$\,$\sim$\,16 galaxy candidates but rather are transients. They faded in F277W, F356W, and F444W (see Figure \ref{z16_candidates}). 

It is unclear if the emission at \tr{101}'s location in Epoch2 is a host galaxy or remaining SN emission, so we do not assign a redshift to \tr{101}. There is no other reasonably nearby potential host galaxy. If the Epoch2 emission is from a host galaxy, \tr{101} could be still at reasonably high redshift. However, the fidelity of a high-redshift solution decreases dramatically because there would no longer be a sharp break in the SED. See Section \ref{subsec:highz_sne} for details on the galaxy SED fitting attempts on the remaining LW emission in Epoch2. 

\tr{107} lies near a galaxy with $\it{z}_{\mathrm{phot}}$\,$=$\,4.24 (JADES-GS+53.20046-27.78476), shown in the right panel of Figure \ref{z16_candidates}. We assign this galaxy as the host, as there is very little emission remaining at \tr{107}'s location in Epoch2. We interpret the small amount of remaining emission as leftover SN emission. \tr{107} is a robust high-redshift ($z$\,$\geq$\,4) SN candidate because it emits only at LW.

See Section \ref{subsec: z16_interpretation} for a discussion about the implications that these $z$\,$\sim$\,16 galaxy candidate interlopers have on high-redshift galaxy surveys.


\section{Discussion} \label{sec:discussion}

\subsection{Classification} \label{subsec:classification}

We attempted to classify each of the SN candidates using the STARDUST2 Bayesian light curve classification tool \citep{Rodney2014TypeUniverse}, which is built on the underlying \texttt{SNCosmo}
framework and was originally designed for classifying SNe using HST. STARDUST2 uses the SALT3-NIR model to represent Type\,Ia SNe \citep[]{Pierel2022SALT3-NIR:Measurements, Guy2007SALT2, Kenworthy2021SALT3} and a collection of spectrophotometric time series templates to represent CC SNe (27 Type\,II and 15 Type\,Ib/c). The CCSN templates are those developed for the SN analysis software \texttt{SNANA} \citep{Kessler2009SNANA:Analysis}, derived from the SN samples of the Sloan Digital Sky Survey \citep{Frieman2008THESUMMARY,Sako2008THEOBSERVATIONS,DAndrea2010TYPEMETHOD}, \textit{Supernova Legacy Survey} \citep{Astier2006TheSet}, and \textit{Carnegie Supernova Project} \citep{Hamuy2006TheSurvey,Stritzinger2009THEWAVELENGTHS,Morrell2012CarnegieSupernovae}. The models produced for \texttt{SNANA} were extended to the NIR by \citet{Pierel2018ExtendingObservations}. Within STARDUST2 a nested sampling algorithm \citep{Skilling2004NestedSampling} measures likelihoods over the SN simulation parameter space, including priors on dust parameters described in \citet{Rodney2014TypeUniverse}. We use all available data for each classification, which is 1-4 light curve epochs depending on the SN location and up to 7 wide-band filters and 2 medium-band filters. STARDUST2 has been successfully implemented for single-epoch classification in the past \citep{Golubchik2023ASupernovae}, and we are able to validate it by comparing single-epoch classifications to spectroscopic classifications for the sub-sample with spectra. 

We performed light-curve fitting for each of the SNe, and 40 of the 45 JADES-SN-22 fits and 30 of the 34 JADES-SN-23 fits converged successfully. The JADES-SN-22 sample contains 9 Type\,Ia, 5 Type\,Ib/c, and 26 Type\,II SN candidates, and the JADES-SN-23 sample contains 2 Type\,Ia, 8 Type\,Ib/c, and 20 Type\,II SN candidates. Refer to Tables \ref{tab_classifications_2023} and \ref{tab_classifications_2022}, respectively, for the JADES-SN-23 and JADES-SN-22 classifications, best-fit models, and best-fit model redshifts compared to the host redshifts. 

Some sources were imaged multiple times with the same filter on the same day through different programs (i.e., the overlapping filters of Epochs5.1/5.2/5.3, taken on UT 2024 Jan 1). For light curve fitting, we averaged this photometry filter-by-filter into a single Epoch5.

If the SN's host had a spec-$z$, this spec-$z$ was set as a fixed parameter in the light curve fitting and thus the best-fit model redshift equals the host spec-$z$. However, in cases where the host only had a photo-$z$, redshift was set as a free parameter with the host photo-$z$ and its associated uncertainties set as priors. The best-fit model redshift, in these cases, corresponds to the redshift that the light curve fitting converged upon for the best-fit SN model. In Tables \ref{tab_classifications_2023} and \ref{tab_classifications_2022}, spectroscopic host redshifts are shown in bold, and the uncertainties listed for the photometric host redshifts are the larger of the $\pm$1$\sigma$ uncertainties listed in Tables \ref{tab_JD23_redshifts} and \ref{tab_JD22_redshifts}. 

To provide a general sense of our confidence in these fits, Tables \ref{tab_classifications_2023} and \ref{tab_classifications_2022} show the probability of Type Ia vs.\ II vs.\ Ib/c based on the \texttt{SNCosmo} analysis, as well as the $\chi^2$ per degrees of freedom (DOF) and number of DOF for the best-fit model. We also list the number of epochs considered in the light curve fitting for the JADES-SN-23 sample, where Epochs5.1-5.3 are considered as one epoch. All JADES-SN-22 sources are covered by one epoch except for \tr{20}, which has 5 epochs of coverage. The relative probabilities used to determine the SN type are Bayesian probabilities from \texttt{SNCosmo} based upon the goodness of fit for individual SN models. They should not be interpreted as the likelihood of the source being Type Ia vs. II vs. Ib/c. For example, a source may have poor fits for both Type\,II and Ib/c models, but the Type\,II model fits are comparatively better, and so the Type\,II probability ends up being very high despite a poor fit to the data. The $\chi^2$/DOF value for the best-fit model is a better indicator of the quality of the classification, with $\sim$\,5 being our assumed limit for a robust classification. There are some sources, however, with multiple epochs of coverage that have $\chi^2$/DOF\,$>$\,5 but the best-fit light curve model fits the data well (e.g., \tr{26}, \tr{28}, and \tr{24}). In these cases, the high-value of $\chi^2$/DOF is likely driven by underestimated photometric uncertainties rather than a poor fit to the data.

The best-fit model is the model with the lowest $\chi^2$/DOF taken from the list of models with the highest-probability SN type. There are a few cases where Type\,Ia is most-favored by the \texttt{SNCosmo} probabilities but the \texttt{x1} parameter, which describes the ``shape'' or ``stretch'' of the fit, is unphysical, so we assign the best-fit model from next highest-probability SN type. We mark these cases in Tables \ref{tab_classifications_2023} and \ref{tab_classifications_2022} with an asterisk. 

Many of the sources, including all of the JADES-SN-22 sources, only have one point along the light curve. To test how reliably we can classify SNe based on just one  epoch, we performed the classification scheme for 6 of the SNe whose types were spectroscopically-confirmed through DDT Program 6541 (\tr{7}, \tr{10}, \tr{22}, \tr{26}, \tr{28}, and \tr{29}) based on just one of their 3 or more imaging epochs. For each epoch, we performed the classification scheme and compared this single-epoch classification result to the multi-epoch and spectroscopic classification result. The comparisons indicate that single-epoch classifications reliably distinguish Type\,Ia versus CC SNe but discern Type\,II vs. Ib/c with only $\sim$\,50\% accuracy. However, we note that the JADES-SN-23 sample, which benefits from multi-epoch light curve fitting, only has two Type\,Ia SN candidates (\tr{9} and \tr{27}), whereas the JADES-SN-22 sample, which relies entirely on single-epoch light curve fitting, has 9 Type\,Ia candidates. It is unlikely that all 9 of these SNe are really Type\,Ia SNe, but their photometry is best-fit by the SALT3-NIR model with a physically reasonable \texttt{x1} parameter (-2\,$\leq$\,\texttt{x1}\,$\leq$\,2), so we list them as Type\,Ia candidates.

The fitted light curves for JADES-SN-23 and JADES-SN-22 SNe that have best-fit models with $\chi^2$/DOF\,$\leq$\,50 
are shown in Appendices \ref{sec:deep_2023} and \ref{sec:deep_2022}, respectively. Note that the error bounds shown for the light curves are only statistical in nature. They do not account for model uncertainties.

We note that \tr{33}'s light curve fitting yields a $z$\,$=$\,6.83 Type\,II solution, but we consider this fit unreliable as it is based on just one epoch of observation.  While \tr{33} is still a high-redshift SN candidate based on its host photometric redshift of $z_{\mathrm{phot}}$\,$=$\,4.82\,$\pm$\,0.49 
(see Figure~\ref{galaxy_seds} and Section~\ref{subsec:highz_sne}),
we do not claim that it is a $z$\,$=$\,6.83 SN.

Below, we highlight the best-fit light curves and classifications of five of the most interesting SNe in the sample: \tr{27}, \tr{26}, \tr{10}, \tr{13}, and \tr{20}.

\begin{deluxetable*}{cCCCccccccL}
\tablecaption{JADES-SN-23 Classifications}
\tablehead{
\colhead{ID} & \colhead{P(Ia)} & \colhead{P(II)} & \colhead{P(Ibc)} & \colhead{Type} & \colhead{Best-Fit Model} & \colhead{Model $\chi^2$/DOF} & \colhead{DOF} & \colhead{\# of Epochs} & \colhead{Best-Fit Model z} & \colhead{Host z}
} 
\label{tab_classifications_2023}
\startdata
\tr{53}  & 0       & 1       & 99      & Ibc     & snana-2006lc & 1.1     & 9       & 3 & 4.35    & 4.35 \pm 0.04  \\
\tr{50}  & 0       & 100     & 0       & II      & snana-2007og & 1.7     & 22      & 4 & 4.117   & \textbf{4.117} \\		 
\tr{88}* & 53      & 16      & 31      & Ibc     & snana-2004gq & 0.8     & 10      & 2 & 3.70    & 3.74 \pm 0.17  \\
\tr{10}  & 0       & 100     & 0       & II      & snana-2006kv & 1.3     & 13      & 3 & 3.61    & \textbf{3.61}  \\		
\tr{44}  & 7       & 91      & 2       & II      & snana-2006gq & 2.5     & 23      & 4 & 2.78    & 3.21 \pm 0.58  \\
\tr{71}* & 97      & 0       & 3       & Ibc     & snana-2006jo & 2.6     & 12      & 3 & 3.090   & \textbf{3.090} \\		
\tr{27}  & \nodata & \nodata & \nodata & Ia      & salt3-nir    & 3.9     & 29      & 4 & 2.90    & \textbf{2.90}  \\	
\tr{36}  & 0       & 100     & 0       & II      & snana-2006ez & 1.4     & 28      & 4 & 3.06    & 2.86 \pm 0.10  \\
\tr{26}  & 0       & 0       & 100     & Ibc     & snana-2004gv & 32.3    & 29      & 4 & 2.83    & \textbf{2.83} \\
\tr{52}  & 15      & 85      & 0       & II      & snana-2007ld & 2.1     & 14      & 3 & 2.63    & 2.78 \pm 0.12  \\
\tr{15}  & \nodata & \nodata & \nodata & \nodata & \nodata      & \nodata & \nodata & 3 & \nodata & 2.77 \pm 0.86  \\
\tr{29}  & 0       & 0       & 100     & Ibc     & snana-2004ib & 3.9     & 18      & 3 & 2.73    & \textbf{2.73}  \\
\tr{6}   & 0       & 100     & 0       & II      & snana-2004hx & 11.7    & 15      & 3 & 2.623   & \textbf{2.623} \\
\tr{11}  & 0       & 100     & 0       & II      & snana-2007lb & 1.8     & 14      & 3 & 2.344   & \textbf{2.344} \\
\tr{19}  & 1       & 99      & 0       & II      & snana-2006kn & 1.6     & 4       & 1 & 2.31    & 2.24 \pm 0.13  \\
\tr{7}   & 0       & 56      & 44      & II      & snana-2007pg & 5.9     & 16      & 3 & 2.06    & \textbf{2.06}  \\
\tr{28}  & 0       & 100     & 0       & II      & snana-2007ms & 10.9    & 28      & 4 & 1.73    & 1.94 \pm 0.12  \\
\tr{87}  & 0       & 81      & 19      & II      & snana-2007pg & 1.5     & 10      & 2 & 1.932   & \textbf{1.932} \\
\tr{60}  & 0       & 100     & 0       & II      & snana-2007kw & 5.7     & 5       & 1 & 1.912   & \textbf{1.912} \\
\tr{5}   & 0       & 0       & 100     & Ibc     & snana-2004gq & 10.6    & 12      & 3 & 1.85    & 1.86 \pm 0.10  \\
\tr{9}   & 100     & 0       & 0       & Ia      & salt3-nir    & 3.1     & 14      & 4 & 1.854   & \textbf{1.854} \\
\tr{83}  & 2       & 96      & 2       & II      & snana-2005gi & 3.0     & 5       & 1 & 1.748   & \textbf{1.748} \\
\tr{22}  & 0       & 100     & 0       & II      & snana-2006iw & 33.1    & 29      & 4 & 1.62    & \textbf{1.62}  \\	
\tr{35}  & 0       & 100     & 0       & II      & snana-2007lz & 1.5     & 9       & 2 & 1.500   & \textbf{1.500} \\
\tr{30}  & 14      & 86      & 0       & II      & snana-2007lz & 181.1   & 4       & 1 & 1.41    & 1.19 \pm 0.11  \\
\tr{81}  & 0       & 100     & 0       & II      & snana-2006gq & 7.3     & 10      & 2 & 1.171   & \textbf{1.171} \\
\tr{48}  & 0       & 26      & 74      & Ibc     & snana-2006jo & 1.3     & 12      & 3 & 1.14    & 1.16 \pm 0.05  \\
\tr{45}  & 0       & 0       & 100     & Ibc     & snana-2004ib & 9.9     & 4       & 2 & 1.139   & \textbf{1.139} \\
\tr{24}  & 0       & 100     & 0       & II      & snana-2006iw & 91.1    & 8       & 2 & 1.01    & \textbf{1.01}  \\	
\tr{82}  & 0       & 90      & 10      & II      & snana-2007ll & 2.4     & 3       & 1 & 0.665   & \textbf{0.665} \\
\tr{14}  & 0       & 100     & 0       & II      & snana-2004hx & 58.8    & 12      & 3 & 0.657   & \textbf{0.657} \\	
\tr{90}  & \nodata & \nodata & \nodata & \nodata & \nodata      & \nodata & \nodata & 1 & \nodata & \textbf{0.533} \\		
\tr{89}  & \nodata & \nodata & \nodata & \nodata & \nodata      & \nodata & \nodata & 1 & \nodata & \textbf{0.210} \\	
\tr{25}  & \nodata & \nodata & \nodata & \nodata & \nodata      & \nodata & \nodata & 1 & \nodata & \nodata        \\
\sidehead{Marginal Detections}
\tr{84}  & 29      & 64      & 7       & II      & snana-2007ll & 2.4     & 24      & 4 & 1.78    & 1.86 \pm 0.11  \\
\tr{59}  & 0       & 100     & 0       & II      & snana-2007nr & 5.9     & 6       & 2 & 0.996   & \textbf{0.996} \\
\enddata
\tablecomments{ * indicates P(Ia) has highest probability value but Ia fit is unphysical}
\tablecomments{ P(Ia) vs. P(II) vs. P(Ibc) should not be interpreted as percentage likelihood of SN types. See text for more details.}
\tablecomments{ \textbf{Bold} redshift = spectroscopic}
\tablecomments{ Model redshift shown simply to compare to host redshift as fit quality indicator. Should not be quoted as SN redshift}
\end{deluxetable*}
\begin{deluxetable*}{cCCCcccccL}
\tablecaption{JADES-SN-22 Classifications}
\tablehead{
\colhead{ID} & \colhead{P(Ia)} & \colhead{P(II)} & \colhead{P(Ibc)} & \colhead{Type} & \colhead{Best-Fit Model} & \colhead{Model $\chi^2$/DOF} & \colhead{DOF} & \colhead{Best-Fit Model z} & \colhead{Host z}
} 
\label{tab_classifications_2022}
\startdata
\tr{77}  & 87      & 2       & 11      & Ia      & salt3-nir        & 0.4     & 2       & 4.80    & 4.82 \pm 0.05  \\
\tr{33}  & 2       & 98      & 0       & II      & snana-2006ix     & 6.4     & 4       & 6.78    & 4.82 \pm 0.49  \\
\tr{39}  & 84      & 3       & 13      & Ia      & salt3-nir        & 1.6     & 3       & 4.504   & \textbf{4.504} \\
\tr{93}* & 83      & 2       & 15      & Ibc     & snana-2004gv     & 3.9     & 5       & 4.471   & \textbf{4.471} \\	
\tr{107} & 62      & 14      & 24      & Ia      & salt3-nir        & 0.6     & 2       & 4.43    & 4.24 \pm 0.09  \\
\tr{102} & 13      & 85      & 2       & II      & snana-2006ez     & 21.3    & 1       & 3.47    & 3.96 \pm 0.14  \\
\tr{103} & 56      & 26      & 18      & Ia      & salt3-nir        & 3.4     & 4       & 3.605   & \textbf{3.605} \\
\tr{13}  & 4       & 0       & 96      & Ibc     & snana-04d1la     & 12.4    & 2       & 3.44    & 3.58 \pm 0.14  \\
\tr{38}  & 87      & 9       & 4       & Ia      & salt3-nir        & 1.7     & 4       & 3.166   & \textbf{3.166} \\
\tr{55}  & 0       & 50      & 50      & Ibc     & snana-2004ib     & 1.8     & 3       & 2.82    & 2.79 \pm 0.11  \\
\tr{21}  & 5       & 95      & 0       & II      & snana-2007ms     & 15.2    & 3       & 2.92    & 2.73 \pm 0.39  \\
\tr{100} & \nodata & \nodata & \nodata & \nodata & \nodata          & \nodata & \nodata & \nodata & 2.62 \pm 1.54  \\
\tr{79}  & 0       & 41      & 59      & Ibc     & snana-2006lc     & 0.8     & 4       & 2.617   & \textbf{2.617} \\
\tr{80}  & 0       & 83      & 17      & II      & snana-2007lz     & 1.0     & 4       & 2.617   & \textbf{2.617} \\
\tr{95}* & 93      & 4       & 3       & II      & snana-2007iz     & 4.1     & 4       & 2.74    & 2.56 \pm 0.39  \\
\tr{8}   & \nodata & \nodata & \nodata & \nodata & \nodata          & \nodata & \nodata & \nodata & 2.48 \pm 0.68  \\
\tr{34}  & 100     & 0       & 0       & Ia      & salt3-nir        & 2.2     & 5       & 2.323   & \textbf{2.323} \\
\tr{23}  & 0       & 100     & 0       & II      & snana-2006jl     & 4.8     & 5       & 2.315   & \textbf{2.315} \\ 	
\tr{66}  & 8       & 84      & 8       & II      & snana-2007nv     & 1.4     & 4       & 2.41    & 2.29 \pm 0.23  \\
\tr{92}* & 79      & 13      & 8       & II      & snana-2005gi     & 3.1     & 4       & 1.79    & 2.02 \pm 0.29  \\
\tr{37}  & 0       & 100     & 0       & II      & snana-2007pg     & 2.4     & 4       & 1.78    & 2.01 \pm 0.16  \\
\tr{20}  & \nodata & \nodata & \nodata & \nodata & \nodata          & \nodata & \nodata & \nodata & 2.00 \pm 0.37  \\
\tr{111} & 54      & 46      & 0       & Ia      & salt3-nir        & 0.4     & 4       & 1.92    & \textbf{1.92}  \\
\tr{2}   & 81      & 19      & 0       & Ia      & salt3-nir        & 554.0   & 4       & 1.85    & 1.79 \pm 0.26  \\
\tr{16}  & 0       & 100     & 0       & II      & snana-2007lj     & 87.2    & 5       & 1.771   & \textbf{1.771} \\
\tr{64}  & 0       & 100     & 0       & II      & snana-2006iw     & 2.5     & 5       & 1.766   & \textbf{1.766} \\
\tr{1}   & 100     & 0       & 0       & Ia      & salt3-nir        & 27.7    & 5       & 1.688   & \textbf{1.688} \\
\tr{69}  & 6       & 93      & 1       & II      & snana-2007ld     & 1.4     & 4       & 1.56    & 1.62 \pm 0.07  \\
\tr{12}  & 0       & 100     & 0       & II      & snana-2006iw     & 1.2     & 5       & 1.567   & \textbf{1.567} \\
\tr{46}  & 0       & 100     & 0       & II      & snana-2006iw     & 6.7     & 4       & 1.40    & 1.42 \pm 0.11  \\
\tr{65}  & 0       & 100     & 0       & II      & snana-2006jl     & 2.8     & 5       & 1.415   & \textbf{1.415} \\
\tr{54}  & 27      & 16      & 57      & Ibc     & snana-2006ep     & 59.2    & 4       & 1.18    & 1.36 \pm 0.21  \\
\tr{67}  & 0       & 100     & 0       & II      & snana-2007ms     & 3.1     & 5       & 1.294   & \textbf{1.294} \\
\tr{56}  & 0       & 100     & 0       & II      & snana-2007lx     & 8.5     & 5       & 1.244   & \textbf{1.244} \\
\tr{68}  & 0       & 100     & 0       & II      & snana-2007lz     & 5.5     & 5       & 1.114   & \textbf{1.114} \\
\tr{18}  & 0       & 100     & 0       & II      & snana-2007kw     & 1.4     & 4       & 1.094   & \textbf{1.094} \\
\tr{17}  & 0       & 100     & 0       & II      & snana-2007kw     & 35.5    & 4       & 0.996   & \textbf{0.996} \\
\tr{31}  & 0       & 100     & 0       & II      & snana-2007lz     & 3.0     & 4       & 0.953   & \textbf{0.953} \\
\tr{61}  & 0       & 100     & 0       & II      & snana-2006gq     & 19.3    & 3       & 0.669   & \textbf{0.669} \\
\tr{109} & 25      & 40      & 35      & II      & snana-2007ll     & 0.7     & 3       & 0.669   & \textbf{0.669} \\
\tr{3}   & 0       & 100     & 0       & II      & snana-2007ld     & 4.9     & 3       & 0.665   & \textbf{0.665} \\
\tr{4}   & 42      & 53      & 5       & II      & snana-2006jl     & 0.8     & 3       & 0.665   & \textbf{0.665} \\
\tr{110} & 15      & 85      & 0       & II      & snana-2007pg     & 2.5     & 2       & 0.540   & \textbf{0.540} \\
\tr{101} & \nodata & \nodata & \nodata & \nodata & \nodata          & \nodata & \nodata & \nodata & \nodata        \\		
\tr{32}  & \nodata & \nodata & \nodata & \nodata & \nodata          & \nodata & \nodata & \nodata & \nodata        \\
\sidehead{Marginal Detections}
\tr{96}* & 67      & 11      & 22      & Ibc     & snana-sdss014475 & 4.8     & 3       & 3.913   & \textbf{3.913} \\
\tr{94}  & 0       & 100     & 0       & II      & snana-2007iz     & 2.5     & 4       & 2.29    & 2.67 \pm 0.20  \\
\enddata
\tablecomments{ Same notes as Table \ref{tab_classifications_2023}}
\end{deluxetable*}

\subsubsection{AT2023adsy: Type\,Ia at $z_{\rm spec}$\,$=$\,2.90}

We fit the light-curve for \tr{27}, which was spectroscopically-confirmed as a Type\,Ia SN at $z$\,$=$\,2.90 in host JADES-GS+53.13485-27.82088 with DDT Program 6541. This is a groundbreaking result, as it is the highest redshift spectroscopically-confirmed Type\,Ia SN discovered thus far. Refer to \citet{Pierel24Ia} for a complete analysis of \tr{27}, which presents the SN spectrum and photometric light curve fitting and provides the first glimpse of Type\,Ia SNe as standard candles at $z$\,$>$\,2. 

\begin{figure}
    \centering
 {\includegraphics[width=8cm]{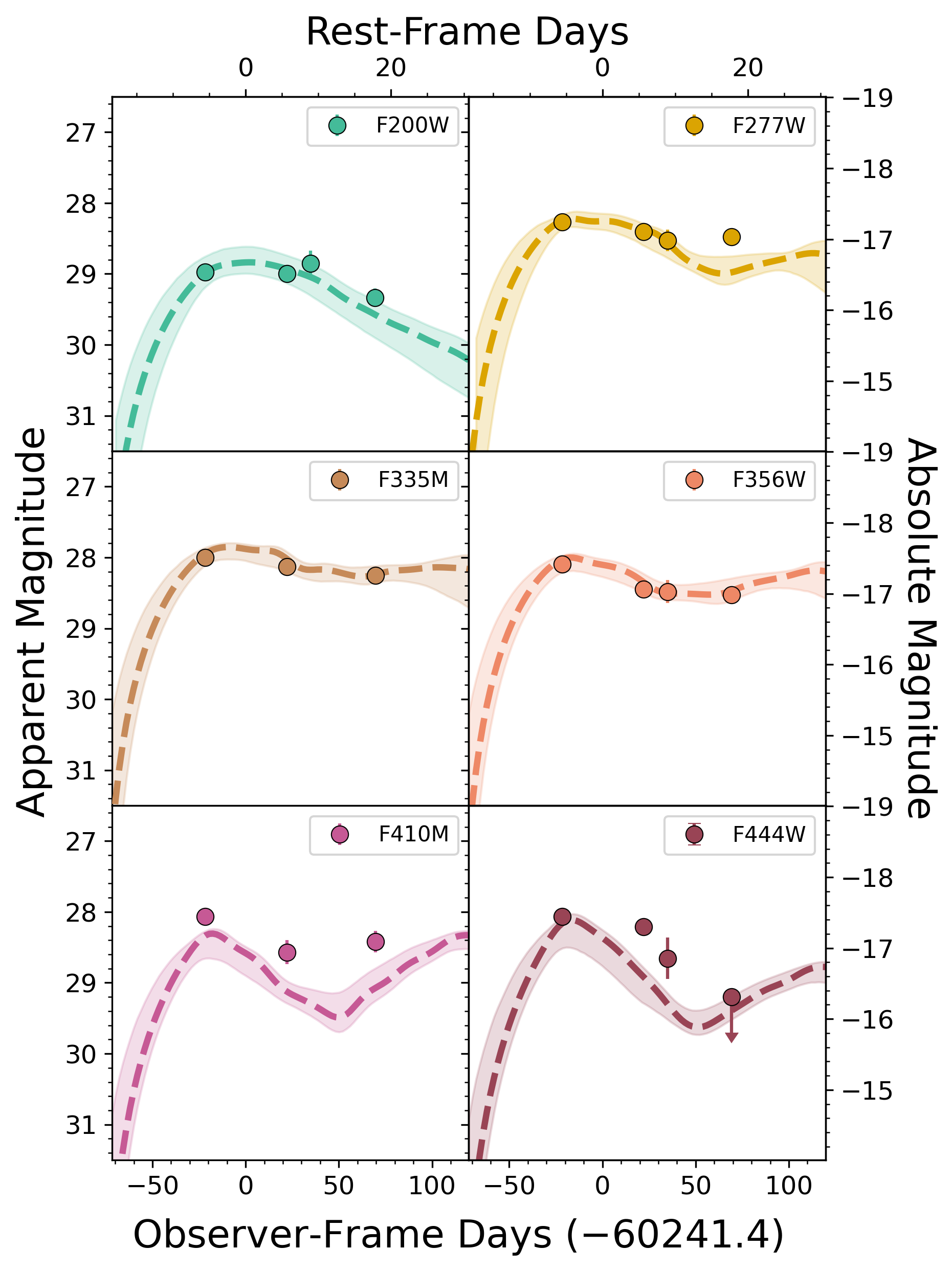}}
    \caption{Type\,Ia SALT3-NIR light curve fit at $z$\,$=$\,2.90 for \tr{27} in F200W, F277W, F335M, F356W, F410M, and F444W. The bottom (top) x-axis is shown in observer-frame (rest-frame) days since peak brightness (minus 60241.4 in the observer-frame). The left (right) y-axis shows apparent (absolute) magnitudes. The circles show measured PSF photometry for \tr{27} and circles with downward arrows indicate 2$\sigma$ upper limits. The thick dashed line shows the SALT3-NIR fit with $\pm1\sigma$ error bars as the shaded region.}
    \label{tr27_lc}
\end{figure}

Figure \ref{tr27_lc} shows the fitted SALT3-NIR light curve. Most of the photometric measurements are within 1$\sigma$ of the $z$\,$=$2.90 Type\,Ia light curve, indicating the robustness of the fit.  Since the SN Ia classification was adopted from \citet{Pierel24Ia}, we do not list \texttt{SNCosmo} SN type probabilities for \tr{27} in Table \ref{tab_classifications_2023}.

\subsubsection{AT2023adta: Type Ic-BL at $z_{\rm spec}$\,$=$\,2.83}
\label{at2023adta}

\begin{figure*}
    \centering
 {\includegraphics[width=8.5cm]{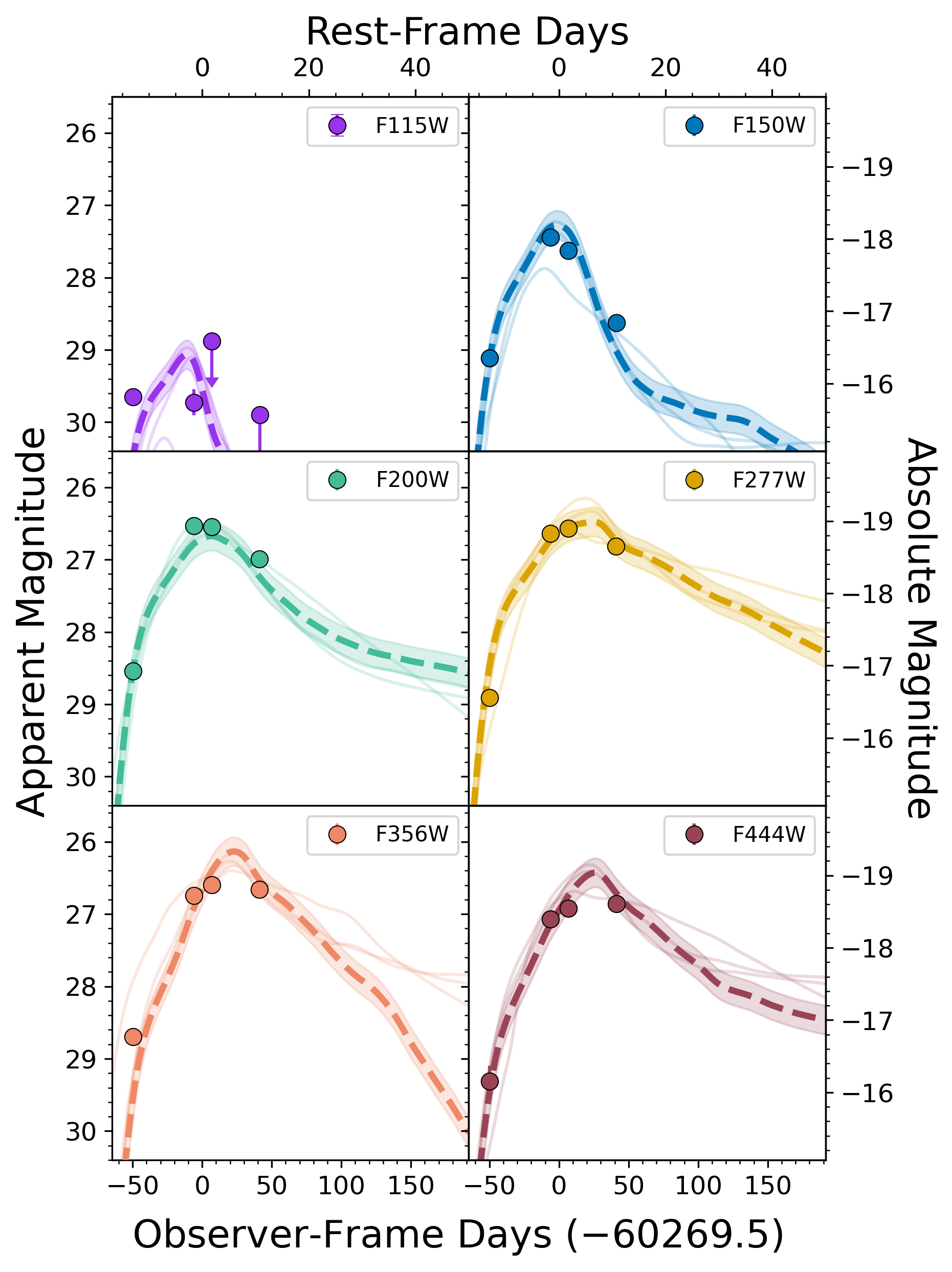}}
    \qquad
{\includegraphics[width=8.5cm]{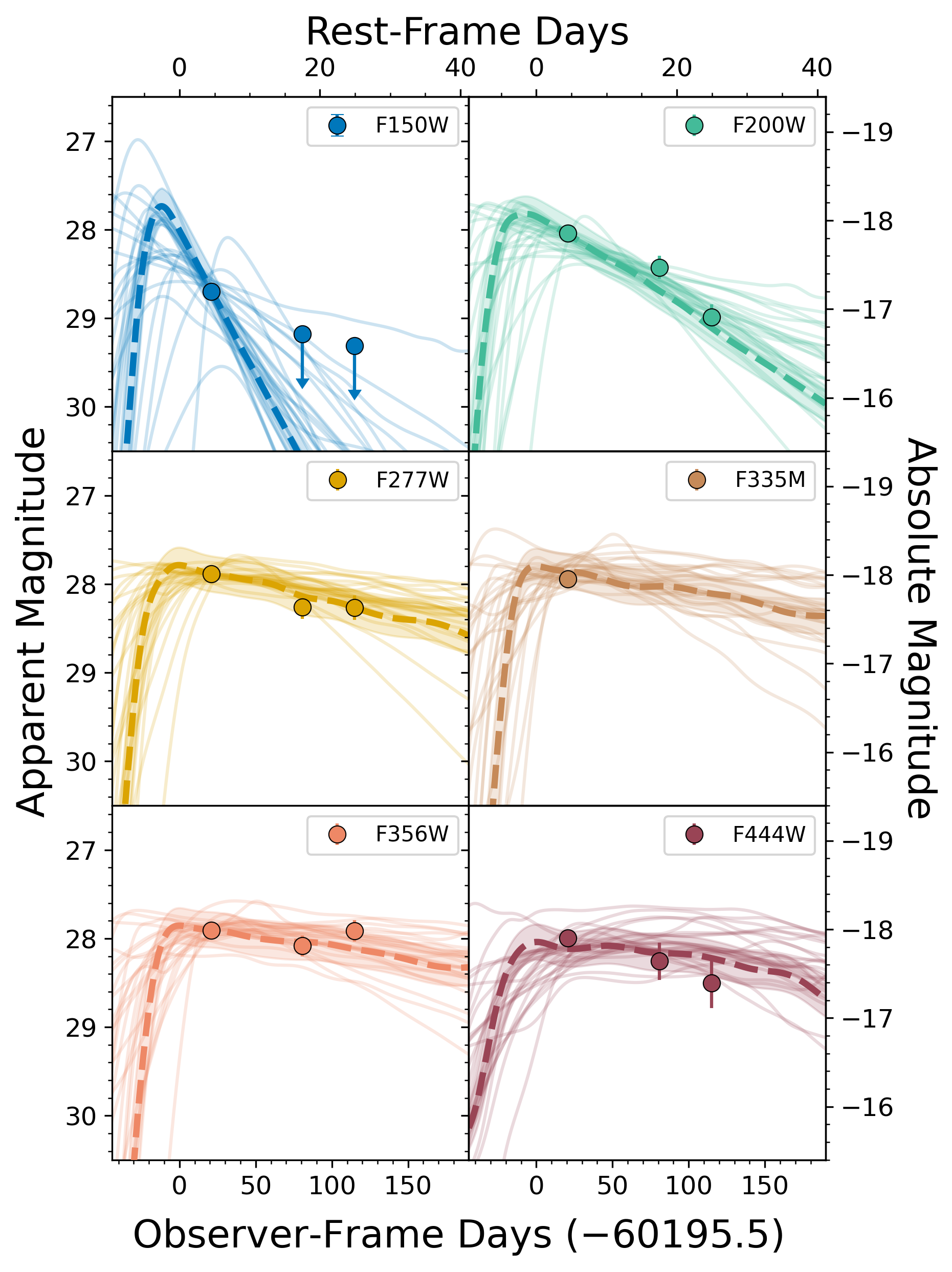} }
    \caption{Best-fit light curves for \tr{26} (left) and \tr{10} (right), where the bottom (top) x-axis is shown in observer-frame (rest-frame) days minus the time of peak brightness (in MJD). The left (right) y-axis shows the apparent (absolute) magnitudes. Both the rest-frame days and absolute magnitudes assume the best-fit model redshift. The measured PSF photometry is shown as circles and 2$\sigma$ upper limits are shown as circles with arrows. The thick dashed lines show the best-fit model with $\pm1\sigma$ error bars as the shaded region, and the faint solid lines show the other models of the same type as the best-fit model with $\chi^2$/DOF\,$\leq$\,50. $\it{Left}$: The best-fit light curve for \tr{26} in F115W, F150W, F200W, F277W, F356W, and F444W. The best-fit model, \texttt{snana-2004gv}, is a Type\,Ib/c model at $z$\,$=$\,2.83. $\it{Right}$: The best-fit light curve for \tr{10} for F150W, F200W, F277W, F335M, F356W, and F444W. The best-fit model, \texttt{snana-2006kv}, is a Type\,IIP model at $z$\,$=$\,3.61.}
    \label{tr26_tr10_lcs}
\end{figure*}

The left-side panel of Figure \ref{tr26_tr10_lcs} shows the best-fit light curve for \tr{26}, which resides in host galaxy JADES-GS+53.13533-27.81457, in F115W, F150W, F200W, F277W, F356W, and F444W. The light curve shows best-fit model \texttt{snana-2004gv}, which corresponds to a Type\,Ib/c SN, at $z$\,$=$\,2.83. \tr{26} is near peak in Epochs 3, 4, 5.1, 5.2, and 5.3, and all of the measured photometry points are within 1$\sigma$ of the best-fit model curve,  so the Type\,Ib/c light curve classification is robust. Light curve fitting alone cannot discern between Type Ib and Type Ic SNe \citep[]{Drout2011TheCurves, Taddia2018TheCurves}. However, a spectrum was taken of \tr{26} near its peak brightness as part of DDT Program 6541, and \tr{26} is spectroscopically-confirmed as a Type Ic-broad line (Ic-BL) SN at $z$\,$=$\,2.83. See \citet{Siebert2024} for a full analysis of \tr{26} and its spectrum. \tr{26} is the highest-redshift spectroscopically-confirmed Type Ic-BL SN. 

\subsubsection{AT2023adsv: Type\,IIP at $z_{\rm spec}$\,$=$\,3.61}

The right panel of Figure \ref{tr26_tr10_lcs} shows the best-fit light curve for \tr{10}, which belongs to host JADES-GS+53.16439-27.83877, in F150W, F200W, F277W, F335M, F356W, and F444W. The best-fit model is \texttt{snana-2006kv}, which is a Type\,IIP model, and the redshift is $z$\,$=$\,3.61. All of the data points fall within 1 $\sigma$ of the best-fit model curve, making the Type\,IIP classification robust. \tr{10} was also targeted by NIRSpec with DDT Program 6541, and it was spectroscopically-confirmed to be at $z$\,$=$\,3.61 based on host galaxy features in the spectrum. See D. Coulter et al. (in preparation) for a complete analysis of \tr{10} and its spectrum. 

\subsubsection{AT2022aevn: Type\,Ib/c at $z_{\rm phot}$\,$\approx$\,3.6 and not a SLSN} \label{tr13}

\begin{figure*}
    \centering
 {\includegraphics[width=13cm]{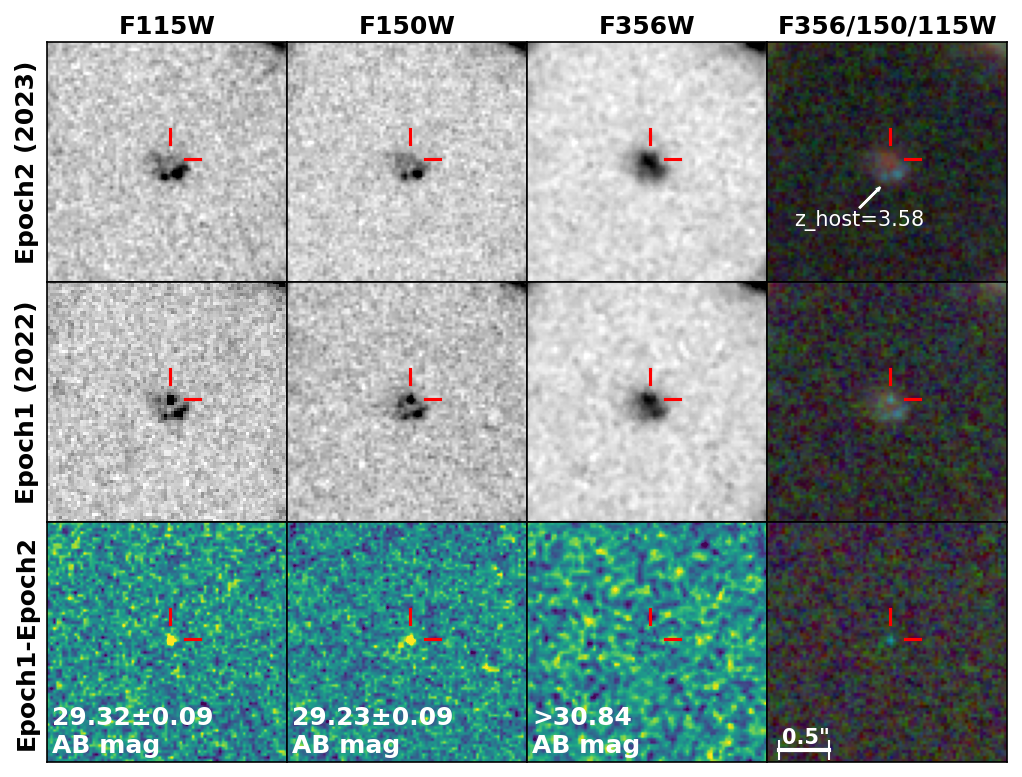}}
    \caption{Images of \tr{13}, which belongs to host galaxy JADES-GS+53.15375-27.82513. The top (middle) row shows F115W, F150W, and F356W single-filter images and three-color image of the Epoch2 (Epoch1) data. The bottom row shows the respective difference images for each filter. \tr{13} clearly appears in each SW filter's difference image but is undetected in the F356W and F444W difference images, making it very blue. At $z_{\mathrm{phot}}$\,$=$\,3.58, which is the redshift of the host galaxy directly below \tr{13}, \tr{13} falls in the SLSN regime of the F150W vs. F150W-F444W CMD (see Figure \ref{f150w_vs_f150w-f444w}).}
    \label{tr13_stamp}
\end{figure*}

\tr{13} is a SLSN candidate due to its position on the F150W vs. F150W-F444W CMD shown in the bottom left panel of Figure \ref{f150w_vs_f150w-f444w}. Figure \ref{tr13_stamp} shows \tr{13} in the Epoch1, Epoch2, and respective difference images in F115W, F150W, and F356W. As expected from a young SLSN, \tr{13} is quite blue. It clearly appears in the SW difference images, but it is not detected in the F356W or F444W difference images, and its F277W magnitude is only m$_{F277W}$\,$=$\,30.78\,$\pm$\,0.35. 

We assign the galaxy directly below \tr{13} as the host (JADES-GS+53.15375-27.82513; see Figure \ref{tr13_stamp}), which has $z_{\mathrm{phot}}$\,$=$\,3.58 (z-rank\,$=$\,2). However, there is emission in the Epoch2 images at \tr{13}'s location. This could either be galaxy emission or SN emission, or a combination of both. Assuming this is host emission, we performed galaxy SED fitting at \tr{13}'s location with photometry measured from the Epoch2 science images (z-rank\,$=$\,3) to determine if \tr{13} occurred within a faint lower redshift host at \tr{13}'s position. This SED fitting yielded $z_{\mathrm{phot}}$\,$=$\,3.37, which is roughly consistent with the redshift of the assigned host. Regardless of the host assignment, \tr{13}  qualifies as a SLSN candidate based its location in the F150W vs.\ F150W-F444W CMD since both host galaxy possibilities are $z$\,$>$\,3. 

There is the possibility that \tr{13} appears in both the Epoch1 and Epoch2 F356W and F444W images without any significant change in brightness, so it is undetected in the F356W and F444W difference images. \tr{13} becomes redder from Epoch1 to Epoch2, and 1 year in the observed frame is $\sim$\,80 days in the rest-frame at $z\sim$\,3.5, so there is a chance that the SN can be $\sim$\,28-29 mag in F356W and F444W. The source of the Epoch1 and Epoch2 LW emission remains unclear.

\begin{figure*}
    \centering
 {\includegraphics[width=8.5cm]{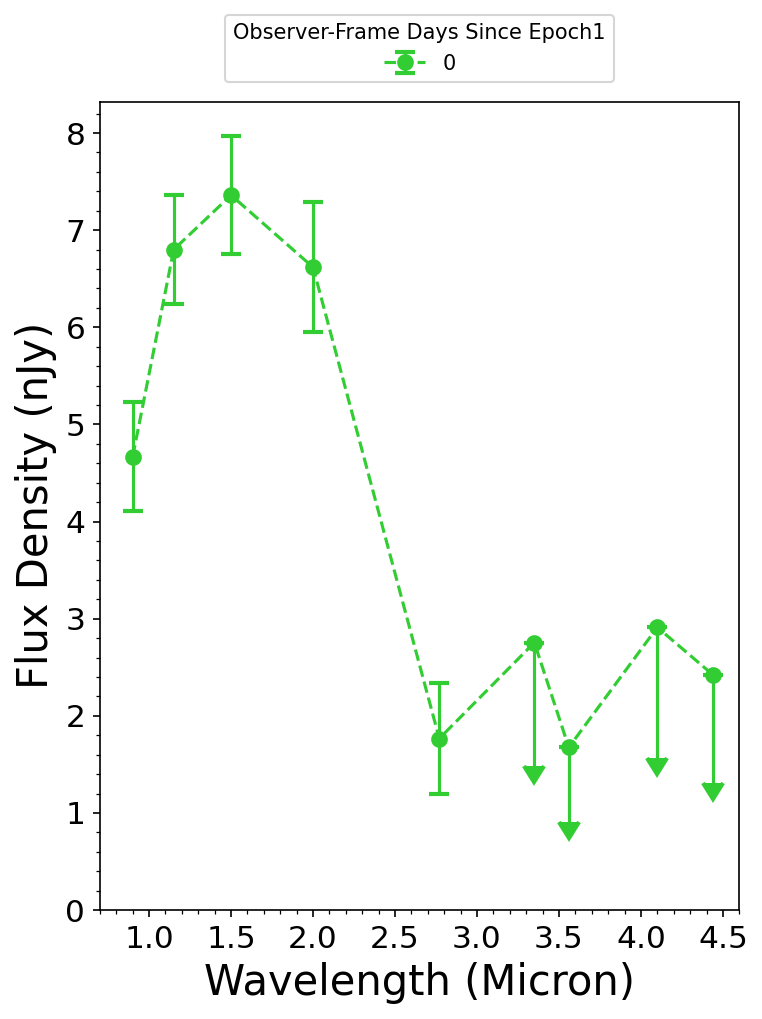}}
    \qquad
{\includegraphics[width=8.5cm]{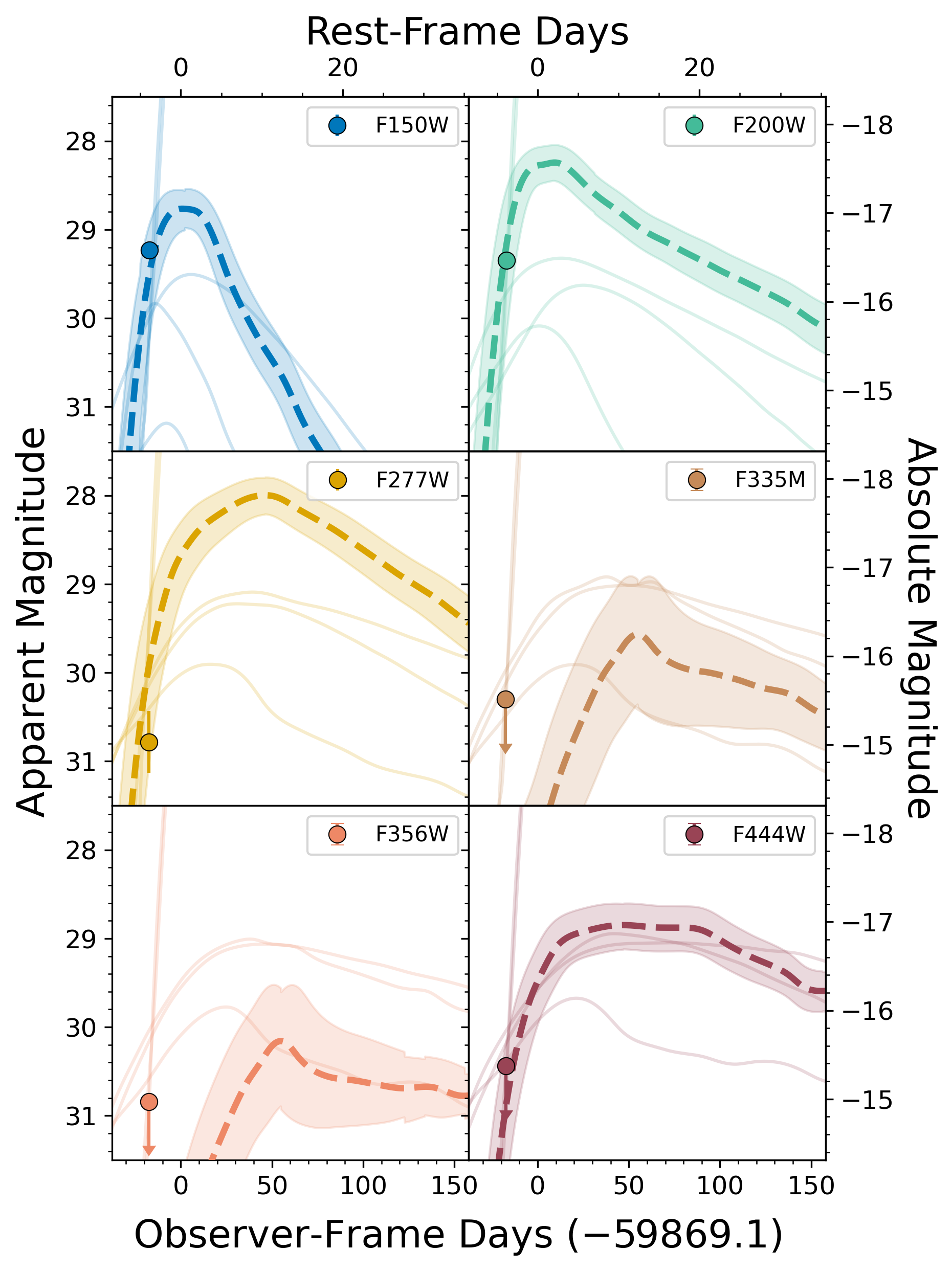} }
    \caption{Epoch1-Epoch2 difference photometry SED (left) and light curve (right) for \tr{13}. $\it{Left}$: \tr{13}'s Epoch1 SED displays its very blue color, peaking in F150W and undetected at wavelengths $>$\,3$\mu$m. $\it{Right}$: \tr{13}'s light curve, with measured PSF photometry shown as circles, 2$\sigma$ upper limits shown as circles with arrows, the best-fit model (\texttt{snana-04d1la}, Type\,Ib/c at $z$\,$=$\,3.44) shown as the thick dashed lines with $\pm1\sigma$ errors (shaded region), and the other Type\,Ib/c models shown as faint solid lines. We show the light curves for F150W, F200W, F277W, F335M, F356W, and F444W. The best-fit curve indicates that \tr{13} is a blue pre-peak Type\,Ib/c SN. However, this classification is based on just one epoch of difference image photometry, so it is not a robust classification.}
    \label{tr13_sed_lc}
\end{figure*}

Figure \ref{tr13_sed_lc} shows \tr{13}'s Epoch1-Epoch2 difference photometry SED (left) and light curve (right).  \tr{13}'s Epoch1 color and magnitudes are consistent with a very young SLSN. However, if \tr{13} were a very young SLSN, it would be significantly brighter in Epoch2 than in Epoch1, which is not observed. \tr{13} appears to be faintly visible in F200W in Epoch2 and may be visible in F356W and F444W. However, it clearly does not brighten in any filter between Epoch1 and Epoch2, so the SLSN classification is ruled out. \tr{13}'s light curve is best-fit by the \texttt{snana-04d1la} model, which is a Type\,Ib/c model. The best-fit redshift is $z$\,$=$\,3.44, which agrees with the assigned host redshift of $z_{\mathrm{phot}}$\,$=$\,3.58\,$\pm$\,0.14. The light curve indicates that \tr{13} is a young blue Type\,Ib/c SN, but this is not a robust classification because it is based on just one epoch of difference image photometry. There is also a possibility that it is some non-SN hot blue object, but that analysis is beyond the scope of this paper.

\subsubsection{AT2022aewb: Type\,II at $z_{\rm phot}$\,$=$\,2.0 that remained visible over 1 Year }

\tr{20}, which resides in host galaxy JADES-GS$+$53.14729$-$27.81047, is another interesting source in the JADES-SN-22 sample, as it is very bright in Epoch1 and continues to be visible in Epoch2 and all of the follow-up epochs. Because of its maintained brightness (only a factor of 4--5 fading over 1 year in the observer's frame), we measured PSF photometry from the science images rather than the difference images and attempted to perform multi-epoch light curve fitting with this photometry. However, the CC models could not fit the follow-up epoch data because the median duration of the CC models is $\sim$\,80 rest-frame days and the maximum duration is $\sim$\,140 rest-frame days. At $z_{\mathrm{phot}}$\,$=$\,2.00 (z-rank\,$=$\,3 host redshift), \tr{20} remained visible for at least $\sim$\,150 rest-frame days (5 months in rest frame; 15 months in the observer frame). We thus do not list \texttt{SNCosmo} classification results for \tr{20}. However, \tr{20} is likely a Type\,II SN because it stayed visible for such a long period of time, and it does not appear to be bright enough to be a SLSN candidate.


\subsection{Other Variable Sources} \label{subsec: other_variable_sources}

There are 9 sources in the JADES Deep Field that show signs of variability but are likely not SNe. Their positions and the reason for their exclusion from the SN sample are listed in Table \ref{tab_other_variable_sources}. 

We note that the source at (03:32:38.8660, $-$27:47:13.465) was reported by \citet{Hayes2024GlimmersVariability} as a $z$\,$>$\,6 variable AGN candidate (object 1052123 in their work). This source fades in Epoch2 but re-brightens in Epoch4, and therefore we removed it from our SN sample. We took a spectrum of this object as part of DDT Program 6541. Although S/N is low, the spectrum does not seem to match any SN, galaxy or AGN template at $z$\,$\simeq$\,5--7. On the contrary, the latest JADES NIRCam SED of this object can be fit with a late-type star at $z$\,$\sim$\,0, which could be intrinsically variable. We also note that the significance of other $z$\,$>$\,6 variable sources in \citet{Hayes2024GlimmersVariability} do not pass our selection criteria.

Any additional analysis of our excluded, non-SN variable sources (e.g., AGN activities) is beyond the scope of this paper.

\begin{deluxetable*}{ccl}
\tablecaption{Other Variable Sources}
\tablehead{
\colhead{RA} & \colhead{Dec} & \colhead{Reason for Exclusion from SN Sample} 
} 
\label{tab_other_variable_sources}
\startdata
03:32:44.4863 & -27:46:42.018 & Brightens then fades below initial brightness; SED fitting indicates AGN presence            \\ 
03:32:38.8660 & -27:47:13.465 & Fades then re-brightens; DDT-6541 spectrum does not match SN spectrum                               \\ 
03:32:43.2401 & -27:49:14.452 & Fades then re-brightens; variability directly over galactic core; previously identified AGN  \\ 
03:32:41.8760 & -27:44:00.214 & Variability directly over galactic core; previously identified AGN                           \\ 
03:32:38.5752 & -27:49:09.357 & Fades then re-brightens; SED fitting indicates AGN presence                                  \\ 
03:32:38.6035 & -27:45:00.285 & Variability directly over galactic core                                                             \\ 
03:32:27.4456 & -27:48:25.316 & SED fitting indicates AGN presence                                                           \\ 
03:32:37.5294 & -27:47:56.522 & Host is obscured MIRI AGN                                                                           \\ 
03:32:43.3201 & -27:49:47.317 & SED fitting indicates AGN presence                                                                  \\ 
\enddata
\end{deluxetable*}


\subsection{Implications of Supernovae Mis-Classified as $z$\,$\sim$\,16 Galaxy Candidates} \label{subsec: z16_interpretation}

The discovery that two $z$\,$\sim$\,16 galaxy candidates were actually transient events (or galaxies hosting transient events) has important implications for high-redshift galaxy surveys\footnote{There was also an earlier discovery of a probable SN reported with the HST HUDF data that could have been mis-identified as a $z$\,$\sim$\,7 “z-band dropout". The source, zD0, is reported in \citet{Bunker2010}.}. With only one deep imaging epoch, it is often difficult to discern real compact high-redshift galaxy candidates from faint transient events. One of the three $z$\,$>$\,15 galaxy candidates in \citet{Hainline2023TheGOODS-N} that fall within the JADES Deep Field is a transient interloper (\tr{101}; see object JADES-GS-53.12692-27.79102 in their Figure 8 for the $z$\,$\sim$\,16 galaxy SED fit), resulting in a transient contamination rate of 1/3. The other two $z$\,$>$\,15 galaxy candidates do not show any sign of brightness variability.  

Note that another $z$\,$\sim$\,16 galaxy candidate that turned out to be a transient (\tr{107}) is not formally part of this $z$\,$>$\,15 galaxy sample, as \citet{Hainline2023TheGOODS-N} rejected this source due to 4$\sigma$ emission in F090W (see object JADES-GS-53.20055-27.78493 in their Figure 11 for the $z$\,$\sim$\,16 SED fit).  This indicates that it is possible to discern between high-redshift galaxy candidates and SNe if the SW data are sufficiently deep because the SNe do not drop out as fast with wavelength as Lyman-break galaxies.  In practice, however, many surveys do not have adequate depth in the dropout filters to rule out SNe. 


The timescale for brightness variation increases as $(1+z)$ in the observer's frame due to the cosmological time dilation. A sufficiently long time baseline is needed to recognize slowly varying high-redshift transient events, which could be mistaken as galaxies at extremely high redshift based on a single-epoch observation because of their faintness in the SW bands.  In the case of the JADES Transient Survey, the two main imaging epochs are separated by one year, allowing high-redshift transients sufficient time to fade significantly ($\sim$\,4 months at $z$\,$\sim$\,2 and $\sim$\,2 months at $z$\,$\sim$\,5 in the SN rest-frame).

With faint LW transient events possibly contaminating the high-redshift galaxy candidate sample, we risk overestimating the galaxy abundance at high-redshift if we base our analysis solely on a single-epoch data. Although the level of such SN contamination is thought to be small ($\sim$\,10\% at $z$\,$>$\,11 according to \citealt{Yan2023PointlikeRedshifts}), it may become more significant in the sample of the highest-redshift galaxies (e.g., $z$\,$>$\,15) as our study seems to indicate.  Since reaching comparable depth in multi-epoch data is very expensive, it is important to explore ways to distinguish high-redshift galaxy candidates and transient events with single-epoch data \citep[e.g.,][]{Yan2023PointlikeRedshifts}. 

In this regard, we also note that although \citet{Hainline2023TheGOODS-N} present \tr{101} as a candidate at $z$\,$>$\,15, they caution that its colors are consistent with a lower redshift solution. 
\tr{101} is bright source with some morphology seen in the individual images in the filters, and its SED is fairly red from 2 to 5 \micron, which is unexpected for a high-redshift galaxy (Figure 8 in \citealt{Hainline2023TheGOODS-N}). These are some key indicators that a $z$\,$>$\,15 galaxy candidate may instead be a transient interloper.

The discovery of these $z$\,$\sim$\,16 galaxy interlopers also raises the question of how much the observed galaxy SEDs are affected by transient phenomena in general.  At high redshift, SN rates are expected to increase due to enhanced star-forming activities in galaxies and potentially more top-heavy IMFs due to lower metallicities and higher gas temperature, partially heated by the cosmic microwave background. SN emission also will remain for longer periods of time due to time dilation.

Since these two $z$\,$\sim$\,16 galaxy interlopers are spatially compact, the possibility remains that what we are seeing may be the light from slowly fading high-redshift SNe from 2022 to 2023. Important future work involves conducting a systematic search of high-redshift galaxies (e.g., $z$\,$\geq$\,5) to determine if JWST has the capability of detecting point-like variability at such high redshift. This will provide information about supernova rates at even higher redshift, and may lend clues for how to discern genuinely high-redshift transient events from those in lower-redshift interlopers.


\subsection{Assessing Supernovae Masquerading as High-Redshift Galaxies} 

To further probe the population of SNe that may masquerade as high-redshift galaxies, we explore the color-color space of theoretical Type\,Ia, Type\,IIP, and Type\,Ib/c phase tracks relative to the color-color space occupied by F115W and F150W dropout galaxies in Figure~\ref{sn_phase_tracks}. F115W dropouts comprise $z$\,$\sim$\,8.5-12 galaxies and F150W dropouts comprise $z$\,$\sim$\,12-16 galaxies.  The F150W dropout color-color space is F150W-F200W vs.\ F200W-F356W and the F115W dropout color-color space is F115W-F150W vs.\ F150W-F277W. Red dashed lines indicate the color-color space where F150W and F115W dropout galaxies are selected, and the selection criteria are adopted from L. Whitler et al. (in preparation). The grey contours show the color-color space occupied by bright JADES galaxies (S/N\,$>$\,20 in non-dropout filters). SN phase tracks are colored by redshift (see color bars), and the phase track start (end) is indicated by black squares (black diamonds).

For Type\,Ia SNe, we show phase tracks for two types of surveys: a JADES-like deep survey ($m_{\rm F150W }$\,$=$\,30.53, $m_{\rm F200W}$\,$=$\,30.42, and $m_{\rm F356W}$\,$=$\,30.84; the top-left panel) and a COSMOS-Web-like wide survey ($m_{\rm F115W }$\,$=$\,$m_{\rm F150W}$\,$=$\,$m_{\rm F277W}$\,$=$\,28; the top-right panel).  For Type\,IIP and Type\,Ib/c SNe, we only show phase tracks for a JADES-like deep survey (the bottom-left and -right panels, respectively) because these types of SNe are fainter than Type\,Ia's and will not produce the colors of high-redshift dropout galaxies in a shallower-depth ($\sim$\,28 mag) survey.

Note that when the model-predicted apparent magnitudes are fainter than the detection limits, those limits are used to calculate the colors shown in Figure~\ref{sn_phase_tracks}.  This means that when the model SN is faint, our ability to measure its dropout color (i.e., F115W-F150W or F150W-F200W) is limited by these detection limits and not by the intrinsic color.  For example, Figure~\ref{sn_phase_tracks} shows a trend that higher-redshift SNe become bluer at later phases in the dropout colors (i.e., phase tracks turn downward after reaching the reddest dropout color).  This does not mean that these SNe are becoming bluer intrinsically, but indicates that their overall fading is reducing the brightness contrast between the dropout and detection bands.

It is worth noting that \tr{101} and \tr{107}, the two SNe masquerading as $z$\,$\sim$\,16 galaxy candidates in this paper, would not have been selected as either F150W or F200W dropouts according to the color-color selection criteria used in L. Whitler et al. This implies that color-color selection criteria should be applied in addition to the SED fitting-based selection criteria outlined in \citet{Hainline2023TheGOODS-N} to select high-redshift galaxy candidates robustly.


\subsubsection{Type\,Ia}

We show Type\,Ia SNe phase tracks (\texttt{x1}\,$=$\,0.0, \texttt{c}\,$=$\,0.0) in the top panels of Figure \ref{sn_phase_tracks}, with F150W dropout criteria shown in the top left panel and F115W dropout criteria shown in the top right panel. The SN phase tracks, generated from the \texttt{SALT3-NIR} model \citep{Pierel2022SALT3-NIR:Measurements}, are shown from 10 days pre-peak to 50 days post-peak. $z$\,$\approx$\,3.5-5 Type\,Ia SNe enter the F150W dropout selection region and $z$\,$\approx$\,2.5-3.0 Type\,Ia SNe enter the F115W dropout selection region. While we display SN\,Ia tracks with \texttt{x1}\,$=$\,0 and \texttt{c}\,$=$\,0, there are other combinations of \texttt{x1} and \texttt{c} parameters whose model phase tracks enter the F150W and F115W dropout regions.


The highest-redshift Type\,Ia SN discovered thus far is \tr{27} at $z$\,$=$\,2.90 \citep{Pierel24Ia}, and we are unsure of the highest redshift at which Type\,Ia SNe exist (such that the white dwarf progenitor has sufficient time to form). It is thus unclear how large of a risk the $z$\,$>$\,3 Type\,Ia SNe pose to single-epoch high-redshift galaxy surveys.

\subsubsection{Type\,IIP}

We show Type\,IIP SNe phase tracks (\texttt{E(B-V)}\,$=$\,0.0) with F150W dropout criteria in the bottom left panel of Figure \ref{sn_phase_tracks}. The SN phase tracks, generated from the ``nugent-sn2p" model \citep{Gilliland1999}, are shown from 40 days post-explosion to 150 days post-explosion, and the peak B-band absolute magnitude is set to -18 AB magnitudes. If the \texttt{E(B-V)} parameter is set to higher values, the SN phase tracks either barely enter or do not enter the F150W dropout selection region at any phase or redshift. The Type\,IIP SNe tracks with \texttt{E(B-V)}\,$=$\,0.0 and $z$\,$\approx$3.5-4.0 just barely enter the F150W dropout selection region in the $\sim$60-70 days post-peak range. 
Based on these plots, deep surveys like JADES are just barely susceptible to mistaking post-peak $z$\,$\approx$\,3.0-4.0 Type\,IIP SNe as F150W dropout galaxies. Notably, we performed this analysis for F115W dropouts with an m$_{\mathrm{F115W}}$\,$=$\,28 detection limit to mimic shallower surveys, but the SN IIP tracks did not enter the F115W dropout selection area at any redshift, phase, or dust value.

\subsubsection{Type\,Ib/c}

\begin{figure*}
    \centering
 {\includegraphics[width=8.5cm]{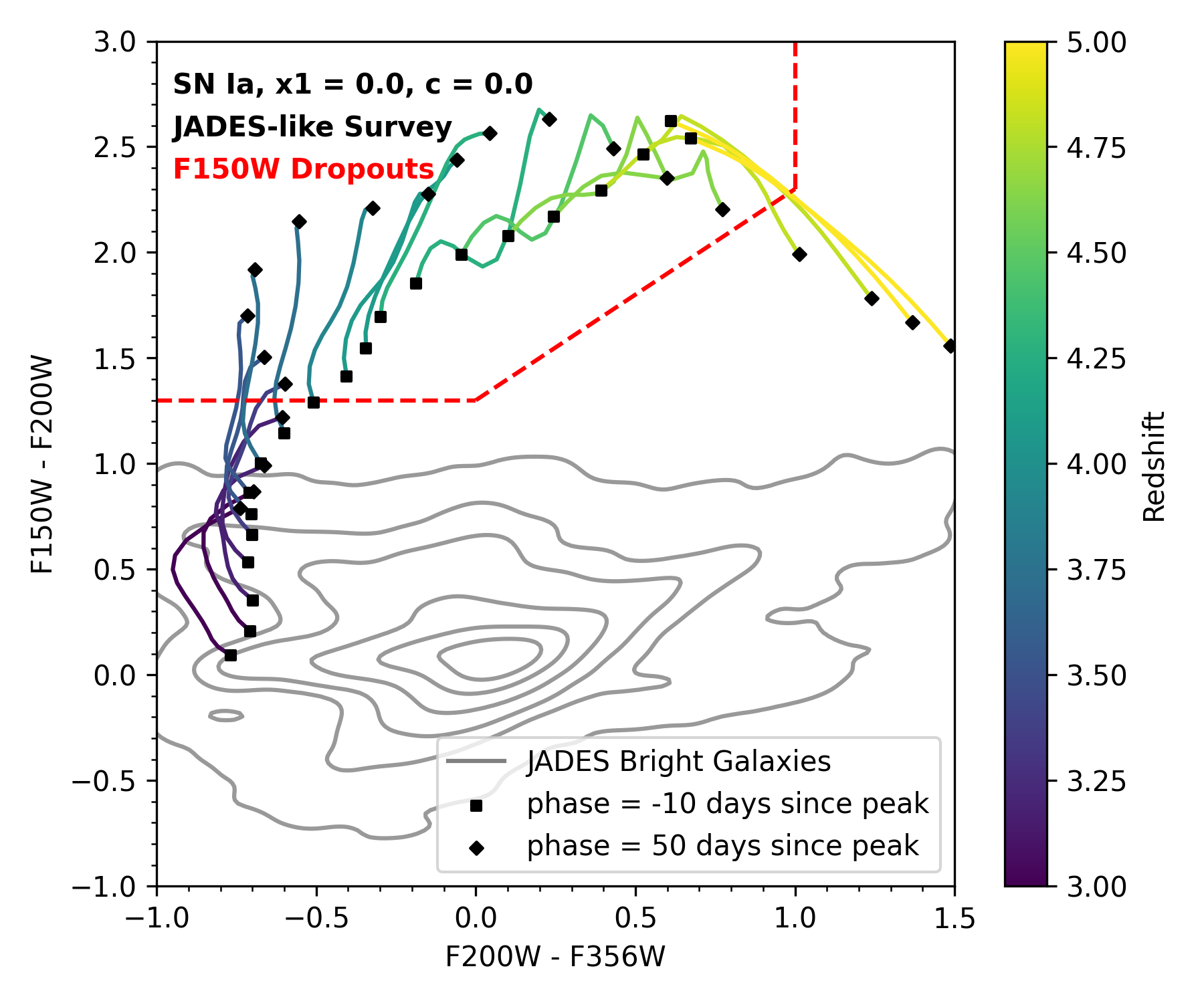}}
    \qquad
{\includegraphics[width=8.5cm]{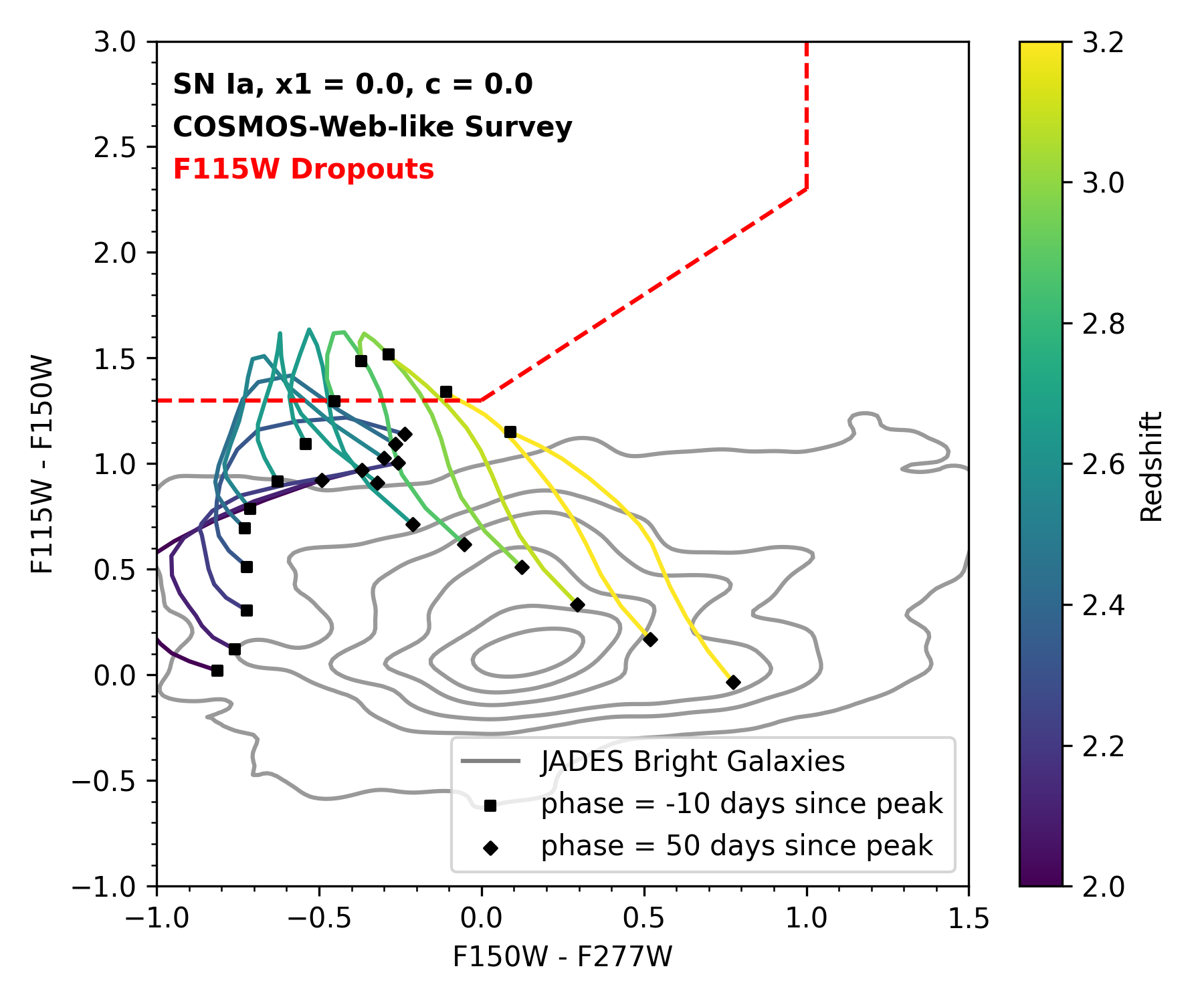}}
    \qquad
{\includegraphics[width=8.5cm]{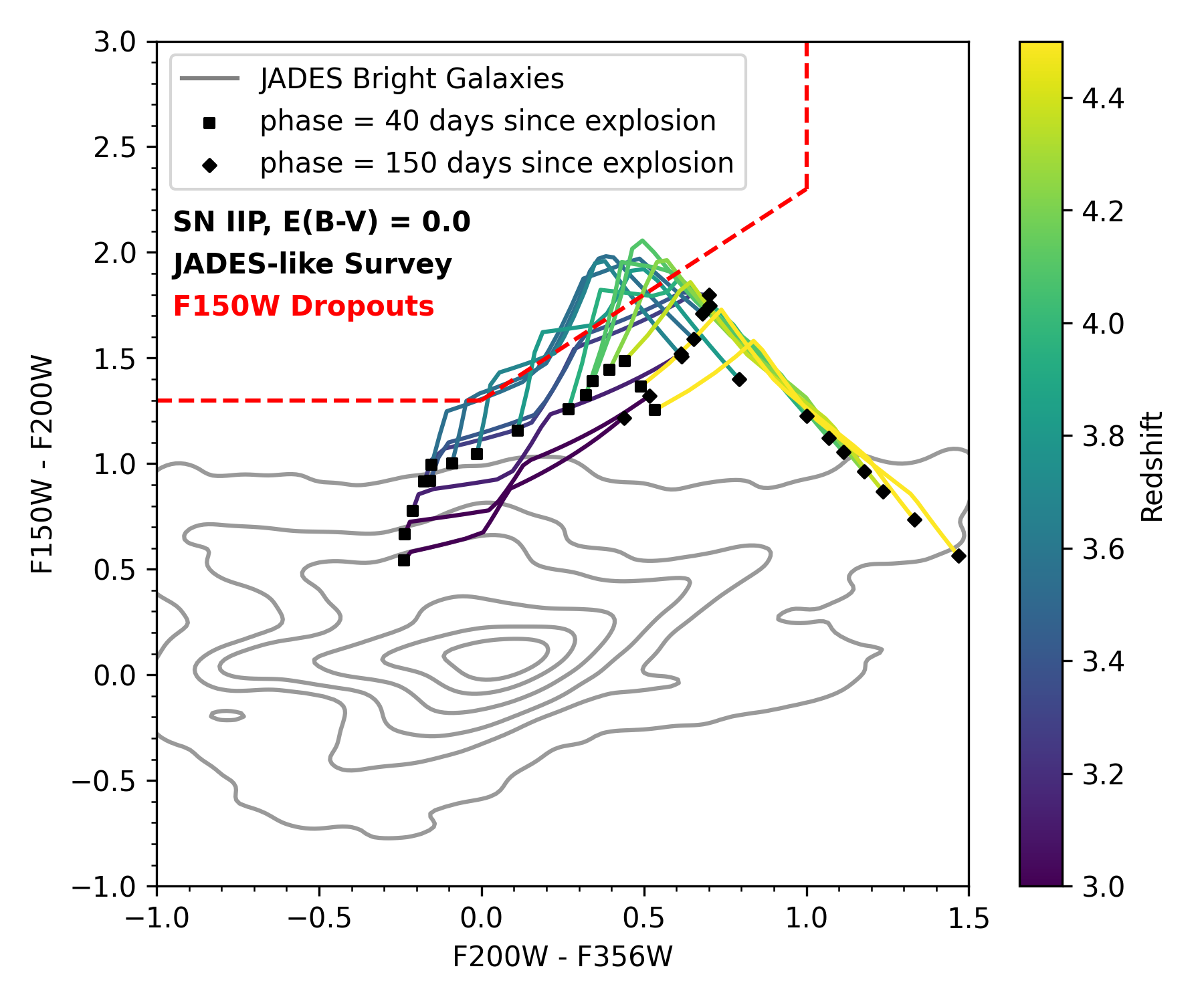}}
    \qquad
{\includegraphics[width=8.5cm]{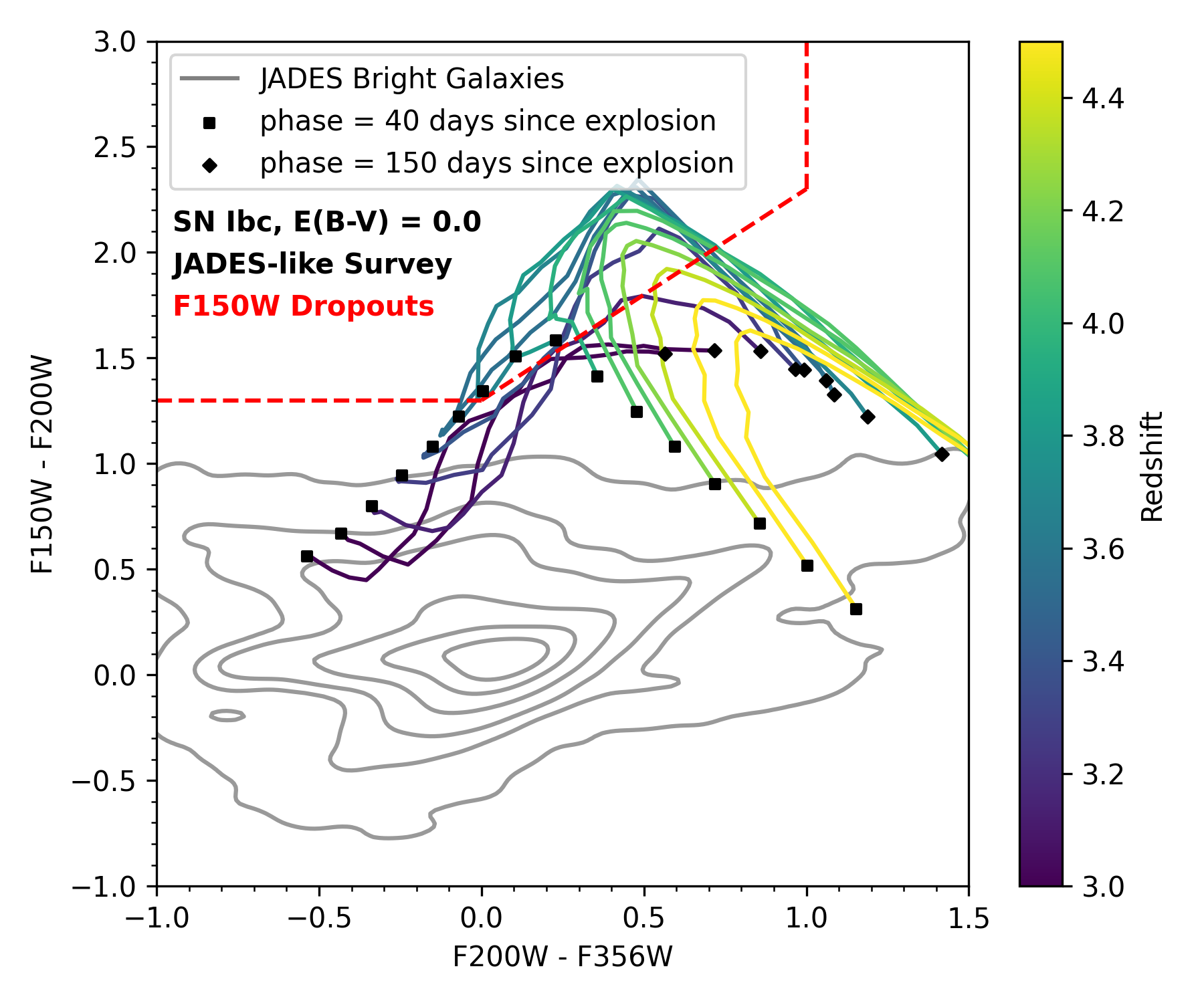}}    
    \caption{Color-color plots showing F150W dropout galaxy and F115W dropout galaxy selection regions (red dashed lines; Whitler et al.\ in prep) overlaid with SN\,Ia (top), SN\,IIP (bottom left), and SN\,Ib/c (bottom right) phase tracks. We assume JADES-like detection limits for the F150W dropout plots ($m_{\rm F150W }$\,$=$\,30.53, $m_{\rm F200W}$\,$=$\,30.42, and $m_{\rm F356W}$\,$=$\,30.84) and COSMOS-Web-like detection limits for the F115W dropout plot ($m_{\rm F115W }$\,$=$\,$m_{\rm F150W}$\,$=$\,$m_{\rm F277W}$\,$=$\,28). The grey contours show the distribution of bright JADES galaxies (SNR\,$>$\,20 in non-dropout filters), and the SN phase track colors indicate the SN redshift as dictated by the color bars. Black squares indicate the start of the SN phase track and black diamonds indicate the phase track end. 
    $\it{Top \: Left}$: F150W-F200W vs. F200W-F356W plot showing F150W dropout selection criteria overlaid with SN Ia (\texttt{x1}\,$=$\,0.0, \texttt{c}\,$=$\,0.0) phase tracks starting 10 days pre-peak and ending 50 days post-peak. We show SN Ia from $z$\,$=$\,3 to $z$\,$=$\,5.0.
    $\it{Top \: Right}$: F115W-F150W vs. F150W-F277W plot showing F115W dropout selection criteria overlaid with SN Ia (\texttt{x1}\,$=$\,0.0, \texttt{c}\,$=$\,0.0) phase tracks starting 10 days pre-peak and ending 50 days post-peak. We show SN Ia from $z$\,$=$\,2 to $z$\,$=$\,3.2.
    $\it{Bottom \: Left}$: F150W-F200W vs. F200W-F356W plot showing F150W dropout selection criteria overlaid with SN IIP (\texttt{E(B-V)}\,$=$\,0.0) phase tracks starting at 40 days post-explosion and ending 150 days post-explosion. We show SN IIP from $z$\,$=$\,3.0 to $z$\,$=$\,4.5.
    $\it{Bottom \: Right}$: F150W-F200W vs. F200W-F356W plot showing F150W dropout selection criteria overlaid with SN Ib/c (\texttt{E(B-V)}\,$=$\,0.0) phase tracks starting 40 days post-explosion and ending 150 days post-explosion. We show SN Ib/c from $z$\,$=$\,3 to $z$\,$=$\,4.5.}
    \label{sn_phase_tracks}
\end{figure*}

We show Type\,Ib/c SNe phase tracks (\texttt{E(B-V)}\,$=$\,0.0) with F150W dropout criteria in the bottom right panel of Figure \ref{sn_phase_tracks}. The SN phase tracks, generated from the ``nugent-sn1bc" model \citep{Levan2005}, are shown from 40 days post-explosion to 150 days post-explosion, and the peak B-band absolute magnitude is set to -18 AB magnitudes. 
Like the SN IIP, if the \texttt{E(B-V)} parameter is set to higher values, the SN phase tracks either barely enter or do not enter the F150W dropout selection region at any phase or redshift. The Type\,IIP SNe tracks with \texttt{E(B-V)}\,$=$\,0.0 and $z$\,$\approx$3.5-4.2 enter the F150W dropout selection region at a variety of phases. Surveys with depths similar to JADES may mistake  $z$\,$\approx$\,3.5-4.2 Type\,Ib/c SNe as F150W dropout galaxies. We generated similar plots for F115W dropouts with an m$_{\mathrm{F115W}}$\,$=$\,28 detection limit and found that SN\,Ib/c do not resemble F115W dropouts at any redshift, phase, or dust value. Thus, shallower surveys are not susceptible to mistaking SN\,Ib/c as F115W dropouts.

\section{Conclusions} \label{sec:conclusion}

The JADES Transient Survey is the deepest systematic transient/SN search with JWST/NIRCam targeting the 9-band multi-epoch JADES Deep Field. It covers $\sim$\,25 arcmin$^2$ and has a depth of $\sim$\,30 magnitude. We discovered transients by differencing the Epoch1 and Epoch2 images and searching for point-like sources of emission. For a subset of the SNe, we had 1-4 multi-band follow-up NIRCam epochs to build light curves (where Epochs5.1-5.3 are considered one epoch). Here, we summarize the main conclusions from the JADES Transient Survey:

\begin{itemize}

    \item The sample we have presented contains 79 SNe in the JADES Deep Field, 34 of which brightened between Epoch1 and Epoch2 and 45 of which faded between Epoch1 and Epoch2. There are 4 additional marginally-detected SNe that are listed in the paper but not included in the main statistical sample. This sample contains many of the highest redshift SNe, 23 at 2\,$<$\,$z$\,$<$\,3, 8 at 3\,$<$\,$z$\,$<$\,4, and 7 at 4\,$<$\,$z$\,$<$\,5.  The difference images clearly show emission down to $\sim$\,30 magnitude, exemplifying JWST's ability to detect distant SNe as well as SNe in dusty regions close to the galaxy centers. Previous SN surveys with HST probed only to $z$\,$\sim$\,2 and $m_{F160W}$\,$\sim$\,26.7 \citep{Grogin2011CANDELS:Survey}, so JWST is exploring new redshift and magnitude space for SNe. We adopt the host redshifts as the SN redshifts, $\sim$\,59\% of which are spectroscopic redshifts, $\sim$\,23\% of which are robust photometric redshifts, and $\sim$\,14\% of which are photometric redshifts whose SEDs may suffer from SN contamination. We were unable to determine the redshift for $\sim$\,4\% of the SNe.
    
    \item At a depth of $\sim$\,30 AB magnitude, the SN detection rate is $\sim$\,1--2 per arcmin$^2$ per observer-frame year, demonstrating JWST's ability to discover SNe essentially anywhere it looks.
    
    \item We have 1-4 follow-up NIRCam observations for the majority of the JADES-SN-23 SNe, which allowed us to build light curves and attempt to classify their SN type. We had to rely on single-epoch light curve fitting for the JADES-SN-22 sources. With varying levels of confidence indicated by the best-fit model's $\chi^2$/DOF, we classified 40 of the 45 JADES-SN-22 sources and 30 of the 34 JADES-SN-23 sources. The JADES-SN-22 sample contains 9 Type\,Ia, 5 Type\,Ib/c, and 26 Type\,II SN candidates, and the JADES-SN-23 sample contains 2 Type\,Ia, 8 Type\,Ib/c, and 20 Type\,II SN candidates. The best-fit SN redshifts generally agree well with the associated host galaxy redshifts.
    
    \item The sample includes a spectroscopically-confirmed Type\,Ia SN at $z$\,$=$\,2.90 (\tr{27}; \citealt{Pierel24Ia}), a spectroscopically-confirmed Type Ic-BL SN at $z$\,$=$\,2.83 (\tr{26}; \citealt{Siebert2024}) and a Type\,IIP SN at $z_{\mathrm{spec}}$\,$=$\,3.61 (\tr{10}; D. Coulter et al. in preparation). 
    \tr{27} and \tr{26} are the highest-redshift spectroscopically-confirmed SNe of their respective types, emphasizing the groundbreaking nature of the JADES-SN-23 and JADES-SN-22 samples. Additionally, \tr{10} is the highest-redshift Type\,IIP SN discovered thus far.
    
    \item Two Epoch1 $z$\,$\sim$\,16 galaxy candidates are actually transients that faded at LW in Epoch2. We treat them as SNe, but there is a possibility that they are some other type of isolated transient events. This result exemplifies how multi-epoch surveys can identify interlopers in high-redshift galaxy surveys, as transients can mimic high-redshift galaxies when they are compact and lack SW emission. 
    
\end{itemize}

The overwhelming result from this paper is that JWST is a SN discovery machine. JWST opens up an entirely new redshift regime for SN discovery, which allows the possibility of finding more exotic SNe, such as Population III SNe that are expected to exist at high-redshift. It is essential to take advantage of JWST's power to discover SNe and plan additional high-redshift transient surveys to add to our sample of JWST transients and SNe.

\begin{center}
    \textbf{Acknowledgements}
\end{center}

We thank the anonymous referee whose comments and suggestions led to significant improvements in the paper.
This work is based on observations made with the NASA/ESA/CSA James Webb Space Telescope. The data were obtained from the Mikulski Archive for Space Telescopes at the Space Telescope Science Institute, which is operated by the Association of Universities for Research in Astronomy, Inc., under NASA contract NAS 5-03127 for JWST. These observations are associated with program \#1180 and 6541. The specific JWST observations analyzed can be accessed via \dataset[DOI: 10.17909/c4qk-xv53]{https://doi.org/10.17909/c4qk-xv53}. This research is based (in part) on observations made with the NASA/ESA Hubble Space Telescope obtained from the Space Telescope Science Institute, which is operated by the Association of Universities for Research in Astronomy, Inc., under NASA contract NAS 5–26555. All the HST data used in this paper can be found in MAST: \dataset[10.17909/T91019]{http://dx.doi.org/10.17909/T91019}.
Additionally, this work made use of the {\it lux} supercomputer at UC Santa Cruz which is funded by NSF MRI grant AST 1828315, as well as the High Performance Computing (HPC) resources at the University of Arizona which is funded by the Office of Research Discovery and Innovation (ORDI), Chief Information Officer (CIO), and University Information Technology Services (UITS).
CD thanks Ofer Yaron for his extensive support in producing the TNS IDs for each source. 
AJB acknowledges funding from the “FirstGalaxies” Advanced Grant from the European Research Council (ERC) under the European Union’s Horizon 2020 research and innovation program (Grant agreement No. 789056). 
PAC, EE, DJE, BDJ, GR, MR, FS, CNAW are supported by JWST/NIRCam contract to the University of Arizona, NAS5-02015.
DJE is also supported as a Simons Investigator.
RM acknowledges support by the Science and Technology Facilities Council (STFC), by the ERC through Advanced Grant 695671 “QUENCH”, and by the UKRI Frontier Research grant RISEandFALL. RM also acknowledges funding from a research professorship from the Royal Society.
BER acknowledges support from the NIRCam Science Team contract to the University of Arizona, NAS5-02015, and JWST Program 3215.
ST acknowledges support by the Royal Society Research Grant G125142.
The research of CCW is supported by NOIRLab, which is managed by the Association of Universities for Research in Astronomy (AURA) under a cooperative agreement with the National Science Foundation.


\facilities{JWST, HST}

\software{astropy \citep{AstropyCollaboration2022ThePackage}, EAZY and eazy-py \citep{Brammer2008EAZY:Code}, STARDUST2 \citep{Rodney2014TypeUniverse}, photutils \citep{Bradley2024Astropy/photutils:1.12.0}}

\appendix


\section{The JADES-SN-23 Sample}\label{sec:deep_2023}

In Appendix~\ref{sec:deep_2023}, we present, (1) multi-band cutout NIRCam images (top), (2) multi-epoch SEDs (lower-left), and (3) multi-band light curves (lower-right) for each of the JADES-SN-23 sources.  At the top of each page, we show the IAU ID (issued by TNS), redshift (spectroscopic $z_{\rm spec}$ or photometric $z_{\rm phot}$), and SN classification based on the light-curve analysis.  

The panel of multi-band images (top) displays NIRCam images in three filters (from left to right in the order of increasing wavelengths).  For each filter, we show Epoch1 (middle), Epoch2 (top), and Epoch2-Epoch1 images (bottom). We also show three-color images for each epoch and the difference image (the right column). 

The SED plot (lower-left) shows the PSF photometry measured as described in Section \ref{subsec:psf_photometry} in nJy. For sources covered by multiple epochs, we color code the multi-epoch SEDs by observer-frame days since the first observation. The SED associated with the first observation is always shown in lime green.

The light-curve plot (lower-right) is presented only when, (i) the fitting converged to a solution, and (ii) the $\chi^2$/DOF\,$\leq$\,50 for the best-fit model (i.e., the model listed in Table \ref{tab_classifications_2023}). See Section \ref{subsec:classification} for more details.
The y-axis on the left shows the measured apparent AB magnitude while the y-axis on the right shows the corresponding absolute AB magnitude calculated with the best-fit model redshift (see Table \ref{tab_classifications_2023}).
Observer-frame days minus the mean Modified Julian Date (MJD) of peak brightness are shown on the bottom x-axis while the top x-axis shows the corresponding rest-frame days assuming the best-fit model redshift. The circles are the measured photometry (or 2$\sigma$ upper limits when shown with a downward arrow), the thick dashed line shows the best-fit model with $\pm 1\sigma$ error bars as the shaded region, and the faint lines show the other models of the same SN type as the best-fit model (i.e., Type\,II, Type\,Ib/c, or Type\,Ia) that also have $\chi^2$/DOF\,$\leq$\,50.

\section{The JADES-SN-22 Sample}\label{sec:deep_2022}

In Appendix~\ref{sec:deep_2022}, we present the same information for each of the JADES-SN-22 sources. The bottom row of the multi-band images show Epoch1-Epoch2 difference images. Note that \tr{20}'s SED has multiple epochs shown because it remained visible in all of the follow-up observations. We measured \tr{20}'s photometry from the science images rather than the difference images to avoid underestimating its brightness. For the light curves, the best-fit models, along with their associated redshifts that were used to convert apparent to absolute AB magnitude and observer-frame days to rest-frame days, are listed in Table \ref{tab_classifications_2022}.

\begin{longrotatetable}
\begin{deluxetable}{cccccccccc}
\tablecaption{JADES-SN-23 Epoch2-Epoch1 SN photometry}
\tablehead{
\colhead{ID} & \colhead{F090W} & \colhead{F115W} & \colhead{F150W} & \colhead{F200W} & \colhead{F277W} & \colhead{F335M} & \colhead{F356W} & \colhead{F410M} & \colhead{F444W} \\
\colhead{} & \colhead{(AB mag)} & \colhead{(AB mag)} & \colhead{(AB mag)} & \colhead{(AB mag)} & \colhead{(AB mag)} & \colhead{(AB mag)} & \colhead{(AB mag)} & \colhead{(AB mag)} & \colhead{(AB mag)}
} 
\startdata
\tr{53} & $>$30.25            & $>$30.57            & $>$30.53            & $>$30.42            & 29.35 $\pm$ 0.16 & 28.99    $\pm$ 0.18 & 28.59 $\pm$ 0.09 & 28.40    $\pm$ 0.12 & 28.43    $\pm$ 0.11 \\
\tr{50} & $>$30.25            & $>$30.57            & $>$30.53            & $>$30.42            & 29.88 $\pm$ 0.18 & 29.00    $\pm$ 0.14 & 29.39 $\pm$ 0.14 & 29.61    $\pm$ 0.28 & 29.84    $\pm$ 0.30 \\
\tr{88} & $>$30.25            & $>$30.57            & $>$30.53            & 30.41    $\pm$ 0.25 & 29.67 $\pm$ 0.14 & 29.47    $\pm$ 0.21 & 29.73 $\pm$ 0.19 & 29.51    $\pm$ 0.27 & 29.45    $\pm$ 0.23 \\
\tr{10} & $>$30.25            & 29.78    $\pm$ 0.10 & 28.70    $\pm$ 0.06 & 28.04    $\pm$ 0.05 & 27.89 $\pm$ 0.04 & 27.94    $\pm$ 0.06 & 27.91 $\pm$ 0.05 & 28.02    $\pm$ 0.08 & 27.99    $\pm$ 0.07 \\
\tr{44} & $>$30.25            & $>$30.57            & $>$30.53            & 30.11    $\pm$ 0.20 & 29.15 $\pm$ 0.09 & 28.90    $\pm$ 0.12 & 29.26 $\pm$ 0.13 & 29.09    $\pm$ 0.17 & 29.44    $\pm$ 0.22 \\
\tr{71} & $>$30.25            & $>$30.57            & $>$30.53            & 30.15    $\pm$ 0.17 & 29.24 $\pm$ 0.09 & 29.67    $\pm$ 0.19 & 29.18 $\pm$ 0.10 & 28.96    $\pm$ 0.13 & 29.36    $\pm$ 0.16 \\
\tr{27} & $>$30.25            & $>$30.57            & $>$30.53	           & 28.98   $\pm$ 0.09 & 28.27 $\pm$ 0.05 & 28.00    $\pm$ 0.07 & 28.09 $\pm$ 0.05 & 28.07    $\pm$ 0.08 & 28.07    $\pm$ 0.07 \\
\tr{36} & $>$30.25            & $>$30.57            & 30.10    $\pm$ 0.11 & 29.52    $\pm$ 0.09 & 28.92 $\pm$ 0.06 & 28.76    $\pm$ 0.08 & 28.75 $\pm$ 0.07 & 28.54    $\pm$ 0.08 & 28.87    $\pm$ 0.10 \\
\tr{26} & $>$30.25            & 29.65    $\pm$ 0.10 & 29.11    $\pm$ 0.08 & 28.54    $\pm$ 0.07 & 28.91 $\pm$ 0.09 & 28.60    $\pm$ 0.12 & 28.69 $\pm$ 0.09 & 29.19    $\pm$ 0.21 & 29.31    $\pm$ 0.21 \\
\tr{52} & $>$30.25            & $>$30.57            & 29.88    $\pm$ 0.11 & 29.60    $\pm$ 0.10 & 29.71 $\pm$ 0.11 & 29.84    $\pm$ 0.21 & 29.73 $\pm$ 0.14 & 29.78    $\pm$ 0.23 & 30.40    $\pm$ 0.34 \\
\tr{15} & 29.52    $\pm$ 0.11 & 29.69    $\pm$ 0.10 & 29.83    $\pm$ 0.14 & 29.69    $\pm$ 0.14 & 30.47 $\pm$ 0.27 & $>$30.30            & 30.32 $\pm$ 0.29 & $>$30.24            & 30.18    $\pm$ 0.38 \\
\tr{29} & $>$30.25            & $>$30.57            & 29.86    $\pm$ 0.16 & 28.00    $\pm$ 0.11 & 27.77 $\pm$ 0.03 & 27.35    $\pm$ 0.03 & 27.47 $\pm$ 0.03 & 27.43    $\pm$ 0.04 & 27.61    $\pm$ 0.04 \\
\tr{6}	& $>$30.25            & 29.59    $\pm$ 0.07 & 28.25    $\pm$ 0.04 & 27.97    $\pm$ 0.04 & 27.92 $\pm$ 0.04 & 28.01    $\pm$ 0.07 & 28.09 $\pm$ 0.06 & 28.04    $\pm$ 0.07 & 28.05    $\pm$ 0.06 \\
\tr{11} & 29.77    $\pm$ 0.15 & 29.16    $\pm$ 0.09 & 28.64    $\pm$ 0.07 & \nodata          & 29.07 $\pm$ 0.08 & 29.27       $\pm$ 0.17 & 29.09 $\pm$ 0.10 & 29.44    $\pm$ 0.23 & 29.46    $\pm$ 0.20 \\
\tr{19} & $>$30.25            & 30.35    $\pm$ 0.17 & 29.05    $\pm$ 0.09 & 28.69    $\pm$ 0.07 & 28.96 $\pm$ 0.10 & 28.65    $\pm$ 0.11 & 28.92 $\pm$ 0.11 & 29.28    $\pm$ 0.23 & 29.18    $\pm$ 0.19 \\
\tr{7}  & $>$30.25            & 30.09    $\pm$ 0.16 & 28.70    $\pm$ 0.07 & 28.31    $\pm$ 0.06 & 28.22 $\pm$ 0.06 & 28.34    $\pm$ 0.10 & 28.58 $\pm$ 0.10 & 28.50    $\pm$ 0.13 & 28.30    $\pm$ 0.10 \\
\tr{28} & $>$30.25            & 29.82    $\pm$ 0.11 & 28.56    $\pm$ 0.05 & 27.78    $\pm$ 0.04 & 27.84 $\pm$ 0.03 & 27.83    $\pm$ 0.05 & 27.81 $\pm$ 0.04 & 28.10    $\pm$ 0.06 & 28.03    $\pm$ 0.05 \\
\tr{87} & $>$30.25            & $>$30.57            & $>$30.53            & 29.61    $\pm$ 0.16 & 29.18 $\pm$ 0.13 & 28.82    $\pm$ 0.17 & 28.70 $\pm$ 0.11 & 29.82    $\pm$ 0.42 & 29.38    $\pm$ 0.25 \\
\tr{60} & $>$30.25            & 30.14    $\pm$ 0.20 & 28.87    $\pm$ 0.10 & 28.29    $\pm$ 0.08 & 28.71 $\pm$ 0.14 & 30.28    $\pm$ 0.75 & 28.69 $\pm$ 0.18 & $>$30.24            & $>$30.44            \\	
\tr{5}	& 29.65    $\pm$ 0.11 & 28.63    $\pm$ 0.04 & 27.82    $\pm$ 0.03 & 27.40    $\pm$ 0.03 & 27.30 $\pm$ 0.03 & 27.39    $\pm$ 0.04 & 27.39 $\pm$ 0.03 & 27.40    $\pm$ 0.04 & 27.52    $\pm$ 0.04 \\
\tr{9}  & $>$30.25            & 29.42    $\pm$ 0.09 & 27.63    $\pm$ 0.04 & 27.07    $\pm$ 0.04 & 26.97 $\pm$ 0.04 & 27.78    $\pm$ 0.14 & 28.05 $\pm$ 0.13 & 28.06    $\pm$ 0.20 & 27.60    $\pm$ 0.12 \\
\tr{83} & $>$30.25            & $>$30.57            & 30.08    $\pm$ 0.24 & 28.73    $\pm$ 0.09 & 28.52 $\pm$ 0.09 & 28.58    $\pm$ 0.16 & 28.76 $\pm$ 0.12 & 28.62    $\pm$ 0.16 & 28.38    $\pm$ 0.11 \\
\tr{22} & 27.15    $\pm$ 0.03 & 27.08    $\pm$ 0.02 & 27.13    $\pm$ 0.03 & 27.22    $\pm$ 0.03 & 27.59 $\pm$ 0.03 & 27.78    $\pm$ 0.06 & 27.83 $\pm$ 0.05 & 27.86    $\pm$ 0.07 & 28.01    $\pm$ 0.07 \\
\tr{35} & $>$30.25            & $>$30.57            & $>$30.53            & 30.25    $\pm$ 0.16 & 29.14 $\pm$ 0.10 & 28.63    $\pm$ 0.10 & 28.63 $\pm$ 0.08 & 28.95    $\pm$ 0.16 & 28.87    $\pm$ 0.14 \\
\tr{30} & 28.43    $\pm$ 0.05 & 27.21    $\pm$ 0.03 & 27.18    $\pm$ 0.03 & 27.32    $\pm$ 0.03 & 28.88 $\pm$ 0.08 & 28.43    $\pm$ 0.09 & 28.22 $\pm$ 0.06 & 28.37    $\pm$ 0.10 & 28.25    $\pm$ 0.08 \\
\tr{81} & $>$30.25            & 30.49    $\pm$ 0.23 & 28.69    $\pm$ 0.08 & 29.26    $\pm$ 0.14 & 29.72 $\pm$ 0.21 & 29.99    $\pm$ 0.45 & 29.89 $\pm$ 0.30 & 28.98    $\pm$ 0.23 & 28.57    $\pm$ 0.13 \\
\tr{48} & $>$30.25            & $>$30.57            & $>$30.53            & $>$30.42            & 30.07 $\pm$ 0.17 & 29.43    $\pm$ 0.15 & 29.29 $\pm$ 0.11 & 28.80    $\pm$ 0.10 & 28.92    $\pm$ 0.11 \\
\tr{45} & \nodata             & \nodata             & \nodata             & \nodata             & 27.39 $\pm$ 0.05 & 27.63    $\pm$ 0.09 & 27.59 $\pm$ 0.07 & 27.44    $\pm$ 0.08 & 27.51    $\pm$ 0.06 \\
\tr{24} & 27.21    $\pm$ 0.03 & 26.80    $\pm$ 0.02 & 26.45    $\pm$ 0.02 & 26.57    $\pm$ 0.02 & 26.77 $\pm$ 0.02 & 27.00    $\pm$ 0.03 & 26.93 $\pm$ 0.02 & 27.19    $\pm$ 0.03 & 27.25    $\pm$ 0.03 \\
\tr{82} & $>$30.25            & $>$30.57            & 30.28    $\pm$ 0.16 & 29.98    $\pm$ 0.13 & 29.93 $\pm$ 0.15 & 30.01    $\pm$ 0.25 & 30.80 $\pm$ 0.34 & 30.16    $\pm$ 0.33 & 30.09    $\pm$ 0.25 \\
\tr{14} & 30.15    $\pm$ 0.23 & 28.12    $\pm$ 0.05 & 26.53    $\pm$ 0.03 & 25.91    $\pm$ 0.03 & 26.29 $\pm$ 0.03 & 26.27    $\pm$ 0.05 & 26.18 $\pm$ 0.04 & 26.61    $\pm$ 0.05 & 26.86    $\pm$ 0.05 \\
\tr{90} & \nodata             & \nodata             & \nodata             & \nodata             & 27.46 $\pm$ 0.06 & 28.22    $\pm$ 0.17 & 27.79 $\pm$ 0.08 & 28.65    $\pm$ 0.20 & 27.99    $\pm$ 0.10 \\
\tr{89} & 29.93    $\pm$ 0.19 & 29.50    $\pm$ 0.13 & 29.18    $\pm$ 0.11 & \nodata             & 29.02 $\pm$ 0.13 & 30.11    $\pm$ 0.45 & 29.97 $\pm$ 0.32 & 29.42    $\pm$ 0.29 & 29.51    $\pm$ 0.27 \\
\tr{25} & 29.61    $\pm$ 0.13 & 29.28    $\pm$ 0.08 & 28.76    $\pm$ 0.07 & 28.63    $\pm$ 0.06 & 28.62 $\pm$ 0.07 & 28.74    $\pm$ 0.11 & 29.20 $\pm$ 0.12 & 28.56    $\pm$ 0.12 & 29.12    $\pm$ 0.18 \\
\sidehead{Marginal Detections} 
\tr{84} & $>$30.25            & $>$30.57            & 30.28    $\pm$ 0.33 & 28.90    $\pm$ 0.12 & 29.33 $\pm$ 0.18 & 29.28    $\pm$ 0.26 &	29.30 $\pm$ 0.21 & $>$30.24            & 29.63    $\pm$ 0.29 \\
\tr{59} & 30.09    $\pm$ 0.25 & 28.89    $\pm$ 0.08 & 28.75    $\pm$ 0.10 & 28.27    $\pm$ 0.09 & 30.52 $\pm$ 0.79 & 27.79    $\pm$ 0.18 &	28.64 $\pm$ 0.28 & 28.94    $\pm$ 0.43 & 29.35    $\pm$ 0.47 \\
\enddata

\label{tab_JD23_photometry}

\end{deluxetable}
\end{longrotatetable}
\begin{longrotatetable}
\begin{deluxetable*}{cccccccccc}
\tablecaption{JADES-SN-22 Epoch1-Epoch2 SN photometry}
\tablehead{
\colhead{ID} & \colhead{F090W} & \colhead{F115W} & \colhead{F150W} & \colhead{F200W} & \colhead{F277W} & \colhead{F335M} & \colhead{F356W} & \colhead{F410M} & \colhead{F444W} \\
\colhead{} & \colhead{(AB mag)} & \colhead{(AB mag)} & \colhead{(AB mag)} & \colhead{(AB mag)} & \colhead{(AB mag)} & \colhead{(AB mag)} & \colhead{(AB mag)} & \colhead{(AB mag)} & \colhead{(AB mag)}
} 
\startdata
\tr{77}  & $>$30.25            & $>$30.57            & $>$30.53            & $>$30.42            & 29.49 $\pm$ 0.14 & 29.27    $\pm$ 0.18 & 29.03    $\pm$ 0.12 & 29.42    $\pm$ 0.24 & 29.31    $\pm$ 0.21 \\
\tr{33}  & $>$30.25            & $>$30.57            & $>$30.53            & 28.94    $\pm$ 0.06 & 27.74 $\pm$ 0.03 & 27.46    $\pm$ 0.04 & 27.56    $\pm$ 0.03 & 27.61    $\pm$ 0.04 & 27.56    $\pm$ 0.04 \\
\tr{39}  & $>$30.25            & $>$30.57            & $>$30.53            & $>$30.42            & 30.17 $\pm$ 0.15 & 30.08    $\pm$ 0.25 & 30.24    $\pm$ 0.22 & $>$30.24            & 30.06    $\pm$ 0.24 \\
\tr{93}  & $>$30.25            & $>$30.57            & $>$30.53            & $>$30.42            & 29.78 $\pm$ 0.11 & 30.07    $\pm$ 0.25 & 29.73    $\pm$ 0.14 & 30.09    $\pm$ 0.28 & 30.05    $\pm$ 0.26 \\
\tr{107} & $>$30.25            & $>$30.57            & $>$30.53            & $>$30.42            & 30.07 $\pm$ 0.15 & 29.74    $\pm$ 0.19 & 29.87    $\pm$ 0.16 & 30.00    $\pm$ 0.27 & 30.20    $\pm$ 0.31 \\
\tr{102} & $>$30.25            & 30.22    $\pm$ 0.43 & 29.14    $\pm$ 0.16 & \nodata             & 28.35 $\pm$ 0.08 & \nodata             & \nodata             & 28.52    $\pm$ 0.16 & 29.30    $\pm$ 0.20 \\
\tr{103} & $>$30.25            & $>$30.57            & $>$30.53            & 30.30    $\pm$ 0.18 & 30.22 $\pm$ 0.19 & 30.24    $\pm$ 0.32 & 30.52    $\pm$ 0.31 & $>$30.24            & 29.69    $\pm$ 0.19 \\
\tr{13}  & 29.73    $\pm$ 0.13 & 29.32    $\pm$ 0.09 & 29.23    $\pm$ 0.09 & 29.35    $\pm$ 0.11 & 30.78 $\pm$ 0.35 & $>$30.30            & $>$30.84            & $>$30.24            & $>$30.44            \\	
\tr{38}  & $>$30.25            & $>$30.57            & 29.86    $\pm$ 0.17 & 28.86    $\pm$ 0.10 & 28.31 $\pm$ 0.07 & 28.82    $\pm$ 0.18 & 28.74    $\pm$ 0.13 & 28.93    $\pm$ 0.25 & 29.14    $\pm$ 0.23 \\
\tr{55}  & $>$30.25            & $>$30.57            & $>$30.53            & 29.04    $\pm$ 0.09 & 28.48 $\pm$ 0.07 & 27.80    $\pm$ 0.08 & 28.04    $\pm$ 0.07 & 27.97    $\pm$ 0.10 & 28.07    $\pm$ 0.09 \\
\tr{21}  & $>$30.25            & 29.99    $\pm$ 0.09 & 28.64    $\pm$ 0.04 & 27.97    $\pm$ 0.03 & 27.79 $\pm$ 0.03 & 27.86    $\pm$ 0.05 & 27.81    $\pm$ 0.04 & 27.93    $\pm$ 0.05 & 27.92    $\pm$ 0.05 \\
\tr{100} & $>$30.25            & $>$30.57            & $>$30.53            & $>$30.42            & 30.25 $\pm$ 0.15 & $>$30.30            & $>$30.84            & $>$30.24            & $>$30.44            \\
\tr{79}  & $>$30.25            & $>$30.57            & $>$30.53            & 29.79    $\pm$ 0.18 & 29.24 $\pm$ 0.13 & 29.03    $\pm$ 0.19 & 28.89    $\pm$ 0.13 & 28.54    $\pm$ 0.14 & 28.60    $\pm$ 0.13 \\
\tr{80}  & $>$30.25            & $>$30.57            & $>$30.53            & $>$30.42            & 30.00 $\pm$ 0.15 & 29.10    $\pm$ 0.12 & 29.18    $\pm$ 0.11 & 29.28    $\pm$ 0.17 & 29.34    $\pm$ 0.16 \\
\tr{95}  & $>$30.25            & $>$30.57            & $>$30.53            & 29.83    $\pm$ 0.17 & 29.24 $\pm$ 0.12 & 30.12    $\pm$ 0.36 & 29.09    $\pm$ 0.13 & 29.07    $\pm$ 0.20 & 29.04    $\pm$ 0.17 \\
\tr{8}   & $>$30.25            & 29.47    $\pm$ 0.07 & 28.68    $\pm$ 0.05 & 28.32    $\pm$ 0.04 & 28.55 $\pm$ 0.07 & \nodata             & \nodata             & 28.97    $\pm$ 0.20 & 28.45    $\pm$ 0.12 \\
\tr{34}  & $>$30.25            & $>$30.57            & 28.73    $\pm$ 0.06 & 28.08    $\pm$ 0.04 & 28.20 $\pm$ 0.06 & 28.66    $\pm$ 0.14 & 28.44    $\pm$ 0.09 & 28.59    $\pm$ 0.14 & 28.62    $\pm$ 0.12 \\
\tr{23}  &	$>$30.25            & 30.13    $\pm$ 0.10 & 28.76   $\pm$ 0.05 & 27.80    $\pm$ 0.03 & 27.16 $\pm$ 0.03 & 26.99    $\pm$ 0.04 & 26.92    $\pm$ 0.03 & 26.91    $\pm$ 0.04 & 26.99    $\pm$ 0.04 \\
\tr{66}  & $>$30.25            & $>$30.57            & 29.86    $\pm$ 0.15 & 29.81    $\pm$ 0.18 & 29.29 $\pm$ 0.11 & 29.42    $\pm$ 0.19 & 29.22    $\pm$ 0.13 & 29.56    $\pm$ 0.29 & 29.85    $\pm$ 0.30 \\
\tr{92}  & $>$30.25            & $>$30.57            & $>$30.53            & 29.95    $\pm$ 0.14 & 29.68 $\pm$ 0.11 & 29.55    $\pm$ 0.16 & 29.82    $\pm$ 0.15 & 29.74    $\pm$ 0.22 & 29.75    $\pm$ 0.22 \\
\tr{37}  & $>$30.25            & $>$30.57            & 30.07    $\pm$ 0.17 & 30.17    $\pm$ 0.22 & 29.67 $\pm$ 0.14 & $>$30.30            & 29.56    $\pm$ 0.16 & 29.16    $\pm$ 0.20 & 29.81    $\pm$ 0.28 \\
\tr{20}  & 29.84    $\pm$ 0.08 & 28.79    $\pm$ 0.03 & 27.93    $\pm$ 0.03 & 27.15    $\pm$ 0.02 & 27.24 $\pm$ 0.02 & 27.14    $\pm$ 0.03 & 27.10    $\pm$ 0.02 & 27.28    $\pm$ 0.03 & 27.29    $\pm$ 0.03 \\
\tr{111} & $>$30.25            & 30.13    $\pm$ 0.53 & 28.79    $\pm$ 0.09 & \nodata             & 28.00 $\pm$ 0.10 & 28.99    $\pm$ 0.33 & 29.38    $\pm$ 0.16 & 29.52    $\pm$ 0.35 & 29.17    $\pm$ 0.22 \\
\tr{2}   & 27.94    $\pm$ 0.05 & 26.93    $\pm$ 0.02 & 26.67    $\pm$ 0.03 & 26.60    $\pm$ 0.02 & 26.67 $\pm$ 0.03 & 26.74    $\pm$ 0.03 & 26.77    $\pm$ 0.03 & 26.95    $\pm$ 0.04 & 26.91    $\pm$ 0.04 \\
\tr{16}  & 26.40    $\pm$ 0.02 & 25.73    $\pm$ 0.02 & 25.27    $\pm$ 0.02 & 25.15    $\pm$ 0.02 & 25.28 $\pm$ 0.02 & 25.42    $\pm$ 0.02 & 25.43    $\pm$ 0.02 & 25.61    $\pm$ 0.03 & 25.69    $\pm$ 0.02 \\
\tr{64}  & $>$30.25            & 30.42    $\pm$ 0.15 & 29.96    $\pm$ 0.12 & 29.78    $\pm$ 0.12 & 29.75 $\pm$ 0.12 & 30.21    $\pm$ 0.29 & $>$30.84            & $>$30.24            & $>$30.4             \\
\tr{1}   & 27.31    $\pm$ 0.03 & 26.46    $\pm$ 0.02 & 26.48    $\pm$ 0.02 & 26.74    $\pm$ 0.03 & 27.19 $\pm$ 0.03 & 27.49    $\pm$ 0.04 & 27.56    $\pm$ 0.03 & 27.83    $\pm$ 0.05 & 27.86    $\pm$ 0.05 \\
\tr{69}  & $>$30.25            & 30.28    $\pm$ 0.15 & 29.37    $\pm$ 0.09 & 28.96    $\pm$ 0.08 & 29.12 $\pm$ 0.10 & 29.04    $\pm$ 0.16 & 29.28    $\pm$ 0.13 & 29.20    $\pm$ 0.18 & 29.38    $\pm$ 0.18 \\
\tr{12}  & 28.31    $\pm$ 0.06 & 28.11    $\pm$ 0.05 & 27.94    $\pm$ 0.05 & 27.91    $\pm$ 0.06 & 28.15 $\pm$ 0.09 & 28.39    $\pm$ 0.20 & 28.10    $\pm$ 0.11 & 28.65    $\pm$ 0.25 & 28.73    $\pm$ 0.22 \\
\tr{46}  & $>$30.25            & 29.14    $\pm$ 0.08 & 28.32    $\pm$ 0.05 & 28.03    $\pm$ 0.05 & 28.01 $\pm$ 0.05 & 28.73    $\pm$ 0.13 & 28.62    $\pm$ 0.09 & 28.31    $\pm$ 0.11 & 28.46    $\pm$ 0.12 \\
\tr{65}  & $>$30.25            & $>$30.57            & 30.48    $\pm$ 0.19 & 29.99    $\pm$ 0.16 & 30.02 $\pm$ 0.14 & $>$30.30            & 30.34    $\pm$ 0.23 & $>$30.24            & 30.01    $\pm$ 0.25 \\
\tr{54}  & $>$30.25            & $>$30.57            & 29.33    $\pm$ 0.10 & 29.34    $\pm$ 0.10 & 29.64 $\pm$ 0.16 & 30.16    $\pm$ 0.38 & 30.33    $\pm$ 0.31 & $>$30.24            & 29.47    $\pm$ 0.22 \\
\tr{67}  & $>$30.25            & $>$30.57            & 29.69    $\pm$ 0.12 & 29.52    $\pm$ 0.13 & 30.03 $\pm$ 0.18 & $>$30.30            & 29.73    $\pm$ 0.19 & 29.28    $\pm$ 0.19 & 29.33    $\pm$ 0.16 \\
\tr{56}  & 29.20    $\pm$ 0.08 & 28.44    $\pm$ 0.04 & 28.08    $\pm$ 0.04 & 28.22    $\pm$ 0.06 & 27.69 $\pm$ 0.05 & 28.35    $\pm$ 0.15 & 28.16    $\pm$ 0.09 & 27.11    $\pm$ 0.06 & 27.26    $\pm$ 0.05 \\
\tr{68}  & 29.30    $\pm$ 0.10 & 28.91    $\pm$ 0.06 & 29.00    $\pm$ 0.07 & 28.99    $\pm$ 0.08 & 29.72 $\pm$ 0.15 & $>$30.30            & $>$30.84            & $>$30.24            & $>$30.44            \\
\tr{18}  & 29.78    $\pm$ 0.19 & 28.37    $\pm$ 0.06 & 27.85    $\pm$ 0.05 & \nodata             & 27.44 $\pm$ 0.05 & 27.63    $\pm$ 0.09 & 27.57    $\pm$ 0.06 & 27.62    $\pm$ 0.08 & 27.77    $\pm$ 0.08 \\
\tr{17}  & 27.32    $\pm$ 0.03 & 26.51    $\pm$ 0.02 & 26.13    $\pm$ 0.02 & 26.17    $\pm$ 0.03 & 26.45 $\pm$ 0.03 & 26.59    $\pm$ 0.06 & 26.64    $\pm$ 0.04 & 26.74    $\pm$ 0.06 & 26.99    $\pm$ 0.05 \\
\tr{31}  & $>$30.25            & $>$30.57            & 29.76    $\pm$ 0.19 & 27.90    $\pm$ 0.06 & 27.09 $\pm$ 0.05 & 26.89    $\pm$ 0.07 & 26.75    $\pm$ 0.04 & 26.94    $\pm$ 0.07 & 26.86    $\pm$ 0.05 \\
\tr{61}  & 29.59    $\pm$ 0.15 & 28.36    $\pm$ 0.05 & 28.29    $\pm$ 0.06 & 28.24    $\pm$ 0.07 & 28.58 $\pm$ 0.10 & 28.05    $\pm$ 0.11 & 27.91    $\pm$ 0.08 & 27.17    $\pm$ 0.06 & 27.35    $\pm$ 0.05 \\
\tr{109} & 30.14    $\pm$ 0.23 & 29.83    $\pm$ 0.15 & 29.36    $\pm$ 0.13 & 29.48    $\pm$ 0.16 & 29.86 $\pm$ 0.26 & 30.23    $\pm$ 0.53 & 29.80    $\pm$ 0.29 & 29.42    $\pm$ 0.31 & 30.29    $\pm$ 0.54 \\
\tr{3}   & 27.47    $\pm$ 0.03 & 26.83    $\pm$ 0.02 & 26.53    $\pm$ 0.02 & 26.49    $\pm$ 0.02 & 26.78 $\pm$ 0.03 & 27.09    $\pm$ 0.05 & 27.01    $\pm$ 0.04 & 27.19    $\pm$ 0.05 & 27.26    $\pm$ 0.04 \\
\tr{4}   & 29.61    $\pm$ 0.09 & 29.35    $\pm$ 0.06 & 29.11    $\pm$ 0.07 & 29.33    $\pm$ 0.09 & 29.66 $\pm$ 0.12 & $>$30.30            & 30.31    $\pm$ 0.25 & $>$30.24            & $>$30.44            \\	
\tr{110} & 29.83    $\pm$ 0.15 & 29.31    $\pm$ 0.08 & 29.16    $\pm$ 0.09 & 29.93    $\pm$ 0.18 & 29.63 $\pm$ 0.16 & 29.91    $\pm$ 0.31 & 29.73    $\pm$ 0.20 & 29.43    $\pm$ 0.26 & 30.29    $\pm$ 0.44 \\
\tr{101} & $>$30.25            & $>$30.57            & $>$30.53            & $>$30.42            & 29.95 $\pm$ 0.13 & 29.72    $\pm$ 0.18 & 29.73    $\pm$ 0.14 & 29.68    $\pm$ 0.20 & 29.75    $\pm$ 0.20 \\
\tr{32}  & $>$30.25            & $>$30.57            & 28.73    $\pm$ 0.05 & 28.06    $\pm$ 0.03 & 28.08 $\pm$ 0.04 & 28.06    $\pm$ 0.05 & 28.10    $\pm$ 0.04 & 28.02    $\pm$ 0.06 & 28.28    $\pm$ 0.06 \\
\sidehead{Marginal Detections} 
\tr{96}  & $>$30.25            & $>$30.57            & $>$30.53            & 30.24    $\pm$ 0.25 & 29.40 $\pm$ 0.15 & 29.78    $\pm$ 0.37 & 29.75    $\pm$ 0.24 & $>$30.24            & 30.15    $\pm$ 0.41 \\
\tr{94}  & $>$30.25            & $>$30.57            & $>$30.53            & 30.00    $\pm$ 0.21 & 29.90 $\pm$ 0.21 & 29.72    $\pm$ 0.27 & 29.25    $\pm$ 0.14 & 29.55    $\pm$ 0.29 & $>$30.44            \\
\enddata

\label{tab_JD22_photometry}

\end{deluxetable*}
\end{longrotatetable}
\begin{longrotatetable}
\begin{deluxetable*}{cccccccccc}
\tablecaption{JADES-SN-23 Epoch3-Epoch1 SN photometry}
\tablehead{
\colhead{ID} & \colhead{F090W} & \colhead{F115W} & \colhead{F150W} & \colhead{F200W} & \colhead{F277W} & \colhead{F335M} & \colhead{F356W} & \colhead{F410M} & \colhead{F444W} \\
\colhead{} & \colhead{(AB mag)} & \colhead{(AB mag)} & \colhead{(AB mag)} & \colhead{(AB mag)} & \colhead{(AB mag)} & \colhead{(AB mag)} & \colhead{(AB mag)} & \colhead{(AB mag)} & \colhead{(AB mag)}
} 
\label{tab_E3_photometry}
\startdata
\tr{50} & $>$29.89            & $>$30.30            & $>$30.14            & $>$30.24            & 29.98    $\pm$ 0.27 & 29.24    $\pm$ 0.23 & 29.37    $\pm$ 0.17 & $>$29.53            & 29.66 $\pm$ 0.35 \\	
\tr{44} & $>$29.89            & $>$30.30            & $>$30.14            & 29.88    $\pm$ 0.25 & 29.69    $\pm$ 0.23 & 29.19    $\pm$ 0.23 & 29.37    $\pm$ 0.18 & 29.16    $\pm$ 0.30 & 29.40 $\pm$ 0.33 \\
\tr{27} & $>$29.89            & $>$30.30            & $>$30.14            & 29.00    $\pm$ 0.12 & 28.41    $\pm$ 0.08 & 28.13    $\pm$ 0.09 & 28.45    $\pm$ 0.09 & 28.57    $\pm$ 0.17 & 28.21 $\pm$ 0.11 \\
\tr{36} & $>$29.89            & $>$30.30            & $>$30.14            & 29.64    $\pm$ 0.16 & 28.98    $\pm$ 0.12 & 29.42    $\pm$ 0.23 & 29.07    $\pm$ 0.13 & 28.58    $\pm$ 0.16 & 29.45 $\pm$ 0.25 \\
\tr{26} & $>$29.89            & 29.73    $\pm$	0.18 & 27.45   $\pm$ 0.05 & 26.53    $\pm$ 0.03 & 26.64    $\pm$ 0.03 & 26.52    $\pm$ 0.03 & 26.74    $\pm$ 0.03 & 26.86    $\pm$ 0.05 & 27.07 $\pm$ 0.05 \\
\tr{28} & $>$29.89            & $>$30.30            & 29.13    $\pm$ 0.12 & 28.22    $\pm$ 0.06 & 28.25    $\pm$ 0.07 & 27.95    $\pm$ 0.07 & 28.27    $\pm$ 0.07 & 28.92    $\pm$ 0.21 & 28.52 $\pm$ 0.13 \\
\tr{9}  & $>$29.89            & $>$30.30            & 28.44    $\pm$ 0.12 & 27.67    $\pm$ 0.08 & 26.42    $\pm$ 0.06 & \nodata             & 26.94    $\pm$ 0.11 & \nodata             & 26.46 $\pm$ 0.09 \\
\tr{22} & 28.74    $\pm$ 0.10 & 27.35    $\pm$ 0.03 & 27.12    $\pm$ 0.03 & 27.13    $\pm$ 0.03 & 27.31    $\pm$ 0.04 & 27.51    $\pm$ 0.06 & 27.64    $\pm$ 0.05 & 27.73    $\pm$ 0.10 & 27.74 $\pm$ 0.08 \\
\sidehead{Marginal Detections} 
\tr{84} & $>$29.89            & $>$30.3             & $>$30.14            & $>$30.24            & $>$30.30 $\pm$      & $>$29.75            & $>$30.26            & $>$29.53            & $>$29.76         \\	
\enddata

\end{deluxetable*}
\end{longrotatetable}
\begin{deluxetable*}{ccccccccc}[h]
\tablecaption{JADES-SN-23 Epoch4-Epoch1 SN photometry}
\tablehead{
\colhead{ID} & \colhead{F115W} & \colhead{F150W} & \colhead{F200W} & \colhead{F277W} & \colhead{F356W} & \colhead{F444W} \\
\colhead{} & \colhead{(AB mag)} & \colhead{(AB mag)} & \colhead{(AB mag)} & \colhead{(AB mag)} & \colhead{(AB mag)} & \colhead{(AB mag)}
} 
\label{tab_E4_photometry}
\startdata
\tr{53} & $>$28.88            & $>$29.18            & $>$29.32            & $>$29.61            & 28.76    $\pm$ 0.23 & 28.67    $\pm$ 0.30 \\
\tr{50} & $>$28.88            & $>$29.18            & $>$29.32            & $>$29.61            & 29.24    $\pm$ 0.30 & $>$29.02            \\
\tr{88} & $>$28.88            & $>$29.18            & $>$29.32            & $>$29.61            & $>$29.60            & $>$29.02            \\
\tr{10} & $>$28.88            & $>$29.18            & 28.43    $\pm$ 0.13 & 28.26    $\pm$ 0.13 & 28.08    $\pm$ 0.12 & 28.26    $\pm$ 0.21 \\
\tr{44} & $>$28.88            & $>$29.18            & $>$29.32            & $>$29.61            & 28.85    $\pm$ 0.18 & 29.00    $\pm$ 0.34 \\
\tr{71} & $>$28.88            & $>$29.18            & $>$29.32            & $>$29.61            & $>$29.60            & $>$29.02            \\
\tr{27} & $>$28.88            & $>$29.18            & 28.85    $\pm$ 0.18 & 28.53    $\pm$ 0.15 & 28.48    $\pm$ 0.16 & 28.65    $\pm$ 0.29 \\
\tr{36} & $>$28.88            & $>$29.18            & $>$29.32            & 28.88    $\pm$ 0.19 & 29.47    $\pm$ 0.33 & $>$29.02            \\
\tr{26} & $>$28.88            & 27.63    $\pm$ 0.08 & 26.55    $\pm$ 0.04 & 26.57    $\pm$ 0.04 & 26.59    $\pm$ 0.05 & 26.92    $\pm$ 0.07 \\
\tr{52} & $>$28.88            & $>$29.18            & $>$29.32            & 29.14    $\pm$ 0.24 & $>$29.60            & $>$29.02            \\
\tr{15} & $>$28.88            & $>$29.18            & $>$29.32            & $>$29.61            & $>$29.60            & $>$29.02            \\
\tr{29} & $>$28.88            & $>$29.18            & 28.42    $\pm$ 0.13 & 28.16    $\pm$ 0.10 & 27.65    $\pm$ 0.09 & 27.92    $\pm$ 0.15 \\
\tr{6}  & $>$28.88            & $>$29.18            & 28.86    $\pm$ 0.21 & 28.88    $\pm$ 0.28 & $>$29.60            & $>$29.02            \\
\tr{11} & $>$28.88            & 28.96    $\pm$ 0.29 & 29.10    $\pm$ 0.26 & 28.68    $\pm$ 0.12 & 28.99    $\pm$ 0.13 & $>$29.02            \\
\tr{7}  & $>$28.88            & 28.70    $\pm$ 0.19 & 28.42    $\pm$ 0.16 & 28.55    $\pm$ 0.20 & 28.77    $\pm$ 0.26 & 28.76    $\pm$ 0.35 \\
\tr{28} & $>$28.88            & $>$29.18            & 28.20    $\pm$ 0.12 & 28.24    $\pm$ 0.13 & 28.45    $\pm$ 0.15 & 28.79    $\pm$ 0.30 \\
\tr{5}  & $>$28.88            & 29.12    $\pm$ 0.25 & 28.62    $\pm$ 0.17 & 28.24    $\pm$ 0.12 & 27.68    $\pm$ 0.08 & 28.34    $\pm$ 0.23 \\
\tr{9}  & $>$28.88            & 28.68    $\pm$ 0.20 & 27.72    $\pm$ 0.10 & 26.49    $\pm$ 0.07 & 27.39    $\pm$ 0.16 & 26.57    $\pm$ 0.11 \\
\tr{22} & 27.50    $\pm$ 0.08 & 27.24    $\pm$ 0.06 & 27.16    $\pm$ 0.06 & 27.29    $\pm$ 0.06 & 27.76    $\pm$ 0.10 & 27.67    $\pm$ 0.14 \\
\tr{35} & $>$28.88            & $>$29.18            & $>$29.32            & 29.38    $\pm$ 0.28 & 28.59    $\pm$ 0.16 & $>$29.02            \\
\tr{48} & $>$28.88            & $>$29.18            & $>$29.32            & 29.43    $\pm$ 0.32 & 29.15    $\pm$ 0.27 & $>$29.02            \\
\tr{14} & 28.34    $\pm$ 0.16 & 26.40    $\pm$ 0.05 & 26.56    $\pm$ 0.06 & 26.00    $\pm$ 0.06 & 26.62    $\pm$ 0.10 & 27.05    $\pm$ 0.14 \\
\sidehead{Marginal Detections}
\tr{84} & $>$28.88            & $>$29.18            & $>$29.32            & $>$29.61            & $>$29.60            & $>$29.02            \\
\enddata

\end{deluxetable*}
\begin{deluxetable*}{ccc}
\tablecaption{JADES-SN-23 Epoch5.1-Epoch1 SN photometry}
\tablehead{
\colhead{ID} & \colhead{F200W} & \colhead{F277W} \\
\colhead{} & \colhead{(AB mag)} & \colhead{(AB mag)}
} 
\label{tab_E5p1_photometry}
\startdata
\tr{53} & $>$30.18            & $>$30.34            \\
\tr{50} & $>$30.18            & 29.75    $\pm$ 0.23 \\
\tr{10} & \nodata\tablenotemark{a} & \nodata\tablenotemark{a} \\
\tr{44} & 29.74    $\pm$ 0.20 & 29.06    $\pm$ 0.13 \\
\tr{71} & $>$30.18            & $>$30.34            \\
\tr{27} & \nodata\tablenotemark{a} & 28.42 $\pm$ 0.09 \\
\tr{36} & 29.88    $\pm$ 0.24 & 29.23    $\pm$ 0.15 \\
\tr{26} & 27.00    $\pm$ 0.03 & 26.78    $\pm$ 0.03 \\
\tr{52} & 29.75    $\pm$ 0.18 & 29.90    $\pm$ 0.23 \\
\tr{15} & $>$30.18            & $>$30.34            \\
\tr{29} & 28.81    $\pm$ 0.10 & 28.09    $\pm$ 0.06 \\
\tr{6}  & 28.32    $\pm$ 0.08 & 28.62    $\pm$ 0.14 \\
\tr{11} & 28.59    $\pm$ 0.07 & 28.82    $\pm$ 0.10 \\
\tr{28} & 28.51    $\pm$ 0.08 & 28.95    $\pm$ 0.11 \\
\tr{87} & $>$30.18            & $>$30.34            \\
\tr{5}  & 29.01    $\pm$ 0.13 & 28.51    $\pm$ 0.10 \\
\tr{9}  & 28.15    $\pm$ 0.08 & 26.58    $\pm$ 0.05 \\
\tr{22} & 27.05    $\pm$ 0.04 & 27.11    $\pm$ 0.04 \\
\tr{81} & 28.52    $\pm$ 0.12 & 27.70    $\pm$ 0.08 \\
\tr{48} & $>$30.18            & $>$30.34            \\
\tr{24} & 26.36    $\pm$ 0.03 & 26.62    $\pm$ 0.03 \\
\sidehead{Marginal Detections} 
\tr{84} & $>$30.18            & $>$30.34            \\
\tr{59} & 28.65    $\pm$ 0.19 & \nodata\tablenotemark{a} \\
\enddata
\tablenotetext{a}{Data dropped because source either too close to image edge or contaminated by subtraction residual}
\end{deluxetable*}
\begin{deluxetable*}{ccccccccc}
\tablecaption{JADES-SN-23 Epoch5.2-Epoch1 SN photometry}
\tablehead{
\colhead{ID} & \colhead{F150W} & \colhead{F200W} & \colhead{F277W} & \colhead{F356W} & \colhead{F444W} \\
\colhead{} & \colhead{(AB mag)} & \colhead{(AB mag)} & \colhead{(AB mag)} & \colhead{(AB mag)} & \colhead{(AB mag)}
} 
\label{tab_E5p2_photometry}
\startdata
\tr{53} & $>$29.31            & $>$29.75            & $>$29.44            & 29.37    $\pm$ 0.39 & $>$28.89            \\
\tr{50} & $>$29.31            & $>$29.75            & 29.15    $\pm$ 0.26 & 28.80    $\pm$ 0.21 & $>$28.89            \\
\tr{10} & $>$29.31            & 28.99    $\pm$ 0.15 & 28.27    $\pm$ 0.14 & 27.92    $\pm$ 0.12 & 28.50    $\pm$ 0.28 \\
\tr{44} & $>$29.31            & $>$29.75            & 29.42    $\pm$ 0.26 & $>$29.41            & $>$28.89            \\
\tr{27} & $>$29.31            & 29.33    $\pm$ 0.19 & 28.58    $\pm$ 0.17 & 28.64    $\pm$ 0.18 & $>$28.89            \\
\tr{36} & $>$29.31            & $>$29.75            & 28.90    $\pm$ 0.19 & 29.28    $\pm$ 0.28 & $>$28.89            \\
\tr{26} & 28.44    $\pm$ 0.13 & 26.98    $\pm$ 0.04 & 26.88    $\pm$ 0.06 & 26.69    $\pm$ 0.05 & 26.94    $\pm$ 0.08 \\
\tr{52} & $>$29.31            & 29.45    $\pm$ 0.19 & $>$29.44            & 28.97    $\pm$ 0.22 & $>$28.89            \\
\tr{15} & $>$29.31            & $>$29.75            & $>$29.44            & $>$29.41            & $>$28.89            \\
\tr{29} & $>$29.31            & 28.60    $\pm$ 0.11 & 28.19    $\pm$ 0.10 & 27.76    $\pm$ 0.09 & 28.29    $\pm$ 0.21 \\
\tr{6}  & 29.27    $\pm$ 0.28 & 28.65    $\pm$ 0.13 & 28.74    $\pm$ 0.23 & 28.41    $\pm$ 0.19 & $>$28.89            \\
\tr{11} & 29.12    $\pm$ 0.19 & 28.66    $\pm$ 0.11 & 28.65    $\pm$ 0.16 & 28.72    $\pm$ 0.18 & $>$28.89            \\
\tr{7}  & $>$29.31            & 28.94    $\pm$ 0.18 &	$>$29.44          & $>$29.41            & 28.69    $\pm$ 0.34 \\
\tr{28} & $>$29.31            & 28.54    $\pm$ 0.11 & 28.55    $\pm$ 0.15 & 28.86    $\pm$ 0.20 & $>$28.89            \\
\tr{87} & $>$29.31            & $>$29.75            & $>$29.44            & $>$29.41            & $>$28.89            \\
\tr{5}  & \nodata\tablenotemark{a} & \nodata\tablenotemark{a} & \nodata             & \nodata             & \nodata             \\
\tr{9}  & 29.28    $\pm$ 0.27 & 28.40    $\pm$ 0.13 & 27.14    $\pm$ 0.11 & 29.26    $\pm$ 0.68 & 27.87    $\pm$ 0.33 \\
\tr{22} & 27.26    $\pm$ 0.06 & 27.09    $\pm$ 0.04 & 27.19    $\pm$ 0.06 & 27.46    $\pm$ 0.08 & \nodata\tablenotemark{a} \\
\tr{81} & $>$29.31            & $>$29.75            & $>$29.44            & $>$29.41            & $>$28.89            \\
\tr{24} & 26.31    $\pm$ 0.04 & 26.35    $\pm$ 0.03 & 26.64    $\pm$ 0.04 & 26.75    $\pm$ 0.05 & 27.06    $\pm$ 0.08 \\
\tr{14} & 27.36    $\pm$ 0.09 & 26.93    $\pm$ 0.06 & 26.64    $\pm$ 0.10 & 27.67    $\pm$ 0.22 & 27.45    $\pm$ 0.20 \\
\sidehead{Marginal Detections}
\tr{84} & $>$29.31            & $>$29.75            & $>$29.44            & $>$29.41            & $>$28.89            \\
\tr{59} & 28.40    $\pm$ 0.19 & 28.39    $\pm$ 0.18 & \nodata\tablenotemark{a} & \nodata\tablenotemark{a} & \nodata\tablenotemark{a} \\
\enddata
\tablenotetext{a}{Data dropped because source either too close to image edge or contaminated by subtraction residual}
\end{deluxetable*}
\begin{longrotatetable}
\begin{deluxetable*}{cccccccccc}
\tablecaption{JADES-SN-23 Epoch5.3-Epoch1 SN photometry}
\tablehead{
\colhead{ID} & \colhead{F090W} & \colhead{F115W} & \colhead{F150W} & \colhead{F200W} & \colhead{F277W} & \colhead{F335M} & \colhead{F356W} & \colhead{F410M} & \colhead{F444W} \\
\colhead{} & \colhead{(AB mag)} & \colhead{(AB mag)} & \colhead{(AB mag)} & \colhead{(AB mag)} & \colhead{(AB mag)} & \colhead{(AB mag)} & \colhead{(AB mag)} & \colhead{(AB mag)} & \colhead{(AB mag)}
} 
\label{tab_E5p3_photometry}
\startdata
\tr{50} & $>$29.40 & $>$29.90            & $>$29.80            & $>$29.80            & 29.46    $\pm$ 0.18 & $>$29.10            & 29.37    $\pm$ 0.17 & $>$29.10            & $>$29.20            \\	
\tr{44} & $>$29.40 & $>$29.90            & $>$29.80            & $>$29.80            & $>$29.70            & $>$29.10            & $>$29.80            & $>$29.10            & $>$29.20            \\
\tr{27} & $>$29.40 & $>$29.90            & $>$29.80            & 29.34    $\pm$ 0.15 & 28.44    $\pm$ 0.09 & 28.26    $\pm$ 0.10 & 28.43    $\pm$ 0.09 & 28.42    $\pm$ 0.15 & $>$29.20            \\
\tr{36} & $>$29.40 & $>$29.90            & $>$29.80            & $>$29.80            & 29.45    $\pm$ 0.19 & 29.04    $\pm$ 0.19 & 29.14    $\pm$ 0.15 & 28.94    $\pm$ 0.24 & $>$29.20            \\
\tr{26} & $>$29.40 & $>$29.90            & 28.85    $\pm$ 0.09 & 27.01    $\pm$ 0.03 & 26.79    $\pm$ 0.03 & 26.42    $\pm$ 0.03 & 26.63    $\pm$ 0.03 & 26.57    $\pm$ 0.04 & 26.79    $\pm$ 0.04 \\
\tr{29} & $>$29.40 & $>$29.90            & $>$29.80            & 28.74    $\pm$ 0.11 & 28.15    $\pm$ 0.09 & 27.72    $\pm$ 0.09 & 27.77    $\pm$ 0.07 & 27.95    $\pm$ 0.16 & 28.01    $\pm$ 0.14 \\
\tr{28} & $>$29.40 & $>$29.90            & 29.19    $\pm$ 0.11 & 28.42    $\pm$ 0.07 & 28.48    $\pm$ 0.08 & 28.70    $\pm$ 0.13 & 28.71    $\pm$ 0.10 & $>$29.10            & 28.87    $\pm$ 0.20 \\
\tr{22} & $>$29.40 & 27.92    $\pm$ 0.04 & 27.28    $\pm$ 0.04 & 27.12    $\pm$ 0.03 & 27.18    $\pm$ 0.04 & 27.48    $\pm$ 0.06 & 27.53    $\pm$ 0.05 & 27.44    $\pm$ 0.07 & 27.49    $\pm$ 0.07 \\
\tr{45} & \nodata  & \nodata             & \nodata             & \nodata             & 27.51    $\pm$ 0.08 & \nodata             & 28.07    $\pm$ 0.13 & \nodata             & 27.75    $\pm$ 0.12 \\
\enddata

\end{deluxetable*}
\end{longrotatetable}




\bibliography{references}{}

\begin{thebibliography}{}
\expandafter\ifx\csname natexlab\endcsname\relax\def\natexlab#1{#1}\fi
\providecommand{\url}[1]{\href{#1}{#1}}
\providecommand{\dodoi}[1]{doi:~\href{http://doi.org/#1}{\nolinkurl{#1}}}
\providecommand{\doeprint}[1]{\href{http://ascl.net/#1}{\nolinkurl{http://ascl.net/#1}}}
\providecommand{\doarXiv}[1]{\href{https://arxiv.org/abs/#1}{\nolinkurl{https://arxiv.org/abs/#1}}}

\bibitem[{{Astier} {et~al.}(2006){Astier}, {Guy}, {Regnault}, {Pain}, {Aubourg}, {Balam}, {Basa}, {Carlberg}, {Fabbro}, {Fouchez}, {Hook}, {Howell}, {Lafoux}, {Neill}, {Palanque-Delabrouille}, {Perrett}, {Pritchet}, {Rich}, {Sullivan}, {Taillet}, {Aldering}, {Antilogus}, {Arsenijevic}, {Balland}, {Baumont}, {Bronder}, {Courtois}, {Ellis}, {Filiol}, {Gon{\c{c}}alves}, {Goobar}, {Guide}, {Hardin}, {Lusset}, {Lidman}, {McMahon}, {Mouchet}, {Mourao}, {Perlmutter}, {Ripoche}, {Tao}, \& {Walton}}]{Astier2006TheSet}
{Astier}, P., {Guy}, J., {Regnault}, N., {et~al.} 2006, \aap, 447, 31

\bibitem[{{Astropy Collaboration} {et~al.}(2022){Astropy Collaboration}, {Price-Whelan}, {Lim}, {Earl}, {Starkman}, {Bradley}, {Shupe}, {Patil}, {Corrales}, {Brasseur}, {N{\"o}the}, {Donath}, {Tollerud}, {Morris}, {Ginsburg}, {Vaher}, {Weaver}, {Tocknell}, {Jamieson}, {van Kerkwijk}, {Robitaille}, {Merry}, {Bachetti}, {G{\"u}nther}, {Aldcroft}, {Alvarado-Montes}, {Archibald}, {B{\'o}di}, {Bapat}, {Barentsen}, {Baz{\'a}n}, {Biswas}, {Boquien}, {Burke}, {Cara}, {Cara}, {Conroy}, {Conseil}, {Craig}, {Cross}, {Cruz}, {D'Eugenio}, {Dencheva}, {Devillepoix}, {Dietrich}, {Eigenbrot}, {Erben}, {Ferreira}, {Foreman-Mackey}, {Fox}, {Freij}, {Garg}, {Geda}, {Glattly}, {Gondhalekar}, {Gordon}, {Grant}, {Greenfield}, {Groener}, {Guest}, {Gurovich}, {Handberg}, {Hart}, {Hatfield-Dodds}, {Homeier}, {Hosseinzadeh}, {Jenness}, {Jones}, {Joseph}, {Kalmbach}, {Karamehmetoglu}, {Ka{\l}uszy{\'n}ski}, {Kelley}, {Kern}, {Kerzendorf}, {Koch}, {Kulumani}, {Lee}, {Ly}, {Ma}, {MacBride}, {Maljaars}, {Muna}, {Murphy}, {Norman},
  {O'Steen}, {Oman}, {Pacifici}, {Pascual}, {Pascual-Granado}, {Patil}, {Perren}, {Pickering}, {Rastogi}, {Roulston}, {Ryan}, {Rykoff}, {Sabater}, {Sakurikar}, {Salgado}, {Sanghi}, {Saunders}, {Savchenko}, {Schwardt}, {Seifert-Eckert}, {Shih}, {Jain}, {Shukla}, {Sick}, {Simpson}, {Singanamalla}, {Singer}, {Singhal}, {Sinha}, {Sip{\H{o}}cz}, {Spitler}, {Stansby}, {Streicher}, {{\v{S}}umak}, {Swinbank}, {Taranu}, {Tewary}, {Tremblay}, {de Val-Borro}, {Van Kooten}, {Vasovi{\'c}}, {Verma}, {de Miranda Cardoso}, {Williams}, {Wilson}, {Winkel}, {Wood-Vasey}, {Xue}, {Yoachim}, {Zhang}, {Zonca}, \& {Astropy Project Contributors}}]{AstropyCollaboration2022ThePackage}
{Astropy Collaboration}, {Price-Whelan}, A.~M., {Lim}, P.~L., {et~al.} 2022, \apj, 935, 167

\bibitem[{{Bradley} {et~al.}(2024){Bradley}, {Sip{\H{o}}cz}, {Robitaille}, {Tollerud}, {Vin{\'\i}cius}, {Deil}, {Barbary}, {Wilson}, {Busko}, {Donath}, {G{\"u}nther}, {Cara}, {Lim}, {Me{\ss}linger}, {Burnett}, {Conseil}, {Droettboom}, {Bostroem}, {Bray}, {Andersen Bratholm}, {Jamieson}, {Ginsburg}, {Barentsen}, {Craig}, {Pascual}, {Rathi}, {Perrin}, {Morris}, \& {Perren}}]{Bradley2024Astropy/photutils:1.12.0}
{Bradley}, L., {Sip{\H{o}}cz}, B., {Robitaille}, T., {et~al.} 2024, {astropy/photutils: 1.12.0}, 1.12.0,  Zenodo

\bibitem[{{Brammer} {et~al.}(2008){Brammer}, {van Dokkum}, \& {Coppi}}]{Brammer2008EAZY:Code}
{Brammer}, G.~B., {van Dokkum}, P.~G., \& {Coppi}, P. 2008, \apj, 686, 1503

\bibitem[{{Brout} {et~al.}(2022){Brout}, {Scolnic}, {Popovic}, {Riess}, {Carr}, {Zuntz}, {Kessler}, {Davis}, {Hinton}, {Jones}, {Kenworthy}, {Peterson}, {Said}, {Taylor}, {Ali}, {Armstrong}, {Charvu}, {Dwomoh}, {Meldorf}, {Palmese}, {Qu}, {Rose}, {Sanchez}, {Stubbs}, {Vincenzi}, {Wood}, {Brown}, {Chen}, {Chambers}, {Coulter}, {Dai}, {Dimitriadis}, {Filippenko}, {Foley}, {Jha}, {Kelsey}, {Kirshner}, {M{\"o}ller}, {Muir}, {Nadathur}, {Pan}, {Rest}, {Rojas-Bravo}, {Sako}, {Siebert}, {Smith}, {Stahl}, \& {Wiseman}}]{Brout2022TheConstraints}
{Brout}, D., {Scolnic}, D., {Popovic}, B., {et~al.} 2022, \apj, 938, 110

\bibitem[{{Bunker} {et~al.}(2010){Bunker}, {Wilkins}, {Ellis}, {Stark}, {Lorenzoni}, {Chiu}, {Lacy}, {Jarvis}, \& {Hickey}}]{Bunker2010}
{Bunker}, A.~J., {Wilkins}, S., {Ellis}, R.~S., {et~al.} 2010, \mnras, 409, 855

\bibitem[{{Bunker} {et~al.}(2023){Bunker}, {Cameron}, {Curtis-Lake}, {Jakobsen}, {Carniani}, {Curti}, {Witstok}, {Maiolino}, {D'Eugenio}, {Looser}, {Willott}, {Bonaventura}, {Hainline}, {Uebler}, {Willmer}, {Saxena}, {Smit}, {Alberts}, {Arribas}, {Baker}, {Baum}, {Bhatawdekar}, {Bowler}, {Boyett}, {Charlot}, {Chen}, {Chevallard}, {Circosta}, {DeCoursey}, {de Graaff}, {Egami}, {Eisenstein}, {Endsley}, {Ferruit}, {Giardino}, {Hausen}, {Helton}, {Hviding}, {Ji}, {Johnson}, {Jones}, {Kumari}, {Laseter}, {Luetzgendorf}, {Maseda}, {Nelson}, {Parlanti}, {Perna}, {Rauscher}, {Rawle}, {Rix}, {Rieke}, {Robertson}, {Rodriguez Del Pino}, {Sandles}, {Scholtz}, {Sharpe}, {Skarbinski}, {Stark}, {Sun}, {Tacchella}, {Topping}, {Villanueva}, {Wallace}, {Williams}, \& {Woodrum}}]{Bunker2023}
{Bunker}, A.~J., {Cameron}, A.~J., {Curtis-Lake}, E., {et~al.} 2023, arXiv e-prints, arXiv:2306.02467

\bibitem[{{Conroy} {et~al.}(2009){Conroy}, {Gunn}, \& {White}}]{Conroy2009TheGalaxies}
{Conroy}, C., {Gunn}, J.~E., \& {White}, M. 2009, \apj, 699, 486

\bibitem[{{Cooke} {et~al.}(2012){Cooke}, {Sullivan}, {Gal-Yam}, {Barton}, {Carlberg}, {Ryan-Weber}, {Horst}, {Omori}, \& {D{\'\i}az}}]{Cooke2012Superluminous3.90}
{Cooke}, J., {Sullivan}, M., {Gal-Yam}, A., {et~al.} 2012, \nat, 491, 228

\bibitem[{{Cooper} {et~al.}(2012){Cooper}, {Yan}, {Dickinson}, {Juneau}, {Lotz}, {Newman}, {Papovich}, {Salim}, {Walth}, {Weiner}, \& {Willmer}}]{Cooper2012TheField-South}
{Cooper}, M.~C., {Yan}, R., {Dickinson}, M., {et~al.} 2012, \mnras, 425, 2116

\bibitem[{{Curtin} {et~al.}(2019){Curtin}, {Cooke}, {Moriya}, {Tanaka}, {Quimby}, {Bernard}, {Galbany}, {Jiang}, {Lee}, {Maeda}, {Morokuma}, {Nomoto}, {Pignata}, {Pritchard}, {Suzuki}, {Takahashi}, {Tanaka}, {Tominaga}, {Yamaguchi}, \& {Yasuda}}]{Curtin2019FirstProperties}
{Curtin}, C., {Cooke}, J., {Moriya}, T.~J., {et~al.} 2019, \apjs, 241, 17

\bibitem[{{D'Andrea} {et~al.}(2010){D'Andrea}, {Sako}, {Dilday}, {Frieman}, {Holtzman}, {Kessler}, {Konishi}, {Schneider}, {Sollerman}, {Wheeler}, {Yasuda}, {Cinabro}, {Jha}, {Nichol}, {Lampeitl}, {Smith}, {Atlee}, {Bassett}, {Castander}, {Goobar}, {Miquel}, {Nordin}, {{\"O}stman}, {Prieto}, {Quimby}, {Riess}, \& {Stritzinger}}]{DAndrea2010TYPEMETHOD}
{D'Andrea}, C.~B., {Sako}, M., {Dilday}, B., {et~al.} 2010, \apj, 708, 661

\bibitem[{{DeCoursey} {et~al.}(2023{\natexlab{a}}){DeCoursey}, {Egami}, {Rieke}, {DeFour-Remy}, {Khairnar}, {Ma}, {Sun}, {\ Eisenstein}, {Robertson}, {Johnson}, \& {Tacchella}}]{DeCoursey2023a}
{DeCoursey}, C., {Egami}, E., {Rieke}, M., {et~al.} 2023{\natexlab{a}}, Transient Name Server AstroNote, 16, 1

\bibitem[{{DeCoursey} {et~al.}(2023{\natexlab{b}}){DeCoursey}, {Egami}, {Rieke}, {DeFour-Remy}, {Khairnar}, {Ma}, {Sun}, {Willmer}, {Rest}, {Piere\ l}, {Engesser}, {Hainline}, {Helton}, {Eisenstein}, {Robertson}, {Johnson}, {Tacchella}, {Hausen}, \& {Williams}}]{DeCoursey2023b}
---. 2023{\natexlab{b}}, Transient Name Server AstroNote, 164, 1

\bibitem[{{DeCoursey} {et~al.}(2024){DeCoursey}, {Egami}, {Pierel}, {Sun}, {Rest}, {Coulter}, {Engesser}, {Siebert}, {Hainline}, {Johnson}, {Bunker}, {Cargile}, {Charlot}, {Chen}, {Curti}, {DeFour-Remy}, {Eisenstein}, {Fox}, {Gezari}, {Gomez}, {Jencson}, {Joshi}, {Khairnar}, {Lyu}, {Maiolino}, {Moriya}, {Quimby}, {Rieke}, {Rieke}, {Robertson}, {Shahbandeh}, {Strolger}, {Tacchella}, {Wang}, {Williams}, {Willmer}, {Willott}, \& {Zenati}}]{DeCoursey2024}
{DeCoursey}, C., {Egami}, E., {Pierel}, J.~D.~R., {et~al.} 2024, Transient Name Server AstroNote, 168, 1

\bibitem[{{D'Eugenio} {et~al.}(2024){D'Eugenio}, {Cameron}, {Scholtz}, {Carniani}, {Willott}, {Curtis-Lake}, {Bunker}, {Parlanti}, {Maiolino}, {Willmer}, {Jakobsen}, {Robertson}, {Johnson}, {Tacchella}, {Cargile}, {Arribas}, {Chevallard}, {Curti}, {Egami}, {Eisenstein}, {Kumari}, {Looser}, {Rieke}, {Rodr{\'\i}guez Del Pino}, {Saxena}, {{\"U}bler}, {Venturi}, {Witstok}, {Baker}, {Bhatawdekar}, {Bonav\ entura}, {Boyett}, {Charlot}, {Danhaive}, {Hainline}, {Hausen}, {Helton}, {Ji}, {Ji}, {Jones}, {Joud{\v{z}}balis}, {Maseda}, {P{\'e}rez-Gonz{\'a}lez}, {Perna}, {Pusk{\'a}s}, {Shivaei}, {Silcoc\ k}, {Simmonds}, {Smit}, {Sun}, {Villanueva}, {Williams}, \& {Zhu}}]{DEugenio2024}
{D'Eugenio}, F., {Cameron}, A.~J., {Scholtz}, J., {et~al.} 2024, arXiv e-prints, arXiv:2404.06531

\bibitem[{{Drout} {et~al.}(2011){Drout}, {Soderberg}, {Gal-Yam}, {Cenko}, {Fox}, {Leonard}, {Sand}, {Moon}, {Arcavi}, \& {Green}}]{Drout2011TheCurves}
{Drout}, M.~R., {Soderberg}, A.~M., {Gal-Yam}, A., {et~al.} 2011, \apj, 741, 97

\bibitem[{{Eisenstein} {et~al.}(2023){Eisenstein}, {Willott}, {Alberts}, {Arribas}, {Bonaventura}, {Bunker}, {Cameron}, {Carniani}, {Charlot}, {Curtis-Lake}, {D'Eugenio}, {Endsley}, {Ferruit}, {Giardino}, {Hainline}, {Hausen}, {Jakobsen}, {Johnson}, {Maiolino}, {Rieke}, {Rieke}, {Rix}, {Robertson}, {Stark}, {Tacchella}, {Williams}, {Willmer}, {Baker}, {Baum}, {Bhatawdekar}, {Boyett}, {Chen}, {Chevallard}, {Circosta}, {Curti}, {Danhaive}, {DeCoursey}, {de Graaff}, {Dressler}, {Egami}, {Helton}, {Hviding}, {Ji}, {Jones}, {Kumari}, {L{\"u}tzgendorf}, {Laseter}, {Looser}, {Lyu}, {Maseda}, {Nelson}, {Parlanti}, {Perna}, {Pusk{\'a}s}, {Rawle}, {Rodr{\'\i}guez Del Pino}, {Sandles}, {Saxena}, {Scholtz}, {Sharpe}, {Shivaei}, {Silcock}, {Simmonds}, {Skarbinski}, {Smit}, {Stone}, {Suess}, {Sun}, {Tang}, {Topping}, {{\"U}bler}, {Villanueva}, {Wallace}, {Whitler}, {Witstok}, \& {Woodrum}}]{Eisenstein2023OverviewJADES}
{Eisenstein}, D.~J., {Willott}, C., {Alberts}, S., {et~al.} 2023, arXiv e-prints, arXiv:2306.02465

\bibitem[{{Frieman} {et~al.}(2008){Frieman}, {Bassett}, {Becker}, {Choi}, {Cinabro}, {DeJongh}, {Depoy}, {Dilday}, {Doi}, {Garnavich}, {Hogan}, {Holtzman}, {Im}, {Jha}, {Kessler}, {Konishi}, {Lampeitl}, {Marriner}, {Marshall}, {McGinnis}, {Miknaitis}, {Nichol}, {Prieto}, {Riess}, {Richmond}, {Romani}, {Sako}, {Schneider}, {Smith}, {Takanashi}, {Tokita}, {van der Heyden}, {Yasuda}, {Zheng}, {Adelman-McCarthy}, {Annis}, {Assef}, {Barentine}, {Bender}, {Blandford}, {Boroski}, {Bremer}, {Brewington}, {Collins}, {Crotts}, {Dembicky}, {Eastman}, {Edge}, {Edmondson}, {Elson}, {Eyler}, {Filippenko}, {Foley}, {Frank}, {Goobar}, {Gueth}, {Gunn}, {Harvanek}, {Hopp}, {Ihara}, {Ivezi{\'c}}, {Kahn}, {Kaplan}, {Kent}, {Ketzeback}, {Kleinman}, {Kollatschny}, {Kron}, {Krzesi{\'n}ski}, {Lamenti}, {Leloudas}, {Lin}, {Long}, {Lucey}, {Lupton}, {Malanushenko}, {Malanushenko}, {McMillan}, {Mendez}, {Morgan}, {Morokuma}, {Nitta}, {Ostman}, {Pan}, {Rockosi}, {Romer}, {Ruiz-Lapuente}, {Saurage}, {Schlesinger}, {Snedden}, {Sollerman},
  {Stoughton}, {Stritzinger}, {Subba Rao}, {Tucker}, {Vaisanen}, {Watson}, {Watters}, {Wheeler}, {Yanny}, \& {York}}]{Frieman2008THESUMMARY}
{Frieman}, J.~A., {Bassett}, B., {Becker}, A., {et~al.} 2008, \aj, 135, 338

\bibitem[{{Frye} {et~al.}(2024){Frye}, {Pascale}, {Pierel}, {Chen}, {Foo}, {Leimbach}, {Garuda}, {Cohen}, {Kamieneski}, {Windhorst}, {Koekemoer}, {Kelly}, {Summers}, {Engesser}, {Liu}, {Furtak}, {Polletta}, {Harrington}, {Willner}, {Diego}, {Jansen}, {Coe}, {Conselice}, {Dai}, {Dole}, {D'Silva}, {Driver}, {Grogin}, {Marshall}, {Meena}, {Nonino}, {Ortiz}, {Pirzkal}, {Robotham}, {Ryan}, {Strolger}, {Tompkins}, {Willmer}, {Yan}, {Yun}, \& {Zitrin}}]{Frye2023TheG165.7+67.0}
{Frye}, B.~L., {Pascale}, M., {Pierel}, J., {et~al.} 2024, \apj, 961, 171

\bibitem[{{Garilli} {et~al.}(2021){Garilli}, {McLure}, {Pentericci}, {Franzetti}, {Gargiulo}, {Carnall}, {Cucciati}, {Iovino}, {Amorin}, {Bolzonella}, {Bongiorno}, {Castellano}, {Cimatti}, {Cirasuolo}, {Cullen}, {Dunlop}, {Elbaz}, {Finkelstein}, {Fontana}, {Fontanot}, {Fumana}, {Guaita}, {Hartley}, {Jarvis}, {Juneau}, {Maccagni}, {McLeod}, {Nandra}, {Pompei}, {Pozzetti}, {Scodeggio}, {Talia}, {Calabr{\`o}}, {Cresci}, {Fynbo}, {Hathi}, {Hibon}, {Koekemoer}, {Magliocchetti}, {Salvato}, {Vietri}, {Zamorani}, {Almaini}, {Balestra}, {Bardelli}, {Begley}, {Brammer}, {Bell}, {Bowler}, {Brusa}, {Buitrago}, {Caputi}, {Cassata}, {Charlot}, {Citro}, {Cristiani}, {Curtis-Lake}, {Dickinson}, {Fazio}, {Ferguson}, {Fiore}, {Franco}, {Georgakakis}, {Giavalisco}, {Grazian}, {Hamadouche}, {Jung}, {Kim}, {Khusanova}, {Le F{\`e}vre}, {Longhetti}, {Lotz}, {Mannucci}, {Maltby}, {Matsuoka}, {Mendez-Hernandez}, {Mendez-Abreu}, {Mignoli}, {Moresco}, {Nonino}, {Pannella}, {Papovich}, {Popesso}, {Roberts-Borsani}, {Rosario},
  {Saldana-Lopez}, {Santini}, {Saxena}, {Schaerer}, {Schreiber}, {Stark}, {Tasca}, {Thomas}, {Vanzella}, {Wild}, {Williams}, \& {Zucca}}]{Garilli2021TheMeasurements}
{Garilli}, B., {McLure}, R., {Pentericci}, L., {et~al.} 2021, \aap, 647, A150

\bibitem[{{Giavalisco} {et~al.}(2004){Giavalisco}, {Ferguson}, {Koekemoer}, {Dickinson}, {Alexander}, {Bauer}, {Bergeron}, {Biagetti}, {Brandt}, {Casertano}, {Cesarsky}, {Chatzichristou}, {Conselice}, {Cristiani}, {Da Costa}, {Dahlen}, {de Mello}, {Eisenhardt}, {Erben}, {Fall}, {Fassnacht}, {Fosbury}, {Fruchter}, {Gardner}, {Grogin}, {Hook}, {Hornschemeier}, {Idzi}, {Jogee}, {Kretchmer}, {Laidler}, {Lee}, {Livio}, {Lucas}, {Madau}, {Mobasher}, {Moustakas}, {Nonino}, {Padovani}, {Papovich}, {Park}, {Ravindranath}, {Renzini}, {Richardson}, {Riess}, {Rosati}, {Schirmer}, {Schreier}, {Somerville}, {Spinrad}, {Stern}, {Stiavelli}, {Strolger}, {Urry}, {Vandame}, {Williams}, \& {Wolf}}]{Giavalisco2004TheImaging}
{Giavalisco}, M., {Ferguson}, H.~C., {Koekemoer}, A.~M., {et~al.} 2004, \apjl, 600, L93

\bibitem[{{Gilliland} {et~al.}(1999){Gilliland}, {Nugent}, \& {Phillips}}]{Gilliland1999}
{Gilliland}, R.~L., {Nugent}, P.~E., \& {Phillips}, M.~M. 1999, \apj, 521, 30, \dodoi{10.1086/307549}

\bibitem[{{Golubchik} {et~al.}(2023){Golubchik}, {Zitrin}, {Pierel}, {Furtak}, {Meena}, {Graur}, {Kelly}, {Coe}, {Andrade-Santos}, {Asif}, {Bradley}, {Chen}, {Frye}, {Gomez}, {Jha}, {Mahler}, {Nonino}, {Strolger}, \& {Su}}]{Golubchik2023ASupernovae}
{Golubchik}, M., {Zitrin}, A., {Pierel}, J., {et~al.} 2023, \mnras, 522, 4718

\bibitem[{{Graur} {et~al.}(2014){Graur}, {Rodney}, {Maoz}, {Riess}, {Jha}, {Postman}, {Dahlen}, {Holoien}, {McCully}, {Patel}, {Strolger}, {Ben{\'\i}tez}, {Coe}, {Jouvel}, {Medezinski}, {Molino}, {Nonino}, {Bradley}, {Koekemoer}, {Balestra}, {Cenko}, {Clubb}, {Dickinson}, {Filippenko}, {Frederiksen}, {Garnavich}, {Hjorth}, {Jones}, {Leibundgut}, {Matheson}, {Mobasher}, {Rosati}, {Silverman}, {U}, {Jedruszczuk}, {Li}, {Lin}, {Mirmelstein}, {Neustadt}, {Ovadia}, \& {Rogers}}]{Graur2014}
{Graur}, O., {Rodney}, S.~A., {Maoz}, D., {et~al.} 2014, \apj, 783, 28

\bibitem[{{Grogin} {et~al.}(2011){Grogin}, {Kocevski}, {Faber}, {Ferguson}, {Koekemoer}, {Riess}, {Acquaviva}, {Alexander}, {Almaini}, {Ashby}, {Barden}, {Bell}, {Bournaud}, {Brown}, {Caputi}, {Casertano}, {Cassata}, {Castellano}, {Challis}, {Chary}, {Cheung}, {Cirasuolo}, {Conselice}, {Roshan Cooray}, {Croton}, {Daddi}, {Dahlen}, {Dav{\'e}}, {de Mello}, {Dekel}, {Dickinson}, {Dolch}, {Donley}, {Dunlop}, {Dutton}, {Elbaz}, {Fazio}, {Filippenko}, {Finkelstein}, {Fontana}, {Gardner}, {Garnavich}, {Gawiser}, {Giavalisco}, {Grazian}, {Guo}, {Hathi}, {H{\"a}ussler}, {Hopkins}, {Huang}, {Huang}, {Jha}, {Kartaltepe}, {Kirshner}, {Koo}, {Lai}, {Lee}, {Li}, {Lotz}, {Lucas}, {Madau}, {McCarthy}, {McGrath}, {McIntosh}, {McLure}, {Mobasher}, {Moustakas}, {Mozena}, {Nandra}, {Newman}, {Niemi}, {Noeske}, {Papovich}, {Pentericci}, {Pope}, {Primack}, {Rajan}, {Ravindranath}, {Reddy}, {Renzini}, {Rix}, {Robaina}, {Rodney}, {Rosario}, {Rosati}, {Salimbeni}, {Scarlata}, {Siana}, {Simard}, {Smidt}, {Somerville}, {Spinrad},
  {Straughn}, {Strolger}, {Telford}, {Teplitz}, {Trump}, {van der Wel}, {Villforth}, {Wechsler}, {Weiner}, {Wiklind}, {Wild}, {Wilson}, {Wuyts}, {Yan}, \& {Yun}}]{Grogin2011CANDELS:Survey}
{Grogin}, N.~A., {Kocevski}, D.~D., {Faber}, S.~M., {et~al.} 2011, \apjs, 197, 35

\bibitem[{{Gupta} {et~al.}(2016){Gupta}, {Kuhlmann}, {Kovacs}, {Spinka}, {Kessler}, {Goldstein}, {Liotine}, {Pomian}, {D'Andrea}, {Sullivan}, {Carretero}, {Castander}, {Nichol}, {Finley}, {Fischer}, {Foley}, {Kim}, {Papadopoulos}, {Sako}, {Scolnic}, {Smith}, {Tucker}, {Uddin}, {Wolf}, {Yuan}, {Abbott}, {Abdalla}, {Benoit-L{\'e}vy}, {Bertin}, {Brooks}, {Carnero Rosell}, {Carrasco Kind}, {Cunha}, {da Costa}, {Desai}, {Doel}, {Eifler}, {Evrard}, {Flaugher}, {Fosalba}, {Gazta{\~n}aga}, {Gruen}, {Gruendl}, {James}, {Kuehn}, {Kuropatkin}, {Maia}, {Marshall}, {Miquel}, {Plazas}, {Romer}, {S{\'a}nchez}, {Schubnell}, {Sevilla-Noarbe}, {Sobreira}, {Suchyta}, {Swanson}, {Tarle}, {Walker}, \& {Wester}}]{Gupta2016HostSurveys}
{Gupta}, R.~R., {Kuhlmann}, S., {Kovacs}, E., {et~al.} 2016, \aj, 152, 154

\bibitem[{{Guy} {et~al.}(2007){Guy}, {Astier}, {Baumont}, {Hardin}, {Pain}, {Regnault}, {Basa}, {Carlberg}, {Conley}, {Fabbro}, {Fouchez}, {Hook}, {Howell}, {Perrett}, {Pritchet}, {Rich}, {Sullivan}, {Antilogus}, {Aubourg}, {Bazin}, {Bronder}, {Filiol}, {Palanque-Delabrouille}, {Ripoche}, \& {Ruhlmann-Kleider}}]{Guy2007SALT2}
{Guy}, J., {Astier}, P., {Baumont}, S., {et~al.} 2007, \aap, 466, 11, \dodoi{10.1051/0004-6361:20066930}

\bibitem[{{Hainline} {et~al.}(2024){Hainline}, {Johnson}, {Robertson}, {Tacchella}, {Helton}, {Sun}, {Eisenstein}, {Simmonds}, {Topping}, {Whitler}, {Willmer}, {Rieke}, {Suess}, {Hviding}, {Cameron}, {Alberts}, {Baker}, {Baum}, {Bhatawdekar}, {Bonaventura}, {Boyett}, {Bunker}, {Carniani}, {Charlot}, {Chevallard}, {Chen}, {Curti}, {Curtis-Lake}, {D'Eugenio}, {Egami}, {Endsley}, {Hausen}, {Ji}, {Looser}, {Lyu}, {Maiolino}, {Nelson}, {Pusk{\'a}s}, {Rawle}, {Sandles}, {Saxena}, {Smit}, {Stark}, {Williams}, {Willott}, \& {Witstok}}]{Hainline2023TheGOODS-N}
{Hainline}, K.~N., {Johnson}, B.~D., {Robertson}, B., {et~al.} 2024, \apj, 964, 71

\bibitem[{{Hamuy} {et~al.}(2006){Hamuy}, {Folatelli}, {Morrell}, {Phillips}, {Suntzeff}, {Persson}, {Roth}, {Gonzalez}, {Krzeminski}, {Contreras}, {Freedman}, {Murphy}, {Madore}, {Wyatt}, {Maza}, {Filippenko}, {Li}, \& {Pinto}}]{Hamuy2006TheSurvey}
{Hamuy}, M., {Folatelli}, G., {Morrell}, N.~I., {et~al.} 2006, \pasp, 118, 2

\bibitem[{{Hayes} {et~al.}(2024){Hayes}, {Tan}, {Ellis}, {Young}, {Cammelli}, {Singh}, {Runnholm}, {Saxena}, {Lunnan}, {Keller}, {Monaco}, {Laporte}, \& {Melinder}}]{Hayes2024GlimmersVariability}
{Hayes}, M.~J., {Tan}, J.~C., {Ellis}, R.~S., {et~al.} 2024, arXiv e-prints, arXiv:2403.16138

\bibitem[{{Hsiao} {et~al.}(2007){Hsiao}, {Conley}, {Howell}, {Sullivan}, {Pritchet}, {Carlberg}, {Nugent}, \& {Phillips}}]{Hsiao2007K-CorrectionsSupernovae}
{Hsiao}, E.~Y., {Conley}, A., {Howell}, D.~A., {et~al.} 2007, \apj, 663, 1187

\bibitem[{{Inami} {et~al.}(2017){Inami}, {Bacon}, {Brinchmann}, {Richard}, {Contini}, {Conseil}, {Hamer}, {Akhlaghi}, {Bouch{\'e}}, {Cl{\'e}ment}, {Desprez}, {Drake}, {Hashimoto}, {Leclercq}, {Maseda}, {Michel-Dansac}, {Paalvast}, {Tresse}, {Ventou}, {Kollatschny}, {Boogaard}, {Finley}, {Marino}, {Schaye}, \& {Wisotzki}}]{Inami2017TheGalaxies}
{Inami}, H., {Bacon}, R., {Brinchmann}, J., {et~al.} 2017, \aap, 608, A2

\bibitem[{{Ji} {et~al.}(2023){Ji}, {Williams}, {Tacchella}, {Suess}, {Baker}, {Alberts}, {Bunker}, {Johnson}, {Robertson}, {Sun}, {Eisenstein}, {Rieke}, {Maseda}, {Hainline}, {Hausen}, {Rieke}, {Willmer}, {Egami}, {Shivaei}, {Carniani}, {Charlot}, {Chevallard}, {Curtis-Lake}, {Looser}, {Maiolino}, {Willott}, {Chen}, {Helton}, {Lyu}, {Nelson}, {Bhatawdekar}, {Boyett}, \& {Sandles}}]{Ji2023JADES4.5}
{Ji}, Z., {Williams}, C.~C., {Tacchella}, S., {et~al.} 2023, arXiv e-prints, arXiv:2305.18518

\bibitem[{{Kasen} {et~al.}(2011){Kasen}, {Woosley}, \& {Heger}}]{Kasen2011PairBreakout}
{Kasen}, D., {Woosley}, S.~E., \& {Heger}, A. 2011, \apj, 734, 102

\bibitem[{{Kenworthy} {et~al.}(2021){Kenworthy}, {Jones}, {Dai}, {Kessler}, {Scolnic}, {Brout}, {Siebert}, {Pierel}, {Dettman}, {Dimitriadis}, {Foley}, {Jha}, {Pan}, {Riess}, {Rodney}, \& {Rojas-Bravo}}]{Kenworthy2021SALT3}
{Kenworthy}, W.~D., {Jones}, D.~O., {Dai}, M., {et~al.} 2021, \apj, 923, 265, \dodoi{10.3847/1538-4357/ac30d8}

\bibitem[{{Kessler} {et~al.}(2009){Kessler}, {Bernstein}, {Cinabro}, {Dilday}, {Frieman}, {Jha}, {Kuhlmann}, {Miknaitis}, {Sako}, {Taylor}, \& {Vanderplas}}]{Kessler2009SNANA:Analysis}
{Kessler}, R., {Bernstein}, J.~P., {Cinabro}, D., {et~al.} 2009, \pasp, 121, 1028

\bibitem[{{Kool} {et~al.}(2018){Kool}, {Ryder}, {Kankare}, {Mattila}, {Reynolds}, {McDermid}, {P{\'e}rez-Torres}, {Herrero-Illana}, {Schirmer}, {Efstathiou}, {Bauer}, {Kotilainen}, {V{\"a}is{\"a}nen}, {Baldwin}, {Romero-Ca{\~n}izales}, \& {Alberdi}}]{Kool2018FirstDetection}
{Kool}, E.~C., {Ryder}, S., {Kankare}, E., {et~al.} 2018, \mnras, 473, 5641

\bibitem[{{Le F{\`e}vre} {et~al.}(2013){Le F{\`e}vre}, {Cassata}, {Cucciati}, {Garilli}, {Ilbert}, {Le Brun}, {Maccagni}, {Moreau}, {Scodeggio}, {Tresse}, {Zamorani}, {Adami}, {Arnouts}, {Bardelli}, {Bolzonella}, {Bondi}, {Bongiorno}, {Bottini}, {Cappi}, {Charlot}, {Ciliegi}, {Contini}, {de la Torre}, {Foucaud}, {Franzetti}, {Gavignaud}, {Guzzo}, {Iovino}, {Lemaux}, {L{\'o}pez-Sanjuan}, {McCracken}, {Marano}, {Marinoni}, {Mazure}, {Mellier}, {Merighi}, {Merluzzi}, {Paltani}, {Pell{\`o}}, {Pollo}, {Pozzetti}, {Scaramella}, {Tasca}, {Vergani}, {Vettolani}, {Zanichelli}, \& {Zucca}}]{LeFevre2013The24.75}
{Le F{\`e}vre}, O., {Cassata}, P., {Cucciati}, O., {et~al.} 2013, \aap, 559, A14

\bibitem[{{Le F{\`e}vre} {et~al.}(2015){Le F{\`e}vre}, {Tasca}, {Cassata}, {Garilli}, {Le Brun}, {Maccagni}, {Pentericci}, {Thomas}, {Vanzella}, {Zamorani}, {Zucca}, {Amorin}, {Bardelli}, {Capak}, {Cassar{\`a}}, {Castellano}, {Cimatti}, {Cuby}, {Cucciati}, {de la Torre}, {Durkalec}, {Fontana}, {Giavalisco}, {Grazian}, {Hathi}, {Ilbert}, {Lemaux}, {Moreau}, {Paltani}, {Ribeiro}, {Salvato}, {Schaerer}, {Scodeggio}, {Sommariva}, {Talia}, {Taniguchi}, {Tresse}, {Vergani}, {Wang}, {Charlot}, {Contini}, {Fotopoulou}, {L{\'o}pez-Sanjuan}, {Mellier}, \& {Scoville}}]{LeFevre2015The6}
{Le F{\`e}vre}, O., {Tasca}, L.~A.~M., {Cassata}, P., {et~al.} 2015, \aap, 576, A79

\bibitem[{{Levan} {et~al.}(2005){Levan}, {Nugent}, {Fruchter}, {Burud}, {Branch}, {Rhoads}, {Castro-Tirado}, {Gorosabel}, {Castro Cer{\'o}n}, {Thorsett}, {Kouveliotou}, {Golenetskii}, {Fynbo}, {Garnavich}, {Holland}, {Hjorth}, {M{\o}ller}, {Pian}, {Tanvir}, {Ulanov}, {Wijers}, \& {Woosley}}]{Levan2005}
{Levan}, A., {Nugent}, P., {Fruchter}, A., {et~al.} 2005, \apj, 624, 880, \dodoi{10.1086/428657}

\bibitem[{{Lyu} {et~al.}(2022){Lyu}, {Alberts}, {Rieke}, \& {Rujopakarn}}]{Lyu2022AGNRadio}
{Lyu}, J., {Alberts}, S., {Rieke}, G.~H., \& {Rujopakarn}, W. 2022, \apj, 941, 191

\bibitem[{{Lyu} {et~al.}(2024){Lyu}, {Alberts}, {Rieke}, {Shivaei}, {P{\'e}rez-Gonz{\'a}lez}, {Sun}, {Hainline}, {Baum}, {Bonaventura}, {Bunker}, {Egami}, {Eisenstein}, {Florian}, {Ji}, {Johnson}, {Morrison}, {Rieke}, {Robertson}, {Rujopakarn}, {Tacchella}, {Scholtz}, \& {Willmer}}]{Lyu2024ActiveJWST/MIRI}
{Lyu}, J., {Alberts}, S., {Rieke}, G.~H., {et~al.} 2024, \apj, 966, 229

\bibitem[{{Mignoli} {et~al.}(2005){Mignoli}, {Cimatti}, {Zamorani}, {Pozzetti}, {Daddi}, {Renzini}, {Broadhurst}, {Cristiani}, {D'Odorico}, {Fontana}, {Giallongo}, {Gilmozzi}, {Menci}, \& {Saracco}}]{Mignoli2005ThePopulation}
{Mignoli}, M., {Cimatti}, A., {Zamorani}, G., {et~al.} 2005, \aap, 437, 883

\bibitem[{{Miralda-Escud{\'e}} \& {Rees}(1997)}]{Escude1997}
{Miralda-Escud{\'e}}, J., \& {Rees}, M.~J. 1997, \apjl, 478, L57

\bibitem[{{Momcheva} {et~al.}(2016){Momcheva}, {Brammer}, {van Dokkum}, {Skelton}, {Whitaker}, {Nelson}, {Fumagalli}, {Maseda}, {Leja}, {Franx}, {Rix}, {Bezanson}, {Da Cunha}, {Dickey}, {F{\"o}rster Schreiber}, {Illingworth}, {Kriek}, {Labb{\'e}}, {Ulf Lange}, {Lundgren}, {Magee}, {Marchesini}, {Oesch}, {Pacifici}, {Patel}, {Price}, {Tal}, {Wake}, {van der Wel}, \& {Wuyts}}]{Momcheva2016TheGalaxies}
{Momcheva}, I.~G., {Brammer}, G.~B., {van Dokkum}, P.~G., {et~al.} 2016, \apjs, 225, 27

\bibitem[{{Moriya} {et~al.}(2022){Moriya}, {Quimby}, \& {Robertson}}]{Moriya2022DiscoveringTelescope}
{Moriya}, T.~J., {Quimby}, R.~M., \& {Robertson}, B.~E. 2022, \apj, 925, 211

\bibitem[{{Moriya} {et~al.}(2019){Moriya}, {Tanaka}, {Yasuda}, {Jiang}, {Lee}, {Maeda}, {Morokuma}, {Nomoto}, {Quimby}, {Suzuki}, {Takahashi}, {Tanaka}, {Tominaga}, {Yamaguchi}, {Bernard}, {Cooke}, {Curtin}, {Galbany}, {Gonz{\'a}lez-Gait{\'a}n}, {Pignata}, {Pritchard}, {Komiyama}, \& {Lupton}}]{Moriya2019FirstProperties}
{Moriya}, T.~J., {Tanaka}, M., {Yasuda}, N., {et~al.} 2019, \apjs, 241, 16

\bibitem[{{Morrell}(2012)}]{Morrell2012CarnegieSupernovae}
{Morrell}, N.~I. 2012, in Death of Massive Stars: Supernovae and Gamma-Ray Bursts, ed. P.~{Roming}, N.~{Kawai}, \& E.~{Pian}, Vol. 279, 361--362

\bibitem[{{O'Brien} {et~al.}(2024){O'Brien}, {Jansen}, {Grogin}, {Cohen}, {Smith}, {Silver}, {Maksym}, {Windhorst}, {Carleton}, {Koekemoer}, {Hathi}, {Willmer}, {Frye}, {Alpaslan}, {Ashby}, {Ashcraft}, {Bonoli}, {Brisken}, {Cappelluti}, {Civano}, {Conselice}, {Dhillon}, {Driver}, {Duncan}, {Dupke}, {Elvis}, {Fazio}, {Finkelstein}, {Gim}, {Griffiths}, {Hammel}, {Hyun}, {Im}, {Jones}, {Kim}, {Ladjelate}, {Larson}, {Malhotra}, {Marshall}, {Milam}, {Pierel}, {Rhoads}, {Rodney}, {R{\"o}ttgering}, {Rutkowski}, {Ryan}, {Ward}, {White}, {van Weeren}, {Zhao}, {Summers}, {D'Silva}, {Ortiz}, {Robotham}, {Coe}, {Nonino}, {Pirzkal}, {Yan}, \& {Acharya}}]{OBrien2024}
{O'Brien}, R., {Jansen}, R.~A., {Grogin}, N.~A., {et~al.} 2024, \apjs, 272, 19

\bibitem[{{Oesch} {et~al.}(2023){Oesch}, {Brammer}, {Naidu}, {Bouwens}, {Chisholm}, {Illingworth}, {Matthee}, {Nelson}, {Qin}, {Reddy}, {Shapley}, {Shivaei}, {van Dokkum}, {Weibel}, {Whitaker}, {Wuyts}, {Covelo-Paz}, {Endsley}, {Fudamoto}, {Giovinazzo}, {Herard-Demanche}, {Kerutt}, {Kramarenko}, {Labbe}, {Leonova}, {Lin}, {Magee}, {Marchesini}, {Maseda}, {Mason}, {Matharu}, {Meyer}, {Neufeld}, {Prieto Lyon}, {Schaerer}, {Sharma}, {Shuntov}, {Smit}, {Stefanon}, {Wyithe}, \& {Xiao}}]{Oesch2023TheFields}
{Oesch}, P.~A., {Brammer}, G., {Naidu}, R.~P., {et~al.} 2023, \mnras, 525, 2864

\bibitem[{{Oke} \& {Gunn}(1983)}]{Oke1983SecondarySpectrophotometry.}
{Oke}, J.~B., \& {Gunn}, J.~E. 1983, \apj, 266, 713

\bibitem[{{Pan} {et~al.}(2017){Pan}, {Foley}, {Smith}, {Galbany}, {D'Andrea}, {Gonz{\'a}lez-Gait{\'a}n}, {Jarvis}, {Kessler}, {Kovacs}, {Lidman}, {Papadopoulos}, {Sako}, {Sullivan}, {Abbott}, {Abdalla}, {Annis}, {Bechtol}, {Benoit-L{\'e}vy}, {Brooks}, {Buckley-Geer}, {Burke}, {Carnero Rosell}, {Carrasco Kind}, {Carretero}, {Castander}, {Cunha}, {da Costa}, {Desai}, {Diehl}, {Doel}, {Eifler}, {Finley}, {Flaugher}, {Frieman}, {Garc{\'\i}a-Bellido}, {Goldstein}, {Gruen}, {Gruendl}, {Gschwend}, {Gutierrez}, {James}, {Kim}, {Krause}, {Kuehn}, {Kuropatkin}, {Lahav}, {Lima}, {Maia}, {March}, {Marshall}, {Martini}, {Miquel}, {Nugent}, {Plazas}, {Romer}, {Sanchez}, {Scarpine}, {Schubnell}, {Sevilla-Noarbe}, {Smith}, {Sobreira}, {Suchyta}, {Swanson}, {Thomas}, {Walker}, \& {DES Collaboration}}]{Pan2017DES15E2mlf:Bang}
{Pan}, Y.~C., {Foley}, R.~J., {Smith}, M., {et~al.} 2017, \mnras, 470, 4241

\bibitem[{{Pierel} {et~al.}(2018){Pierel}, {Rodney}, {Avelino}, {Bianco}, {Filippenko}, {Foley}, {Friedman}, {Hicken}, {Hounsell}, {Jha}, {Kessler}, {Kirshner}, {Mandel}, {Narayan}, {Scolnic}, \& {Strolger}}]{Pierel2018ExtendingObservations}
{Pierel}, J.~D.~R., {Rodney}, S., {Avelino}, A., {et~al.} 2018, \pasp, 130, 114504

\bibitem[{{Pierel} {et~al.}(2022){Pierel}, {Jones}, {Kenworthy}, {Dai}, {Kessler}, {Ashall}, {Do}, {Peterson}, {Shappee}, {Siebert}, {Barna}, {Brink}, {Burke}, {Calamida}, {Camacho-Neves}, {de Jaeger}, {Filippenko}, {Foley}, {Galbany}, {Fox}, {Gomez}, {Hiramatsu}, {Hounsell}, {Howell}, {Jha}, {Kwok}, {P{\'e}rez-Fournon}, {Poidevin}, {Rest}, {Rubin}, {Scolnic}, {Shirley}, {Strolger}, {Tinyanont}, \& {Wang}}]{Pierel2022SALT3-NIR:Measurements}
{Pierel}, J.~D.~R., {Jones}, D.~O., {Kenworthy}, W.~D., {et~al.} 2022, \apj, 939, 11

\bibitem[{{Pierel} {et~al.}(2024{\natexlab{a}}){Pierel}, {Newman}, {Dhawan}, {Gu}, {Joshi}, {Li}, {Schuldt}, {Strolger}, {Suyu}, {Caminha}, {Cohen}, {Diego}, {D{\'S}ilva}, {Ertl}, {Frye}, {Granata}, {Grillo}, {Koekemoer}, {Li}, {Robotham}, {Summers}, {Treu}, {Windhorst}, {Zitrin}, {Agarwal}, {Agrawal}, {Arendse}, {Belli}, {Burns}, {Ca{\~n}ameras}, {Chakrabarti}, {Chen}, {Collett}, {Coulter}, {Ellis}, {Engesser}, {Foo}, {Fox}, {Gall}, {Garuda}, {Gezari}, {Gomez}, {Glazebrook}, {Hjorth}, {Huang}, {Jha}, {Kamieneski}, {Kelly}, {Larison}, {Moustakas}, {Pascale}, {P{\'e}rez-Fournon}, {Petrushevska}, {Poidevin}, {Rest}, {Shahbandeh}, {Shajib}, {Siebert}, {Storfer}, {Talbot}, {Wang}, {Wevers}, \& {Zenati}}]{Pierel2024LensedGalaxy}
{Pierel}, J.~D.~R., {Newman}, A.~B., {Dhawan}, S., {et~al.} 2024{\natexlab{a}}, \apjl, 967, L37

\bibitem[{{Pierel} {et~al.}(2024{\natexlab{b}}){Pierel}, {Frye}, {Pascale}, {Caminha}, {Chen}, {Dhawan}, {Gilman}, {Grayling}, {Huber}, {Kelly}, {Thorp}, {Arendse}, {Birrer}, {Bronikowski}, {Ca{\~n}ameras}, {Coe}, {Cohen}, {Conselice}, {Driver}, {D{\'S}ilva}, {Engesser}, {Foo}, {Gall}, {Garuda}, {Grillo}, {Grogin}, {Henderson}, {Hjorth}, {Jansen}, {Johansson}, {Kamieneski}, {Koekemoer}, {Larison}, {Marshall}, {Moustakas}, {Nonino}, {Ortiz}, {Petrushevska}, {Pirzkal}, {Robotham}, {Ryan}, {Schuldt}, {Strolger}, {Summers}, {Suyu}, {Treu}, {Willmer}, {Windhorst}, {Yan}, {Zitrin}, {Acebron}, {Chakrabarti}, {Coulter}, {Fox}, {Huang}, {Jha}, {Li}, {Mazzali}, {Meena}, {P{\'e}rez-Fournon}, {Poidevin}, {Rest}, \& {Riess}}]{Pierel2024JWST1.78}
{Pierel}, J.~D.~R., {Frye}, B.~L., {Pascale}, M., {et~al.} 2024{\natexlab{b}}, \apj, 967, 50

\bibitem[{{Pierel} {et~al.}(2024{\natexlab{c}}){Pierel}, {Engesser}, {Coulter}, {Decoursey}, {Siebert}, {Rest}, {Egami}, {Chen}, {Fox}, {Jones}, {Joshi}, {Moriya}, {Zenati}, {Bunker}, {Cargile}, {Curti}, {Eisenstein}, {Gezari}, {Gomez}, {Guolo}, {Johnson}, {Karmen}, {Maiolino}, {Quimby}, {Robertson}, {Shahbandeh}, {Strolger}, {Sun}, {Wang}, \& {Wevers}}]{Pierel24Ia}
{Pierel}, J.~D.~R., {Engesser}, M., {Coulter}, D.~A., {et~al.} 2024{\natexlab{c}}, arXiv e-prints, arXiv:2406.05089

\bibitem[{{Postman} {et~al.}(2012){Postman}, {Coe}, {Ben{\'\i}tez}, {Bradley}, {Broadhurst}, {Donahue}, {Ford}, {Graur}, {Graves}, {Jouvel}, {Koekemoer}, {Lemze}, {Medezinski}, {Molino}, {Moustakas}, {Ogaz}, {Riess}, {Rodney}, {Rosati}, {Umetsu}, {Zheng}, {Zitrin}, {Bartelmann}, {Bouwens}, {Czakon}, {Golwala}, {Host}, {Infante}, {Jha}, {Jimenez-Teja}, {Kelson}, {Lahav}, {Lazkoz}, {Maoz}, {McCully}, {Melchior}, {Meneghetti}, {Merten}, {Moustakas}, {Nonino}, {Patel}, {Reg{\"o}s}, {Sayers}, {Seitz}, \& {Van der Wel}}]{Postman2012}
{Postman}, M., {Coe}, D., {Ben{\'\i}tez}, N., {et~al.} 2012, \apjs, 199, 25

\bibitem[{{Rieke} {et~al.}(2023){Rieke}, {Robertson}, {Tacchella}, {Hainline}, {Johnson}, {Hausen}, {Ji}, {Willmer}, {Eisenstein}, {Pusk{\'a}s}, {Alberts}, {Arribas}, {Baker}, {Baum}, {Bhatawdekar}, {Bonaventura}, {Boyett}, {Bunker}, {Cameron}, {Carniani}, {Charlot}, {Chevallard}, {Chen}, {Curti}, {Curtis-Lake}, {Danhaive}, {DeCoursey}, {Dressler}, {Egami}, {Endsley}, {Helton}, {Hviding}, {Kumari}, {Looser}, {Lyu}, {Maiolino}, {Maseda}, {Nelson}, {Rieke}, {Rix}, {Sandles}, {Saxena}, {Sharpe}, {Shivaei}, {Skarbinski}, {Smit}, {Stark}, {Stone}, {Suess}, {Sun}, {Topping}, {{\"U}bler}, {Villanueva}, {Wallace}, {Williams}, {Willott}, {Whitler}, {Witstok}, \& {Woodrum}}]{Rieke2023JADESImaging}
{Rieke}, M.~J., {Robertson}, B., {Tacchella}, S., {et~al.} 2023, \apjs, 269, 16

\bibitem[{{Rodney} {et~al.}(2014){Rodney}, {Riess}, {Strolger}, {Dahlen}, {Graur}, {Casertano}, {Dickinson}, {Ferguson}, {Garnavich}, {Hayden}, {Jha}, {Jones}, {Kirshner}, {Koekemoer}, {McCully}, {Mobasher}, {Patel}, {Weiner}, {Cenko}, {Clubb}, {Cooper}, {Filippenko}, {Frederiksen}, {Hjorth}, {Leibundgut}, {Matheson}, {Nayyeri}, {Penner}, {Trump}, {Silverman}, {U}, {Azalee Bostroem}, {Challis}, {Rajan}, {Wolff}, {Faber}, {Grogin}, \& {Kocevski}}]{Rodney2014TypeUniverse}
{Rodney}, S.~A., {Riess}, A.~G., {Strolger}, L.-G., {et~al.} 2014, \aj, 148, 13

\bibitem[{{Sako} {et~al.}(2008){Sako}, {Bassett}, {Becker}, {Cinabro}, {DeJongh}, {Depoy}, {Dilday}, {Doi}, {Frieman}, {Garnavich}, {Hogan}, {Holtzman}, {Jha}, {Kessler}, {Konishi}, {Lampeitl}, {Marriner}, {Miknaitis}, {Nichol}, {Prieto}, {Riess}, {Richmond}, {Romani}, {Schneider}, {Smith}, {SubbaRao}, {Takanashi}, {Tokita}, {van der Heyden}, {Yasuda}, {Zheng}, {Barentine}, {Brewington}, {Choi}, {Dembicky}, {Harnavek}, {Ihara}, {Im}, {Ketzeback}, {Kleinman}, {Krzesi{\'n}ski}, {Long}, {Malanushenko}, {Malanushenko}, {McMillan}, {Morokuma}, {Nitta}, {Pan}, {Saurage}, \& {Snedden}}]{Sako2008THEOBSERVATIONS}
{Sako}, M., {Bassett}, B., {Becker}, A., {et~al.} 2008, \aj, 135, 348

\bibitem[{{Siebert} {et~al.}(2024){Siebert}, {Decoursey}, {Coulter}, {Engesser}, {Pierel}, {Rest}, {Egami}, {Shahbandeh}, {Chen}, {Fox}, {Zenati}, {Moriya}, {Bunker}, {Cargile}, {Curti}, {Eisenstein}, {Gezari}, {Gomez}, {Guolo}, {Johnson}, {Joshi}, {Karmen}, {Maiolino}, {Quimby}, {Robertson}, {Strolger}, {Sun}, {Wang}, \& {Wevers}}]{Siebert2024}
{Siebert}, M.~R., {Decoursey}, C., {Coulter}, D.~A., {et~al.} 2024, arXiv e-prints, arXiv:2406.05076

\bibitem[{{Skilling}(2004)}]{Skilling2004NestedSampling}
{Skilling}, J. 2004, in American Institute of Physics Conference Series, Vol. 735, Bayesian Inference and Maximum Entropy Methods in Science and Engineering: 24th International Workshop on Bayesian Inference and Maximum Entropy Methods in Science and Engineering, ed. R.~{Fischer}, R.~{Preuss}, \& U.~V. {Toussaint} (AIP), 395--405

\bibitem[{{Smith} {et~al.}(2018){Smith}, {Sullivan}, {Nichol}, {Galbany}, {D'Andrea}, {Inserra}, {Lidman}, {Rest}, {Schirmer}, {Filippenko}, {Zheng}, {Cenko}, {Angus}, {Brown}, {Davis}, {Finley}, {Foley}, {Gonz{\'a}lez-Gait{\'a}n}, {Guti{\'e}rrez}, {Kessler}, {Kuhlmann}, {Marriner}, {M{\"o}ller}, {Nugent}, {Prajs}, {Thomas}, {Wolf}, {Zenteno}, {Abbott}, {Abdalla}, {Allam}, {Annis}, {Bechtol}, {Benoit-L{\'e}vy}, {Bertin}, {Brooks}, {Burke}, {Carnero Rosell}, {Carrasco Kind}, {Carretero}, {Castander}, {Crocce}, {Cunha}, {da Costa}, {Davis}, {Desai}, {Diehl}, {Doel}, {Eifler}, {Flaugher}, {Fosalba}, {Frieman}, {Garc{\'\i}a-Bellido}, {Gaztanaga}, {Gerdes}, {Goldstein}, {Gruen}, {Gruendl}, {Gschwend}, {Gutierrez}, {Honscheid}, {James}, {Johnson}, {Kuehn}, {Kuropatkin}, {Li}, {Lima}, {Maia}, {Marshall}, {Martini}, {Menanteau}, {Miller}, {Miquel}, {Ogando}, {Petravick}, {Plazas}, {Romer}, {Rykoff}, {Sako}, {Sanchez}, {Scarpine}, {Schindler}, {Schubnell}, {Sevilla-Noarbe}, {Smith}, {Soares-Santos}, {Sobreira},
  {Suchyta}, {Swanson}, {Tarle}, {Walker}, \& {DES Collaboration}}]{Smith2018StudyingTwo}
{Smith}, M., {Sullivan}, M., {Nichol}, R.~C., {et~al.} 2018, \apj, 854, 37

\bibitem[{{Stetson}(1987)}]{Stetson1987DAOPHOT:Photometry}
{Stetson}, P.~B. 1987, \pasp, 99, 191

\bibitem[{{Stritzinger} {et~al.}(2009){Stritzinger}, {Mazzali}, {Phillips}, {Immler}, {Soderberg}, {Sollerman}, {Boldt}, {Braithwaite}, {Brown}, {Burns}, {Contreras}, {Covarrubias}, {Folatelli}, {Freedman}, {Gonz{\'a}lez}, {Hamuy}, {Krzeminski}, {Madore}, {Milne}, {Morrell}, {Persson}, {Roth}, {Smith}, \& {Suntzeff}}]{Stritzinger2009THEWAVELENGTHS}
{Stritzinger}, M., {Mazzali}, P., {Phillips}, M.~M., {et~al.} 2009, \apj, 696, 713

\bibitem[{{Strolger} {et~al.}(2015){Strolger}, {Dahlen}, {Rodney}, {Graur}, {Riess}, {McCully}, {Ravindranath}, {Mobasher}, \& {Shahady}}]{Strolger2015TheSurveys}
{Strolger}, L.-G., {Dahlen}, T., {Rodney}, S.~A., {et~al.} 2015, \apj, 813, 93

\bibitem[{{Taddia} {et~al.}(2018){Taddia}, {Stritzinger}, {Bersten}, {Baron}, {Burns}, {Contreras}, {Holmbo}, {Hsiao}, {Morrell}, {Phillips}, {Sollerman}, \& {Suntzeff}}]{Taddia2018TheCurves}
{Taddia}, F., {Stritzinger}, M.~D., {Bersten}, M., {et~al.} 2018, \aap, 609, A136

\bibitem[{{Urrutia} {et~al.}(2019){Urrutia}, {Wisotzki}, {Kerutt}, {Schmidt}, {Herenz}, {Klar}, {Saust}, {Werhahn}, {Diener}, {Caruana}, {Krajnovi{\'c}}, {Bacon}, {Boogaard}, {Brinchmann}, {Enke}, {Maseda}, {Nanayakkara}, {Richard}, {Steinmetz}, \& {Weilbacher}}]{Urrutia2019TheRelease}
{Urrutia}, T., {Wisotzki}, L., {Kerutt}, J., {et~al.} 2019, \aap, 624, A141

\bibitem[{{Vanzella} {et~al.}(2008){Vanzella}, {Cristiani}, {Dickinson}, {Giavalisco}, {Kuntschner}, {Haase}, {Nonino}, {Rosati}, {Cesarsky}, {Ferguson}, {Fosbury}, {Grazian}, {Moustakas}, {Rettura}, {Popesso}, {Renzini}, {Stern}, \& {GOODS Team}}]{Vanzella2008TheIII}
{Vanzella}, E., {Cristiani}, S., {Dickinson}, M., {et~al.} 2008, \aap, 478, 83

\bibitem[{{Xue} {et~al.}(2011){Xue}, {Luo}, {Brandt}, {Bauer}, {Lehmer}, {Broos}, {Schneider}, {Alexander}, {Brusa}, {Comastri}, {Fabian}, {Gilli}, {Hasinger}, {Hornschemeier}, {Koekemoer}, {Liu}, {Mainieri}, {Paolillo}, {Rafferty}, {Rosati}, {Shemmer}, {Silverman}, {Smail}, {Tozzi}, \& {Vignali}}]{Xue2011TheCatalogs}
{Xue}, Y.~Q., {Luo}, B., {Brandt}, W.~N., {et~al.} 2011, \apjs, 195, 10

\bibitem[{{Yan} {et~al.}(2023{\natexlab{a}}){Yan}, {Wang}, {Ma}, \& {Hu}}]{Yan2023PointlikeRedshifts}
{Yan}, H., {Wang}, L., {Ma}, Z., \& {Hu}, L. 2023{\natexlab{a}}, \apjl, 947, L1

\bibitem[{{Yan} {et~al.}(2023{\natexlab{b}}){Yan}, {Ma}, {Sun}, {Wang}, {Kelly}, {Diego}, {Cohen}, {Windhorst}, {Jansen}, {Grogin}, {Beacom}, {Conselice}, {Driver}, {Frye}, {Coe}, {Marshall}, {Koekemoer}, {Willmer}, {Robotham}, {D'Silva}, {Summers}, {Nonino}, {Pirzkal}, {Ryan}, {Ortiz}, {Tompkins}, {Bhatawdekar}, {Cheng}, {Zitrin}, \& {Willner}}]{Yan2023JWSTsField}
{Yan}, H., {Ma}, Z., {Sun}, B., {et~al.} 2023{\natexlab{b}}, \apjs, 269, 43

\end{thebibliography}
\bibliographystyle{aasjournal}

\end{document}